\documentclass[aip,jcp,amsmath,amssymb,preprint]{revtex4-1}

\usepackage{graphicx}
\usepackage{dcolumn}
\usepackage{bm}
\usepackage{color}
\usepackage{textcomp}

\usepackage[utf8]{inputenc}
\usepackage[T1]{fontenc}
\usepackage{mathptmx}

\newcommand{\half}{\frac{1}{2}}
\newcommand{\naux}{N_{\mathrm{aux}}}
\newcommand{\neig}{N_{\mathrm{eig}}}
\newcommand{\ntri}{N_{\mathrm{trip}}}
\newcommand{\nmo}{N_{\mathrm{MO}}}

\begin{document}

\preprint{AIP/123-QED}

\title[Rank-reduced coupled-cluster theory]{Quintic-scaling rank-reduced coupled cluster theory 
with single and double excitations}

\author{Micha\l\ Lesiuk}
\email{m.lesiuk@uw.edu.pl}
\affiliation{\sl Faculty of Chemistry, University of Warsaw, Pasteura 1, 02-093 Warsaw~Poland}

\begin{abstract}
We consider the rank-reduced coupled-cluster theory with single and double excitations 
(RR-CCSD) introduced recently [Parrish \emph{et al.}, J. Chem. Phys. {\bf 150}, 164118 (2019)]. The 
main feature of this method is the decomposed form of the doubly-excited amplitudes which are 
expanded in the basis of largest magnitude eigenvectors of the MP2 or MP3 amplitudes. This approach 
enables a substantial compression of the amplitudes with only minor loss of accuracy. However, the 
formal scaling of the computational costs with the system size ($N$) is unaffected in comparison 
with the conventional CCSD theory ($\propto N^6$) due to presence of some terms quadratic in the 
amplitudes which do not naturally factorize to a simpler form even within the rank-reduced 
framework. We show how to solve this problem, exploiting the fact that their effective rank increases only linearly 
with the system size. 
We provide a systematic way to approximate the problematic terms using the singular value 
decomposition and reduce the scaling of the RR-CCSD iterations down to the level of $N^5$. This is 
combined with an iterative method of finding dominant eigenpairs of the MP2 or MP3 amplitudes which 
eliminates the necessity to perform the complete diagonalization and making the cost of this step 
proportional to the fifth power of the system size, as well. Next, we consider the evaluation of the 
perturbative corrections to the CCSD energies resulting from triply excited configurations. The 
triply-excited amplitudes present in the CCSD(T) method are decomposed to the Tucker-3 format using 
the higher-order orthogonal iteration (HOOI) procedure. This enables to compute the energy 
correction due to triple excitations non-iteratively with $N^6$ cost. The accuracy of the resulting 
rank-reduced CCSD(T) method is studied both for total and relative correlation energies of a 
diverse set of molecules. Accuracy levels better than 99.9\% can be achieved with a substantial 
reduction of the computational costs. Concerning the computational timings, break-even point between the rank-reduced 
and conventional CCSD implementations occurs for systems with about $30-40$ active electrons.
\end{abstract}

\maketitle

\section{\label{sec:intro} Introduction}

With the coupled-cluster (CC) theory~\cite{crawford07,bartlett07} firmly established as a powerful electronic structure 
method, applying it to large molecules remains a considerable challenge. Such applications are limited by 
unfavorable scaling of the computational costs with the size of the system. For example, the ``gold standard'' 
electronic structure method -- CC model with single and double excitations (CCSD) augmented with perturbative triples 
correction [CCSD(T)] -- scales as the seventh power of the system size~\cite{ragha89}. This makes canonical CCSD(T) 
calculations for 
molecules larger than $20-30$ atoms extremely expensive, assuming that a basis set of at least triple-zeta quality is 
used. To an extent, this boundary can be pushed by massive parallelization of the 
code~\cite{kobayashi97,hirata03,auer06,olson07,janowski07,janowski08,vandam11,deumens11,anisimov14,solomonik14,calvin15,
peng16,lyakh19,nagy20,peng2020,datta21,nagy21,kowalski21,calvin21} and/or by 
employing graphical processing units (GPU) to speed up the 
computations~\cite{deprince11,ma11,deprince14,kaliman17,deprince16,peng19,wang20,seritan20}. Other techniques designed 
to reduce the cost of CC calculations rely on optimization of the virtual space (either 
globally\cite{adam87,adam88,neo05,pitoniak06,kumar17} or 
for individual orbital pairs\cite{yang11,kura12,yang12,schutz13}) or employ local correlation 
techniques\cite{li02,li06,li09,neese09b,li10a,li10b,rolik11,rolik13,riplinger13a,riplinger13b,liakos15,schwilk17}. The 
latter family of methods is especially powerful and achieves linear scaling of the computational costs for sufficiently 
large systems.

The unfavorable scaling of the canonical CCSD(T) calculations results from contractions between high-order tensors that 
represent the wavefunction amplitudes and/or the Hamiltonian parameters. The main idea of the tensor decomposition 
techniques~\cite{kolda09} is to approximate these tensors as combinations of lower-rank quantities without compromising 
the accuracy. 
A widely known examples of such procedure are the density 
fitting\cite{whitten73,baerends73,dunlap79,alsenoy88,vahtras93} and Cholesky 
decomposition\cite{beebe77,koch03,pedersen04,folkestad19} of the electron repulsion 
integrals (ERI) where, in essence, the four-index ERI tensor is rewritten as a combination of only two-index and 
three-index objects. As each index represents a quantity with dimension proportional to the system 
size, this leads to significant savings. In recent years, more 
thorough decomposition schemes for ERI have been proposed such as the 
pseudospectral/chain-of-spheres 
approximation\cite{martinez92,martinez93,martinez94,martinez95,martinez96,neese09,kossmann10,izsak11,taras11,izsak12,
izsak13,dutta16}, tensor 
hypercontraction\cite{hohenstein12,parrish12,parrish13a,parrish13b} (THC) or canonical 
decomposition format\cite{benedikt11,benedikt13}. In the latter two methods only two-index quantities are required to 
approximate ERI. Aside from the reduced storage requirements, the aforementioned techniques allow 
to decrease the scaling of methods such as MP2 and MP3 with the system 
size~\cite{hohenstein12,hohenstein13a,schumacher15,lee20,matthews21}. Unfortunately, even 
with the most thorough ERI decomposition it is impossible to reduce the scaling of CC 
calculations as long as the high-order cluster amplitudes tensors are explicitly present.

The evidence that even without locality assumptions CC amplitudes can be efficiently compressed by representing them as 
combinations of low-order tensors is substantial, see, for example, the papers of Bell \emph{et al.}~\cite{bell10}, 
Kinoshita \emph{et al.}~\cite{kinoshita03,hino04}, and Scuseria and collaborators~\cite{scuseria08,schutski17}. Quite 
recently, these findings were exploited to reduce the 
cost 
of various conventional CC models\cite{hohenstein12b,hohenstein13a,hohenstein13b,parrish14,lesiuk20}. In this work we 
focus on the rank-reduced CCSD 
method (RR-CCSD) introduced by 
Parrish and collaborators~\cite{parrish19}, where the doubly-excited CC amplitudes are represented as (details of the 
notation are given 
in the next section)
\begin{align}
\label{t2comp}
 t_{ij}^{ab} = U_{ia}^X\;t_{XY}\,U_{jb}^Y.
\end{align}
The practical advantage of this decomposition is that the length of the summation over $X$, $Y$ has 
to scale (asymptotically) only linearly with the system size to maintain a constant level of 
relative accuracy in the correlation energy. In effect, the four-index amplitudes $t_{ij}^{ab}$ are 
rewritten 
as a combination of only two- and three-dimensional tensors, with each dimension being proportional 
to the system size. Unfortunately, this reduction of storage requirements is not accompanied by a 
commensurate decrease of the overall computational complexity of the RR-CCSD method. While the 
scaling of all 
terms linear in the CC amplitudes (in particular, the dreaded particle-particle ladder diagram) can 
indeed be reduced by a 
factor of $N$ by appropriate ordering of elementary tensor contractions, some terms 
quadratic in the amplitudes resist such factorization attempts. Therefore, the scaling of the 
RR-CCSD method remains formally the same ($N^6$) as the exact CCSD theory. The second problem 
encountered in the RR-CCSD theory is related to the choice of the quantities 
$U_{ia}^X$ present in Eq. (\ref{t2comp}). Following Ref. \onlinecite{parrish19} we adopt 
eigenvectors of the MP2 
or 
MP3 amplitudes for this purpose. While the MP2 amplitudes have a distinctive advantage that their 
diagonalization can be performed rapidly, inclusion of a large number of eigenvectors in Eq. 
(\ref{t2comp}) is required to achieve accuracy levels sufficient for general-purpose 
applications. The MP3 
amplitudes perform much better in this respect and are preferred in practice, but their computation 
requires $\propto N^6$ computational effort which constitutes a considerable overhead.

In this paper we modify the RR-CCSD theory of Parrish and collaborators~\cite{parrish19} in order to 
remove the 
aforementioned roadblocks that prevent the scaling reduction to $\propto N^5$. First, we show that 
the non-factorizable quadratic terms in the RR-CCSD working equations can be eliminated by proper 
definition of certain four-index intermediates and noting that their rank scales linearly (rather 
than quadratically) 
with the system size. This property is demonstrated numerically for realistic systems using the 
singular 
value decomposition procedure. Next, we exploit this finding by expanding the new intermediates in a 
separate 
basis (with a dimension proportional to the system size) which is fixed during the RR-CCSD 
iterations. This approach eliminates the non-factorizable $\propto N^6$ terms from the RR-CCSD 
equations; the error resulting from truncation of the intermediates expansion basis is small and 
controllable.

To solve the problem of efficient determination of the MP3 expansion basis $U_{ia}^X$, we adopt an 
iterative diagonalization method that avoids explicit construction of the amplitudes tensor. 
Instead, only products of the amplitudes with some trial vectors are necessary. Since we need to 
find only a small subset of the eigenvectors, i.e. proportional to the system size, the cost of the 
procedure scales rigorously as $N^5$. Note that in Ref. \onlinecite{parrish19} the authors suggested 
that such an 
approach 
is possible. By combining the proper handling of the intermediates described in the 
previous paragraph with iterative determination of the MP3 eigenvectors, we arrive at the variant 
of the RR-CCSD theory with quintic scaling of the total computational costs with the system size. 
The accuracy of the resulting approach in terms of both total and relative correlation 
energies is accessed by systematic comparison with the exact CCSD results for a large and diverse 
set of polyatomic molecules.

Further in the paper we move to the calculation of perturbative triples correction on top of the 
RR-CCSD method. The conventional implementation of the (T) correction~\cite{ragha89} scales as $N^7$ 
with the 
system size which would constitute a significant bottleneck in comparison with $N^5$ cost of 
RR-CCSD. One may pragmatically argue that this is not a major issue in actual applications as the 
(T) correction can simply be computed with a smaller basis set (and possibly scaled) at a 
significantly reduced cost. 
As the energy corrections resulting from triple excitations typically converge 
faster\cite{helgaker97,karton07,martin99} to the basis set limit than the CCSD contribution, this 
approach is certainly 
adequate 
in many situations. On the other hand, perturbative corrections calculated with a small basis, e.g. 
of double-zeta quality, are not always reliable and may require an independent verification.

In this work we propose a reduced-scaling ($N^6$) method of calculating the (T) correction on top 
of 
the RR-CCSD method. The crucial aspect of the method is the representation of the triply-excited 
amplitudes in the Tucker-3 format\cite{tucker66}
\begin{align}
\label{t3comp_intro}
 t_{ijk}^{abc} = t_{ABC}\,V_{ia}^A\,V_{jb}^B\,V_{kc}^C.
\end{align}
The above decomposition of the $t_{ijk}^{abc}$ tensor has been previously applied to the full CCSDT 
theory\cite{lesiuk20}, as well as to some of its approximate variants\cite{hino04,lesiuk19}, with 
the optimal expansion 
basis 
$V_{ia}^A$ found by the higher-order singular value decomposition procedure 
(HOSVD)\cite{delath00,vannie12}. While 
HOSVD is a robust and 
general method for acquiring the Tucker decomposition, its computational costs are too high to be 
workable in the 
present context. To circumvent this difficulty, we put forward a new scheme of obtaining the optimal 
expansion 
(\ref{t3comp_intro}) for the second-order 
triply-excited amplitudes encountered in the calculation of the (T) correction. It is based on the 
higher-order orthogonal iteration (HOOI) procedure\cite{delath00b,elden09} -- a straightforward 
iterative method of 
finding 
$V_{ia}^A$ by minimization of least-squares error of Eq. (\ref{t3comp_intro}). While the use of 
HOOI 
is widespread in fields of study such as signal processing\cite{cichocki15}, machine 
learning\cite{liu14} or data 
mining\cite{morup11}, applications of this procedure in quantum chemistry are, to the best of our 
knowledge, almost 
non-existent\cite{bell10}. This can be contrasted with another method of tensor decomposition, namely the alternating 
least squares (ALS), which has been thoroughly studied~\cite{hohenstein12,hummel17,schutski17,pierce21}. In this work 
we show that the HOOI enables to compute the decomposition in Eq. (\ref{t3comp_intro}) 
with $N^5$ complexity and is numerically stable and rapidly convergent. Once the decomposition of 
the triply-excited amplitudes given by Eq. (\ref{t3comp_intro}) is available, calculation of the 
(T) correction is a non-iterative step with $N^6$ complexity.

\section{\label{sec:prelim} Preliminaries}

\subsection{\label{subsec:notation} Definitions and notation}

The notation adopted in this paper is as follows. The canonical 
Hartree-Fock (HF) determinant, denoted $|\phi_0\rangle$, is the reference wavefunction. The 
orbitals occupied in the reference are denoted by the symbols $i$, $j$, $k$, etc., and the 
unoccupied (virtual) orbitals by the symbols $a$, $b$, $c$, etc. General indices $p$, $q$, $r$, 
etc. are used when the occupation of the orbital is not specified. We additionally introduce 
the following conventions: $\langle A\rangle \stackrel{\mbox{\tiny def}}{=} \langle \phi_0 | A 
\phi_0 \rangle$ and $\langle A|B\rangle \stackrel{\mbox{\tiny def}}{=} \langle A \phi_0|B \phi_0 
\rangle$ for general operators $A$, $B$. The Einstein convention for summation over repeated 
indices is employed unless explicitly stated otherwise. The electronic Hamiltonian is partitioned 
into a sum of the Fock operator, $F$, and the fluctuation potential, $W$. The number of occupied 
and virtual orbitals in the (molecular) basis set is denoted by $O$ and $V$, respectively. Formulas 
given in this work are valid for a spin-restricted closed-shell reference wavefunction.

All theoretical methods introduced in this work were implemented in a locally modified version 
of the \textsc{Gamess} program package\cite{gamess1,gamess2}. The exact CCSD(T) results, used as a 
reference in some 
calculations, were generated with the help of \textsc{NWChem} program\cite{nwchem20}, version 6.8.

\subsection{\label{subsec:density} Density-fitting approximation}

Unless explicitly stated otherwise, in all CC calculations reported in this work the 
electron repulsion integrals (ERI), $(pq|rs)$, are decomposed with help of the robust 
variant of the density fitting approximation\cite{whitten73,baerends73,dunlap79,alsenoy88,vahtras93}
(Coulomb metric)
\begin{align}
 \label{dfint}
 (pq|rs) = B_{pq}^Q\,B_{rs}^Q.
\end{align}
The capital letters $P$, $Q$ denote the elements of the auxiliary basis set and
\begin{align}
 B_{pq}^Q = (pq|P)\,[\mathbf{V}^{-1/2}]_{PQ},
\end{align}
where $(pq|P)$ and $V_{PQ}=(P|Q)$ are the three-center and two-center ERI as defined in 
Ref.~\onlinecite{katouda09}. For the purposes of subsequent analysis we note that the size of the 
auxiliary basis set, denoted $\naux$ further in the paper, scales linearly with the system size. 
Let us also point out that the accuracy offered by the density-fitting approximation with the 
standard pre-optimized auxiliary basis sets is satisfactory even in accurate CC calculations. As a 
matter of fact, extensive benchmark calculations\cite{epifanovsky13,deprince13,deprince14,lesiuk20b} 
revealed that the 
errors in the CC correlation 
energies resulting from the decomposition (\ref{dfint}) are negligible in comparison with the 
inherent orbital basis set incompleteness errors, at least as long as molecules are not far away 
from their equilibrium structures. Moreover, all equations derived in the present work remain valid 
also for the Cholesky decomposition~\cite{beebe77,koch03,pedersen04,folkestad19} of ERI, where the 
accuracy can be controlled more rigorously. The only necessary change in the replacement of the 
quantities $B_{pq}^Q$ in Eq.~(\ref{dfint}) by the appropriate Cholesky vectors. Finally, we stress 
that the density-fitting approximation is not used at the stage of self-consistent field 
calculations. Due to 
relatively minor computational costs, the Hartree-Fock equations are solved using the exact 
four-index ERI.

\subsection{\label{subsec:bidiag} Truncated singular value decomposition}

Throughout this work we shall repeatedly encounter the problem of calculating singular value 
decomposition (SVD) of some intermediate quantities. The necessary decomposition schemes assume one 
of three possible patterns
\begin{align}
\label{patterns}
\begin{split}
 M_{ia,jb} = U_{ia}^r\,\sigma_r\,V_{jb}^r, \\
 M_{ij,ab} = U_{ij}^r\,\sigma_r\,V_{ab}^r, \\
 M_{ij,kl} = U_{ij}^r\,\sigma_r\,V_{kl}^r, \\
\end{split}
\end{align}
for matrices of size $OV\times OV$, $O^2\times V^2$, and $O^2\times O^2$, respectively. In a 
special case where the matrix under consideration is square symmetric, the SVD can be replaced by 
the usual eigendecomposition for simplicity. The quantities $\mathbf{U}$ and $\mathbf{V}$ then 
coincide, but the eigenvalues $\sigma_r$ can be of an arbitrary sign, unlike the singular values 
which are strictly 
non-negative.

Since the dimension of each matrix in Eq. (\ref{patterns}) is quadratic in the number of orbitals, 
the computational 
cost of determining the complete SVD is proportional to the sixth power of the system size. However, 
in every situation 
encountered in this work only a small subset of singular vectors has to be found that correspond to 
the largest 
singular 
values (or the largest \emph{absolute} eigenvalues in the case of the eigendecomposition). Moreover, 
the number of 
elements of this subset increases only linearly with the system size. Under these conditions it is 
possible to find the 
required subset of singular value/vector pairs with the cost proportional to the fifth power of the 
system size by a 
proper choice of the decomposition algorithm. 

For this purpose we adopt a scheme based on partial Golub-Kahan bidiagonalization\cite{golub65} that 
has been 
previously used to find 
singular vectors of the triply-excited amplitudes tensor\cite{lesiuk19}. The details of the 
procedure are described in 
Ref.~\onlinecite{lesiuk19} and in earlier works in the numerical analysis 
literature\cite{simon00,baglama05}. The most 
important aspect of the algorithm is that the 
matrix under consideration is never formed explicitly. Instead, one needs to evaluate only 
left- and right-hand-side products of the matrix with some trial vectors. Within this setup, the 
desired subset of singular vectors can be found with $N^5$ complexity provided that the left- and 
right-hand-side products with an arbitrary trial vector can be computed with $N^4$ scaling. The 
latter property shall be demonstrated separately for each matrix under consideration in this work. 
Note that the truncated SVD algorithm described here is reminiscent of the Davidson 
diagonalization\cite{davidson75} 
method which has found widespread use in the configuration interaction (CI) calculations, among 
others. 

\section{\label{sec:rrcc} Rank-reduced formalism}

\subsection{\label{subsec:rrccsd} Rank-reduced CCSD method}

In this section we summarize the key aspects of the rank-reduced CCSD method as introduced by 
Parrish \emph{et 
al.}~\cite{parrish19}. Next, we describe some technical aspects and practical limitations of this 
formulation. Finally, 
we propose a modification of this theory that enables to reduce its scaling, as elaborated in 
subsequent sections.

The coupled-cluster theory~\cite{crawford07,bartlett07} employs the exponential parametrization of 
the electronic 
wavefunction
\begin{align} 
|\Psi\rangle = e^T\,|\phi_0\rangle,
\end{align}
where $T$ is the cluster operator. In this work we consider the CCSD method where the 
cluster operator includes only single and double excitations ($T=T_1+T_2$) with respect to the 
reference determinant
\begin{align}
\label{t12}
 T_1 = t_i^a\,E_{ai}, \;\;\;
 T_2 = \frac{1}{2}\,t_{ij}^{ab} \,E_{ai}\,E_{bj},
\end{align}
where $t_i^a$, $t_{ij}^{ab}$ are the cluster amplitudes, and $E_{pq}=p^\dagger_\alpha q_\alpha + 
p^\dagger_\beta q_\beta$ are the spin-adapted singlet orbital replacement 
operators~\cite{paldus88}. The cluster amplitudes are the wavefunction parameters and are found by 
solving non-linear equations
\begin{align}
\begin{split}
 &\langle_i^a|e^{-T}He^T\rangle = 0, \\
 &\langle_{ij}^{ab}|e^{-T}He^T\rangle = 0,
\end{split}
\end{align}
where $\langle_i^a|$ and $\langle_{ij}^{ab}|$ denote projection onto the singly- and doubly-excited 
configurations. Finally, the correlation energy is calculated from the formula 
$E_{\mathrm{corr}}=\langle e^{-T}He^T\rangle$.

In the rank-reduced CCSD (RR-CCSD) theory introduced by Parrish \emph{et al.}~\cite{parrish19} the 
doubly-excited amplitudes are represented by Eq. (\ref{t2comp}), where the quantities $U_{ia}^X$ 
generate the necessary excitation subspace and the core matrix $t_{XY}$ plays the role 
of ``compressed`` amplitudes. While it is, in principle, possible to optimize both $U_{ia}^X$ and 
$t_{XY}$ during the CC iterations, this choice is rather impractical. Instead, the basis vectors 
$U_{ia}^X$ are found 
upfront by diagonalizing the MP2 or MP3 amplitudes and collecting the eigenvectors that 
correspond 
to the eigenvalues of the largest magnitude. The quantities $U_{ia}^X$ are then fixed in the CC 
iterative process where the compressed amplitudes $t_{XY}$ are solved for. Further details of this 
procedure are thoroughly discussed Ref.~\onlinecite{parrish19}. In the present work we do not 
attempt to compress the 
singly-excited amplitudes -- they are treated in exactly the same way as in the exact CCSD theory.

Throughout this paper, the dimension of the excitation subspace, i.e. the length of the summation 
summation over $X$, $Y$ in Eq. (\ref{t2comp}), is denoted by $N_{\mathrm{eig}}$ and we have 
$N_{\mathrm{eig}}\leq OV$. Moreover, in 
the limit $N_{\mathrm{eig}}=OV$ the expansion becomes exact independently of the source of the 
approximate 
amplitudes employed generate $U_{ia}^X$. However, the practical advantage of Eq. (\ref{t2comp}) is 
that to maintain a 
constant relative accuracy in the correlation energy, the quantity $N_{\mathrm{eig}}$ has to grow 
only linearly with 
the system size, rather than quadratically as in the exact CCSD limit ($N_{\mathrm{eig}}=OV$). 
Besides the 
advantage of reducing the storage requirements, this property also opens up a window for reducing 
the scaling of the 
RR-CCSD calculations.

To simplify the task of solving the RR-CCSD equations to obtain the compressed amplitudes $t_{XY}$, 
it is helpful to 
enforce some constraints on the basis vectors $U_{ia}^X$. First, note that as a byproduct of the 
diagonalization, the 
quantities $U_{ia}^X$ are automatically orthonormal in the sense of the following formula
\begin{align}
\label{ortho1}
 U_{ia}^X\,U_{ia}^Y = \delta_{XY}.
\end{align}
By an orthogonal transformation of $U_{ia}^X$ it is possible to simultaneously satisfy the second 
equality
\begin{align}
\label{ortho2}
 U_{ia}^X\,\epsilon_i^a\,U_{ia}^Y = \delta_{XY}\,\epsilon_X,
\end{align}
where $\epsilon_X$ are some real-valued constants, and $\epsilon_i^a=\epsilon_i-\epsilon_a$.
Once the basis vectors $U_{ia}^X$ satisfy the constraints (\ref{ortho1}) and (\ref{ortho2}), 
application of the 
Lagrangian formalism from Ref.~\onlinecite{parrish19} leads to a straightforward prescription for an 
update of the 
compressed amplitudes, namely
\begin{align}
 -\frac{r_{XY}}{\epsilon_X+\epsilon_Y} \longrightarrow t_{XY},
\end{align}
where $r_{XY}$ is the compressed residual defined as
\begin{align}
\label{rxydef}
 r_{XY} = U_{ia}^X\,U_{jb}^Y\,\langle_{ij}^{ab}|e^{-T}He^T\rangle.
\end{align}
These formulas are iterated until convergence, i.e. until the norm of the residual $r_{XY}$ falls 
below a certain 
threshold. Due to the striking similarity of this procedure to the standard CC iterations, various 
techniques 
designed to accelerate the CC convergence~\cite{pulay80,scuseria86,purvis81,ziolo08,ettenhuber15} 
can be 
straightforwardly applied at this point.

As mentioned in the introduction, there are two major problems that limit the applicability of the 
RR-CCSD theory outlined above. 
The first is related to the choice of approximate doubly-excited amplitudes as a source of the 
basis vectors $U_{ia}^X$. Natural candidates for this task are the MP2 or MP3 amplitudes since they 
constitute the first- and second-order approximations to the exact coupled-cluster amplitudes in 
the 
framework of the conventional M\o{}ller-Plesset perturbation theory. However, as demonstrated in 
Ref.~\onlinecite{parrish19}, the MP2 amplitudes require rather large $N_{\mathrm{eig}}$ to achieve 
satisfactory accuracy levels. This poor performance is understandable from a purely mathematical 
point of view: the MP2 amplitudes are negative-definite while the CCSD amplitudes are 
indefinite. This means that the MP2 amplitudes lack the entire portion of the spectrum 
that corresponds to the positive eigenvalues. Despite the negative portion of the spectrum is 
dominant, eigenvectors from the positive part are needed to achieve accurate results. This 
deficiency is rectified by the MP3 amplitudes which are also indefinite. Unfortunately, the 
computation of the MP3 amplitudes is an $N^6$ process which is unacceptable from the present point 
of view. This bottleneck can be removed by noticing that we have to find only a certain subset of 
eigenvectors that correspond to the largest singular values and the dimension of this subset is 
proportional to the system size. In Sec. \ref{subsec:diag} we discuss how this partial 
diagonalization can be accomplished with $N^5$ cost.

The second bottleneck that prevents the scaling reduction of the RR-CCSD method is the computation 
of the residual $r_{XY}$ defined in Eq. (\ref{rxydef}). Many terms present in $r_{XY}$ can be 
computed with $N^5$ scaling by proper arrangement of elementary tensor contractions (in particular, 
all terms linear in the amplitudes). However, there are two terms quadratic in the 
amplitudes that are resistant to such treatment and require $N^6$ operations to compute. In Sec. 
\ref{subsec:scale1} we show that the problem of apparently non-factorizable terms can be solved by 
defining certain intermediate quantities and subjecting them to the singular-value decomposition 
procedure. Similarly as in the case of the amplitudes, we prove numerically that the singular 
vectors corresponding to small singular values can be dropped without significant impact on the 
accuracy. More importantly, a constant relative error in the correlation energy can be maintained 
with 
a number of singular values scaling only linearly with the system size. This paves the way for a 
modified formulation of the RR-CCSD theory with $N^5$ overall scaling.

\subsection{\label{subsec:diag} Efficient determination of the excitation subspace}

The practical usefulness of the RR-CCSD theory hinges upon the assumption that the optimal 
excitation subspace can be found efficiently. In this section we show that the product of the MP2 
and MP3 amplitudes with an arbitrary set of trial vectors with dimension proportional to the system 
size can be assembled with the $N^4$ and $N^5$ cost, respectively. We begin by defining
\begin{align}
\label{t2mp2}
 t_{ij}^{ab}(\mbox{MP2}) = \big( \epsilon_{ij}^{ab} \big)^{-1}\langle_{ij}^{ab}|W\rangle
 = \big( \epsilon_{ij}^{ab} \big)^{-1} (ia|jb),
\end{align}
and
\begin{align}
\label{t2mp3_1}
\begin{split}
 &t_{ij}^{ab}(\mbox{MP3}) = t_{ij}^{ab}(\mbox{MP2}) + \big( \epsilon_{ij}^{ab} \big)^{-1}
 \langle_{ij}^{ab}|\Big[W,T_2^{\mathrm{MP2}}\Big]\rangle
\end{split}
\end{align}
with
\begin{align}
\label{t2mp3_2}
\begin{split}
 &\langle_{ij}^{ab}|\Big[W,T_2^{\mathrm{MP2}}\Big]\rangle = 
 P_{ij}^{ab}\Big[ -\half\,(ki|lj)\,t_{kl}^{ab}(\mbox{MP2}) + (ac|ki)\,t_{kj}^{cb}(\mbox{MP2}) \\
 &- (ai|kc)\,\big[ 2t_{kj}^{cb}(\mbox{MP2}) - t_{kj}^{bc}(\mbox{MP2})\big] + 
(bc|ki)\,t_{kj}^{ac}(\mbox{MP2}) - 
\half\,(ac|bd)\,t_{ij}^{cd}(\mbox{MP2}) \Big],
\end{split}
\end{align}
where $\epsilon_{ij}^{ab} = \epsilon_i^a + \epsilon_j^b$ is the two-particle energy denominator, 
and $P_{ij}^{ab}$ is a permutation operator that simultaneously exchanges the indices 
$i\leftrightarrow j$ and $a\leftrightarrow b$. To enable an efficient handling of the amplitudes 
defined above one has to remove the denominator from both formulas. This is achieved with help of 
the Laplace transformation technique
\begin{align}
\label{laplace}
 (\epsilon_{ij}^{ab})^{-1} = \sum_g^{N_g} w_g\,e^{-t_g \left( \epsilon_i^a + \epsilon_j^b \right)},
\end{align}
where $t_g$ and $w_g$ are the quadrature nodes and weights, respectively, and $N_g$ is the size of 
the 
quadrature. Further in the text we remove the symbol of the sum $\sum_g^{N_g}$ wherever its
presence is clear from the context. The Laplace transformation technique was first proposed by 
Alml\"{o}f~\cite{almlof91} to 
simplify the MP2 calculations, but since then it has been successfully used in combination with 
other electronic structure 
methods~\cite{haser92,ayala99,lambert05,nakajima06,jung04,kats08}. In this work we employ the 
min-max quadrature 
proposed by Takatsuka and 
collaborators~\cite{takatsuka08,braess05,paris16} for the choice of $t_g$ and $w_g$. The number of 
quadrature points in Eq. (\ref{laplace}) is independent of the system size, that~is $N_g\propto 
N^0$. 

Using the Laplace transformation technique and the density-fitting decomposition of the two-electron 
integrals, the 
product of MP2 amplitudes with an arbitrary trial vector $\omega_{ia}$ can be rewritten as
\begin{align}
 t_{ij}^{ab}(\mbox{MP2})\,\omega_{jb} = w_g\,e^{-t_g\epsilon_i^a}
 \bigg[B_{ia}^Q\,\Big( B_{jb}^Q\,\tilde{\omega}_{jb}^g \Big)\bigg],
\end{align}
where $\tilde{\omega}_{jb}^g = \omega_{jb}\,e^{-t_g\epsilon_j^b}$.
By carrying the contractions in the order indicated by the parentheses, the cost of the 
operations is proportional to $OV\naux N_g\propto N^3$. Therefore, the task of obtaining $\neig$ 
dominant 
eigenpairs can be accomplished with $N^4$ cost, because both $\neig$ and the number of trial vectors 
is asymptotically 
linear in the system size. The fact that this is possible has also been demonstrated in 
Ref.~\onlinecite{parrish19}, 
albeit using 
a somewhat different approach.

In order to perform the diagonalization of the MP3 amplitudes efficiently, the 
product $t_{ij}^{ab}(\mbox{MP3})\,\omega_{jb}$ has to evaluated with $N^4$ complexity. To show that 
this is possible, we first introduce a handful of intermediates that combine the density-fitted 
integrals with the expansion vectors $U_{ia}^{X(\mathrm{MP2})}$ obtained previously for the MP2 
amplitudes, namely 
\begin{align}
\label{intd}
 D_{ia}^{QX} = B_{ki}^Q\,U_{ka}^{X(\mathrm{MP2})}
             - B_{ac}^Q\,U_{ic}^{X(\mathrm{MP2})},
\end{align}
\begin{align}
\begin{split}
\label{intg}
 \Gamma_{ia}^X &= \Big( B_{ac}^Q\,U_{kc}^{X(\mathrm{MP2})}\,d_X^{\mathrm{MP2}} \Big)\,B_{ki}^Q \\
               &- 2 B_{ia}^Q\,\Big( B_{kc}^Q\,U_{kc}^{X(\mathrm{MP2})}\,d_X^{\mathrm{MP2}} \Big), \\
\end{split}
\end{align}
\begin{align}
\label{intw}
 &W_{jb}^Q = U_{kb}^{X(\mathrm{MP2})}\,\Big( 
 B_{kc}^Q\,U_{jc}^{X(\mathrm{MP2})}\,d_X^{\mathrm{MP2}} \Big).
\end{align}
Evaluation of each intermediate has $N^5$ complexity, but they are computed only once before the 
diagonalization and stored. With help of Eqs. (\ref{intd})--(\ref{intw}) the contraction of the MP3 
amplitudes with the trial vector $\omega_{jb}$ is rewritten as
\begin{align}
\label{mp3mul}
\begin{split}
 &t_{ij}^{ab}(\mbox{MP3})\,\omega_{jb} = 
 w_g\,e^{-t_g\epsilon_i^a} \bigg[
 D_{ia}^{QY}\,d_Y^{\mathrm{MP2}}\,\Big( D_{jb}^{QY}\,\tilde{\omega}_{jb}^g \Big) \\
 &+ U_{ia}^{Y(\mathrm{MP2})}\,\Big( \Gamma_{jb}^Y\,\tilde{\omega}_{jb}^g \Big)
  + \Gamma_{ia}^Y\,\Big( U_{jb}^{Y(\mathrm{MP2})}\,\tilde{\omega}_{jb}^g \Big) \\
 &+ B_{ia}^Q\,\Big( W_{jb}^Q\,\tilde{\omega}_{jb}^g \Big)
  + W_{ia}^Q\,\Big( B_{jb}^Q\,\tilde{\omega}_{jb}^g \Big) \bigg].
\end{split}
\end{align}
None of the elementary steps in the above formula involve more than four indices at the same 
time (the grid index $g$ does not count since $N_g\propto N^0$). The first term in the above 
formula typically dominates the workload with the scaling $OV\naux\neig N_g\propto N^4$.
This shows that the multiplication $t_{ij}^{ab}(\mbox{MP3})\,\omega_{jb}$ can be accomplished with 
$N^4$ cost and enables efficient ($N^5$) determination of the basis vectors 
$U_{ia}^{Y(\mathrm{MP3})}$ for the MP3 excitation subspace using an iterative eigensolver.

The remaining issue that has to be discussed is an adequate choice of the number of quadrature 
points 
in the Laplace transformation formula, Eq. (\ref{laplace}). In the case of the MP2 amplitudes, Eq. 
(\ref{t2mp2}), we found that ten quadrature points are sufficient to reach relative 
accuracy of a few parts per million in the RR-CCSD correlation energy. This deviation is negligible 
in comparison to other sources of error. Considering the MP3 amplitudes we note that the second 
term in Eq. (\ref{t2mp3_1}) is typically by an order of magnitude smaller than the first. 
Therefore, the efficiency of the diagonalization can be improved without degrading the accuracy if 
a 
smaller number of quadrature points is used for decomposition of the denominator in the second term 
of Eq. (\ref{t2mp3_1}). We found that three points of the min-max quadrature are 
sufficient for 
this task. A numerical illustration of the impact of the $N_g$ parameter on the accuracy of the 
RR-CCSD correlation energy is included in the supplementary material.

\subsection{\label{subsec:scale1} Non-factorizable terms in the RR-CCSD residual}

A complete formula for the RR-CCSD residual, Eq. (\ref{rxydef}), expressed explicitly through the 
basic two-electron integrals and cluster amplitudes is given in the supplementary material for the 
sake of brevity. Here we concentrate only on two terms that do not naturally factorize to a 
form that can be evaluated with $N^5$ cost and write the residual shortly as
\begin{align}
\label{rxynonfact}
\begin{split}
 r_{XY} = \half\,P_{XY}\Big[ &U_{ka}^Z\;t_{ZW}\,U_{lb}^W\,O_{kl}^{ij}\,U_{ia}^X\,U_{jb}^Y \\
       +\,&U_{ka}^Z\;t_{ZW}\,U_{jc}^W\,Z_{ki}^{bc}\,U_{ia}^X\,U_{jb}^Y \Big] \\
       +\,&\mbox{factorizable terms},
\end{split}
\end{align}
where $P_{XY}$ is a permutation operator that exchanges the indices $X$ and $Y$. The intermediate 
quantities 
$O_{kl}^{ij}$ and $Z_{ki}^{bc}$ are defined as
\begin{align}
\label{odef}
 O^{ij}_{kl} = (kc|ld)\,t_{ij}^{cd},
\end{align}
and
\begin{align}
\label{zdef}
 Z_{ij}^{ab} = (ic|kb)\,t_{jk}^{ca}.
\end{align}
The terms in Eq. (\ref{rxynonfact}) that involve the intermediates $O_{kl}^{ij}$ and $Z_{ki}^{bc}$ 
require $\propto O^4V^2$ and $\propto O^3V^3$ operations, respectively, to evaluate. To eliminate 
this bottleneck we decompose the intermediates using the following format
\begin{align}
\label{osvd}
 O^{ij}_{kl} = \alpha_{ik}^F\,o_F\,\alpha_{jl}^F, \\
\label{zsvd}
 Z_{ij}^{ab} = \beta_{ij}^F\,z_F\,\beta_{ab}^F.
\end{align}
The first intermediate obeys the symmetry relation $O^{ij}_{kl}=O^{ji}_{lk}$. Therefore, the 
decomposition (\ref{osvd}) is obtained by rewriting it as $O^2\times O^2$ matrix $O_{ik,jl}$, 
followed by diagonalization. The second decomposition is obtained by SVD of the $Z$ 
intermediate reshaped as a $O^2\times V^2$ matrix, $Z_{ij,ab}$. Consequently, the quantities $z_F$ 
are non-negative while $o_F$ can have an arbitrary sign.

We conjecture that for any fixed threshold $\varepsilon$, the number of singular values (or 
absolute 
eigenvalues) larger than $\varepsilon$ in Eqs. (\ref{osvd}) and (\ref{zsvd}), i.e. $z_F>\varepsilon$ 
or 
$|o_F|>\varepsilon$, grows asymptotically only linearly with the system size, not quadratically as 
the dimensions of 
$O^{ij}_{kl}$ and $Z_{ij}^{ab}$. We did not manage to prove this statement rigorously and hence we 
demonstrate it numerically for two representative model systems: linear alkanes 
C$_n$H$_{2n+2}$ with increasing chain length $n$, and water clusters $\big($H$_2$O$\big)_n$. The 
former system is an idealized, quasi-1D structure with strong covalent bonds, while the latter is a 
fully three-dimensional structure with more diverse bonding character which is more demanding from 
the practical point of view. The geometries of the model systems were taken from 
Refs.~\onlinecite{lesiuk20} 
and~\onlinecite{parrish19}, respectively.

\begin{figure}[ht!]
\includegraphics[scale=0.7]{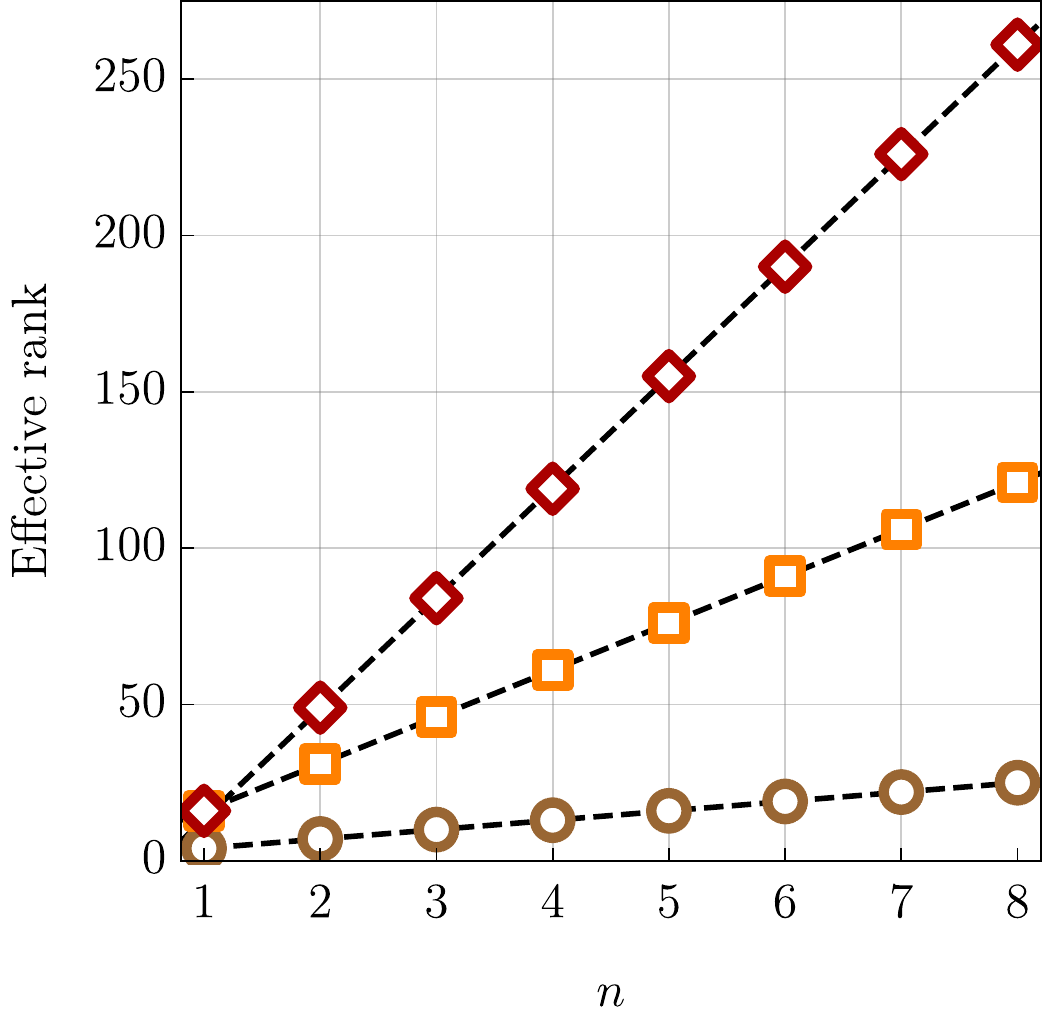} \hspace{0.5cm}
\includegraphics[scale=0.7]{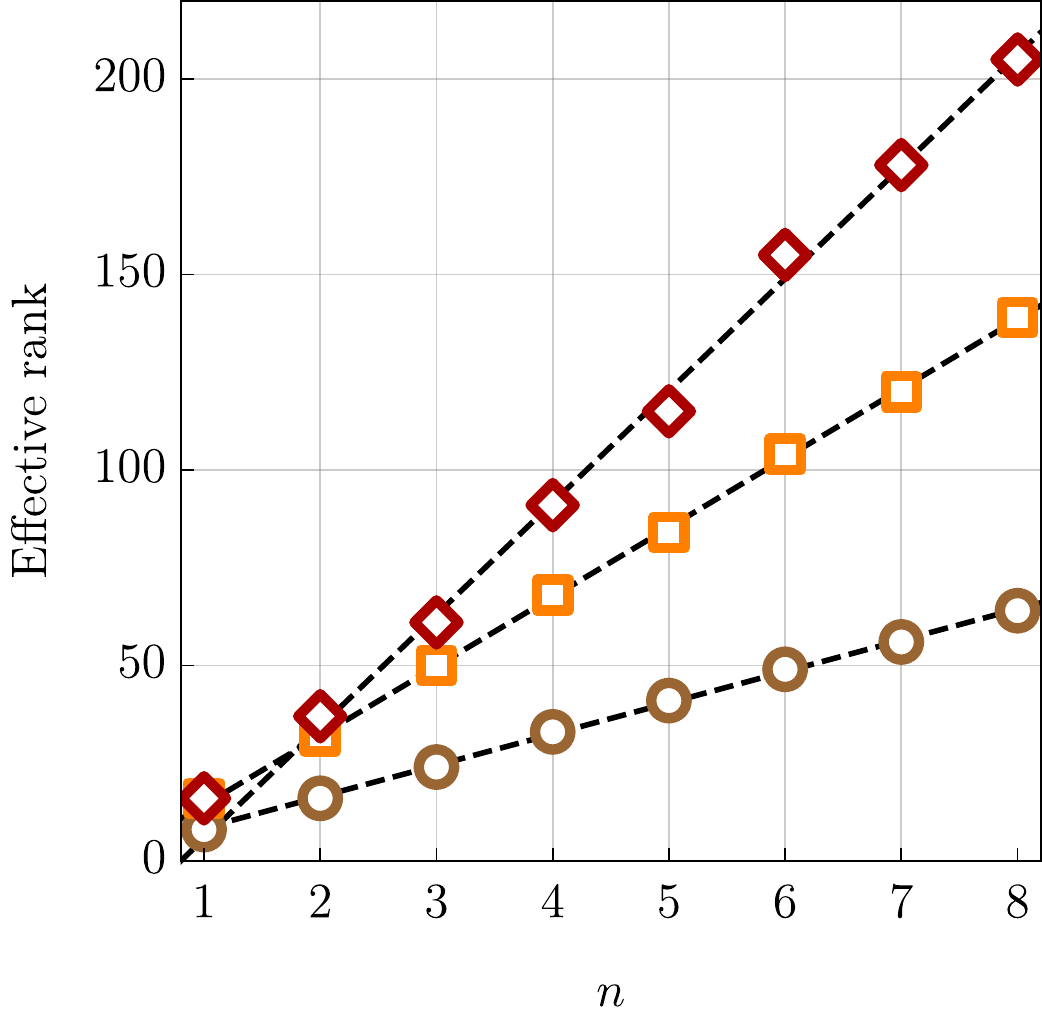}
\caption{\label{fig:o-scaling} Effective rank of the $O^{ij}_{kl}$ intermediate for the linear 
alkanes 
C$_n$H$_{2n+2}$ (left panel) and water clusters $\big($H$_2$O$\big)_n$ (right panel) extracted 
from the CCSD/cc-pVTZ calculations. The brown circles, orange squares and red diamonds indicate the 
effective rank obtained with the thresholds $\varepsilon=10^{-2}$, $10^{-3}$, and $10^{-4}$, 
respectively. The black dashed lines were obtained by least-squares fitting to the corresponding 
data points ($n=2,\ldots,8$).}

\end{figure}
\begin{figure}[ht!]
\includegraphics[scale=0.7]{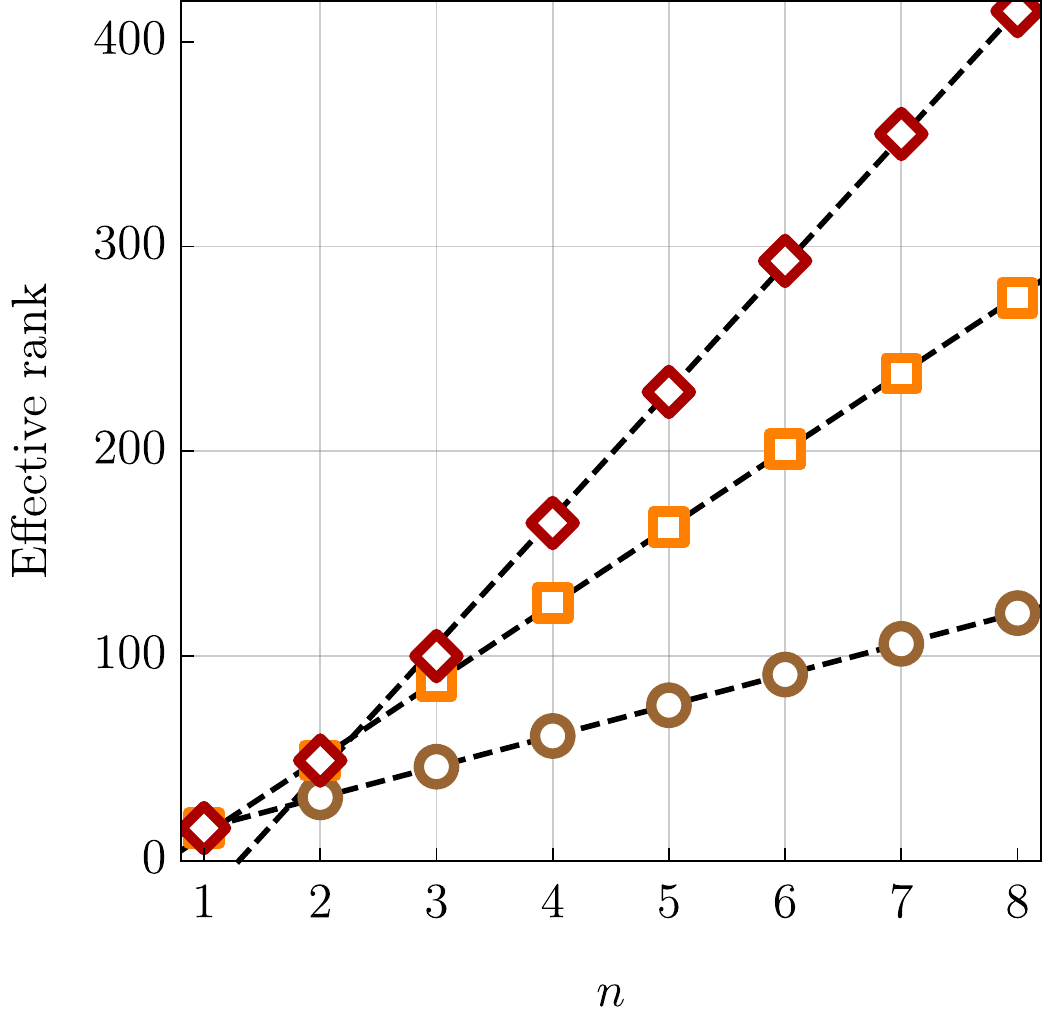} \hspace{0.5cm}
\includegraphics[scale=0.7]{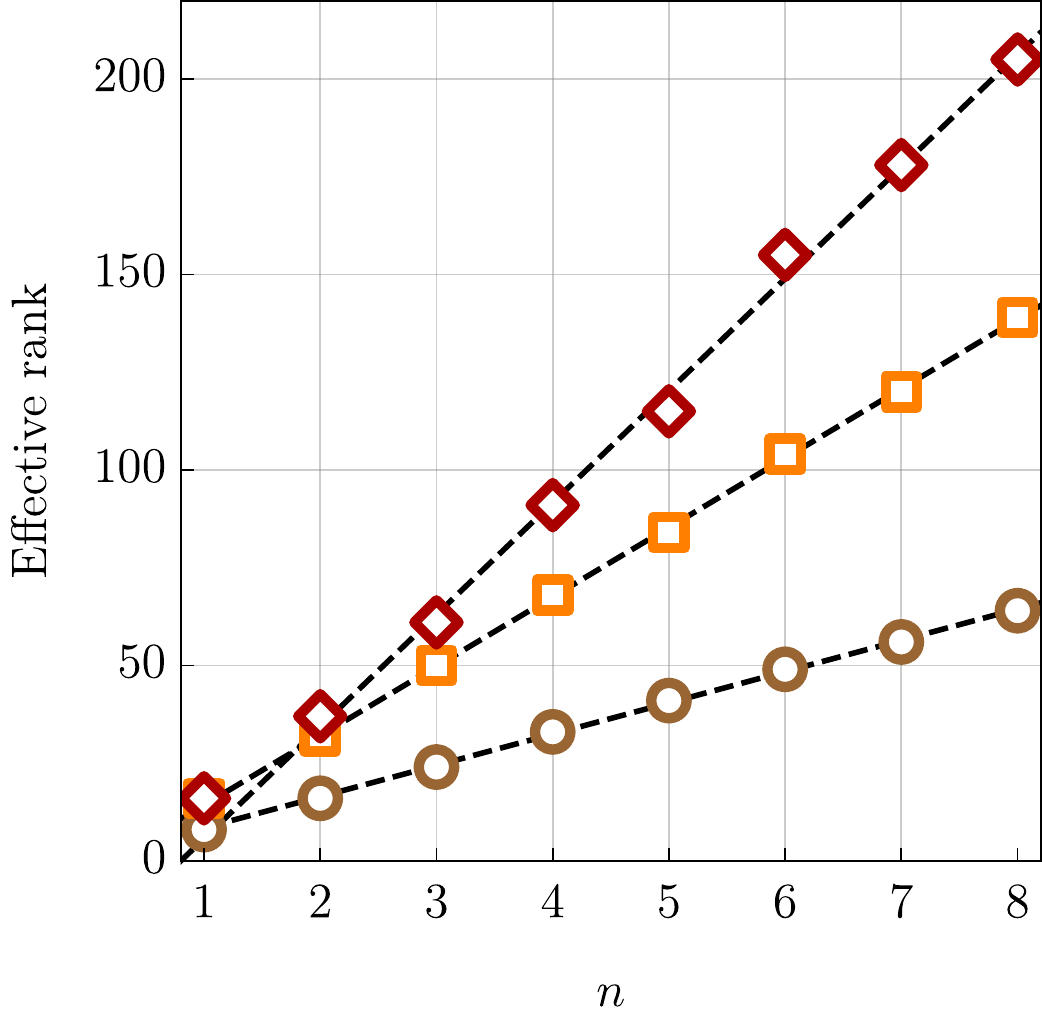}
\caption{\label{fig:z-scaling} Effective rank of the $Z_{ij}^{ab}$ intermediate for the linear 
alkanes 
C$_n$H$_{2n+2}$ (left panel) and water clusters $\big($H$_2$O$\big)_n$ (right panel) extracted 
from the CCSD/cc-pVTZ calculations. The brown circles, orange squares and red diamonds indicate the 
effective rank obtained with the thresholds $\varepsilon=10^{-3}$, $10^{-4}$, and $10^{-5}$, 
respectively. The black dashed lines were obtained by least-squares fitting to the corresponding 
data points ($n=2,\ldots,8$).}
\end{figure}

For both model systems we performed the exact CCSD calculations using the cc-pVTZ orbital basis 
set\cite{dunning89} and 
the 
corresponding cc-pVTZ-RIFIT density-fitting basis\cite{weigend02b}. The $1s$ core orbitals of 
carbon and oxygen 
atoms were frozen in these calculations. Next, we computed the $O^{ij}_{kl}$ and 
$Z_{ij}^{ab}$ intermediates with the converged doubly-excited amplitudes and performed the 
decompositions (\ref{osvd}) and (\ref{zsvd}). Finally, for each system size $n$ and threshold value 
$\varepsilon$ we recorded the number of singular values (or absolute eigenvalues) larger than 
$\varepsilon$. This number is referred to further in 
the text as the effective rank. The results are illustrated in Fig. \ref{fig:o-scaling} for the 
$O^{ij}_{kl}$ 
intermediate 
and in Fig. \ref{fig:z-scaling} for the $Z_{ij}^{ab}$ intermediate. For the former quantity we 
considered the thresholds $\varepsilon=10^{-2}$, $10^{-3}$, and $10^{-4}$. For the latter we 
replaced $\varepsilon=10^{-2}$ by $\varepsilon=10^{-5}$ because the effective ranks for 
$\varepsilon=10^{-2}$ were too small ($\leq6$) for a meaningful comparison. The results 
presented in Fig. \ref{fig:o-scaling} and Fig. \ref{fig:z-scaling} confirm the conjecture that the 
effective ranks of both intermediates increase only linearly with the system size. This statement 
is true to a good degree of approximation for every truncation threshold $\varepsilon$ considered 
here. Some deviations from the trend line are observed for the more challenging test case of water 
clusters, but only for the smallest value of the threshold ($\varepsilon=10^{-5}$). It is also 
noteworthy that a decrease of $\varepsilon$ by an order of magnitude leads to an increase of the 
slope of the linear trend line by approximately a factor of two. 

To address the question whether the results represented graphically in Figs. \ref{fig:o-scaling} 
and \ref{fig:z-scaling} can be reproduced also in a smaller basis set, we performed analogous 
calculations with the cc-pVDZ basis. The plots of effective ranks analogous to Fig. 
\ref{fig:o-scaling} and \ref{fig:z-scaling} are given in supplementary material. In summary, 
the effective ranks change by no more than 5\% when going from the cc-pVDZ to the cc-pVTZ 
basis. The only exception occurs for the water clusters with the smallest threshold 
($\varepsilon=10^{-5}$), where the changes are slightly larger. Nonetheless, the linear growth of 
the effective ranks with the system size is confirmed in every case.

\subsection{\label{subsec:scale2} Quintic-scaling formulation}

Having shown that the effective ranks of the $O^{ij}_{kl}$ and $Z_{ij}^{ab}$ 
intermediates increase only linearly (rather than quadratically) with the system size, we now 
describe how this observation can be exploited to decrease the scaling of the RR-CCSD calculations 
to the level of $N^5$. For simplicity, we consider the $O^{ij}_{kl}$ intermediate first; extension 
of this approach to $Z_{ij}^{ab}$ is presented further in the text.

In the modified RR-CCSD theory the $O^{ij}_{kl}$ intermediate is represented as
\begin{align}
\label{osvd2}
 O^{ij}_{kl} = \alpha_{ik}^F\,o_{FG}\,\alpha_{jl}^G,
\end{align}
The expansion vectors $\alpha_{ij}^F$ are obtained before the iterations and are fixed thereafter, 
while the core matrix $o_{FG}$ changes from iteration to iteration. The length of this expansion, 
i.e. the summations 
over $F$, $G$, is denoted $N_{\mathrm{O}}$ further in the text and it scales linearly with the 
system size. A suitable 
expansion basis $\alpha_{ik}^F$ is obtained by diagonalization of Eq. (\ref{osvd}) reshaped as a 
symmetric 
$O^2\times O^2$ matrix $O_{ik,jl}$, and taking $N_{\mathrm{O}}$ eigenvectors that correspond to the 
largest absolute 
eigenvalues ($N_{\mathrm{O}}$ dominant eigenvectors). Because the exact CCSD amplitudes entering Eq. 
(\ref{osvd}) 
are not known before the iterations, they are approximated by their MP2 or MP3 counterparts in the 
rank-reduced form
\begin{align}
\label{t2diag}
 t_{ij}^{ab} = U_{ia}^X\;d_X\,U_{jb}^X,
\end{align}
that are 
obtained according to the scheme presented in Sec. \ref{subsec:diag}. For brevity, we no longer 
distinguish the 
MP2 and MP3 amplitudes here [$t_{ij}^{ab}(\mbox{MP2})$ and $t_{ij}^{ab}(\mbox{MP3})$], because the 
treatment of the 
$O^{ij}_{kl}$ and $Z_{ij}^{ab}$ intermediates described below is the same in both cases.

Asymptotically, the $N_{\mathrm{O}}$ parameter is much smaller than the dimension ($O^2$) of the 
$O_{ik,jl}$ matrix.
Therefore, the full diagonalization of the matrix $O_{ik,jl}$ can be avoided and only a subset of 
$N_{\mathrm{O}}$ dominant eigenpairs has to be found. This task can be accomplished efficiently 
($N^5$ 
overall scaling) provided that the product $O^{ij}_{kl}\,q_{jl}$, where $q_{jl}$ is an arbitrary 
trial vector, can be 
calculated with $N^4$ cost. To prove that we first define an auxiliary quantity
\begin{align}
 \label{bijqx}
 B_{ki}^{QX} = B_{kc}^Q\,U_{ic}^X,
\end{align}
which can be calculated before the diagonalization ($O^2V \naux\neig$ cost) and stored. Next, we
combine the initial formula (\ref{osvd}) with Eqs. (\ref{t2diag}) -- (\ref{bijqx}) and rearrange 
the order of elementary operations as follows
\begin{align}
 O^{ij}_{kl}\,q_{jl} = B_{ki}^{QX}\Big(d_X\,B_{lj}^{QX} q_{jl}\Big).
\end{align}
The $O^2\naux\neig\propto N^4$ cost of the two contraction steps becomes evident.

Because the expansion basis $\alpha_{ik}^F$ is fixed, in each RR-CCSD iteration one has to find an 
updated core matrix $o_{FG}$, taking into account that the compressed amplitudes $t_{XY}$ from Eq. 
(\ref{t2comp}) 
change. To simplify this task we note that as a byproduct of the diagonalization procedure, the expansion vectors 
obey the orthonormality relation $\alpha_{ij}^F\,\alpha_{ij}^G=\delta_{FG}$. Therefore, in every iteration the core 
matrix is given by an explicit expression
\begin{align}
 o_{FG} = \alpha_{ik}^F\,O^{ij}_{kl}\,\alpha_{jl}^G.
\end{align}
By using the definition (\ref{odef}) and inserting the formulas (\ref{t2diag}) and (\ref{bijqx}) we 
arrive at
\begin{align}
\label{ocore}
 o_{FG} = \Big(B_{ki}^{QX}\,\alpha_{ik}^F\Big)t_{XY}\,\Big(B_{lj}^{QY}\,\alpha_{jl}^G\Big).
\end{align}
Finally, the contribution of the $O^{ij}_{kl}$ intermediate to the RR-CCSD residual is calculated 
by inserting Eqs. (\ref{osvd2}) into (\ref{rxynonfact})
\begin{align}
\label{orxy}
 \half\,P_{XY}\Big[ 
 \big( \alpha_{ik}^F\,U_{ka}^Z\,U_{ia}^X \big)
 o_{FG}\,t_{ZW}
 \big( \alpha_{jl}^G\,U_{lb}^W\,U_{jb}^Y \big) \Big] 
 \rightarrow
 r_{XY}.
\end{align}
It is straightforward to show that both Eq. (\ref{ocore}) and (\ref{orxy}) can be evaluated with 
$N^5$ computational cost. Note that the quantity in the 
round brackets in Eq. (\ref{orxy}) does not change during the RR-CCSD iterations and hence it can 
be precomputed and stored.

The treatment of the $Z_{ij}^{ab}$ intermediate is based on the following representation
\begin{align}
\label{zsvd2}
 Z_{ij}^{ab} = \beta_{ij}^F\,z_{FG}\,\beta_{ab}^G.
\end{align}
with the expansion length (denoted $N_{\mathrm{Z}}$) proportional to the system size. The expansion vectors 
$\beta_{ij}^F$ and $\beta_{ab}^G$ are obtained from SVD of Eq. (\ref{zsvd}) 
reshaped as a rectangular $O^2\times V^2$ matrix $Z_{ij,ab}$, taking $N_{\mathrm{Z}}$ left- and right-singular vectors 
corresponding to the largest singular values. Similarly as for the $O^{ij}_{kl}$ 
intermediate, MP2 or MP3 amplitudes are used in Eq. (\ref{zsvd}), so that the expansion vectors do 
not have to be updated in every RR-CCSD iteration. To guarantee that the SVD can be calculated efficiently, we consider 
the left-hand- and right-hand-side multiplications, $Z_{ij}^{ab}\,y_{ij}$ and $Z_{ij}^{ab}\,y_{ab}$, by an arbitrary 
pair of trial vectors, $y_{ij}$ and $y_{ab}$. The necessary factorized formulas read
\begin{align}
 &Z_{ij}^{ab}\,y_{ij} = B_{kb}^Q\,U_{ka}^X\,\big(d_X\,B_{ij}^{QX}\,y_{ij}\big), \\
 &Z_{ij}^{ab}\,y_{ab} = B_{ij}^{QX}\,d_X\,\big(B_{kb}^Q\,U_{ka}^X\,y_{ab}\big).
\end{align}
Each elementary contraction in the above formulas can be computed with $N^4$ cost. As a result, 
the overall computational cost of the truncated SVD (with the rank $N_{\mathrm{Z}}$) of the $Z_{ij}^{ab}$ intermediate 
scales as the fifth power of the system size.

During each RR-CCSD iteration the core matrix $z_{FG}$ is calculated from the explicit formula
\begin{align}
 z_{FG} = \big(\beta_{ij}^F\,B_{ij}^{QX}\,t_{XY}\big)\,\big(B_{kb}^Q\,U_{ka}^Y\,\beta_{ab}^G\big),
\end{align}
exploiting the orthonormality relations $\beta_{ij}^F\,\beta_{ij}^G=\delta_{FG}$ 
and $\beta_{ab}^F\,\beta_{ab}^G=\delta_{FG}$ which result from properties of singular value decomposition.
Finally, the contribution to the residual is obtained from
\begin{align}
 \half P_{XY}\Big[ \big(\beta_{ki}^F\,U_{ka}^Z\,U_{ia}^X\big)\,
 t_{ZW}\,z_{FG}\,
 \big(U_{jc}^W\,\beta_{bc}^G\,U_{jb}^Y\big)
 \Big] \rightarrow r_{XY}.
\end{align}
The computational costs of evaluating the above expressions scale as $OV\naux\neig 
N_{\mathrm{Z}}\propto N^5$ and $OV\naux\neig^2\propto N^5$ in the rate determining steps. 
This proves that by exploiting the compressed formats of the intermediates (\ref{osvd2}) and 
(\ref{zsvd2}), and noting their effective rank scales only linearly with the system size, the 
overall cost of RR-CCSD iterations can be reduced to the level of $N^5$.

The issue that has not been discussed yet is the practical choice of the expansion lengths in Eqs. 
(\ref{osvd2}) and (\ref{zsvd2}), denoted by the symbols $N_{\mathrm{O}}$ and $N_{\mathrm{Z}}$. 
For convenience, we express both of them as multiples of the number of occupied orbitals in the system, i.e. 
$N_{\mathrm{O}}=mO$ and $N_{\mathrm{Z}}=m'O$, where the parameters $m$ and $m'$ are asymptotically 
independent of the system size. 
Clearly, the parameters $m$ and $m'$ should be chosen to provide an optimal balance between the 
truncation 
error and the computational overhead of performing the decompositions (\ref{osvd2}) and 
(\ref{zsvd2}). To recommend suitable value of $m$ and $m'$ we require a larger and a more diverse 
test set 
of molecules than the model systems considered previously in the paper. For this purpose, we employ 
the Adler-Werner benchmark set developed in Ref.~\onlinecite{adler11}. From this set we removed the hydrogen molecule 
as it is too small to be useful for the present purposes. This leaves $70$ molecules ranging in 
size from two to about twenty light atoms (H, C, N, O, S, Cl). The original geometries from Ref.~\onlinecite{adler11}
were used throughout. The $1s$ core orbitals were frozen in all correlated calculations; for 
the second row atoms the $2s$ and $2p$ orbitals were also excluded.

For all molecules in the Adler-Werner benchmark set we performed two groups of RR-CCSD 
calculations, both within the cc-pVDZ orbital basis. In the first group adopted no approximations 
to 
the $O^{ij}_{kl}$ and $Z_{ij}^{ab}$ intermediates. Therefore, the scaling of these calculations is 
$N^6$ and their purpose is only to provide the reference results for a given $\neig$. In the second 
group of the RR-CCSD calculations we employ the decomposed form of the $O^{ij}_{kl}$ and 
$Z_{ij}^{ab}$ 
intermediates and hence the scaling is $N^5$, but the approximations (\ref{osvd2}) and 
(\ref{zsvd2}) 
introduce an error. The magnitude of this error is quantified by comparing the corresponding 
results from the first and second group with the same $\neig$. This means that 
the error resulting from approximation of the doubly-excited amplitudes, Eq. (\ref{t2comp}), is 
not considered at this point. The only source of the error is the incompleteness of the 
representation of the $O^{ij}_{kl}$ and $Z_{ij}^{ab}$ intermediates themselves. The expansion lengths in Eqs. 
(\ref{osvd2}) and (\ref{zsvd2}) are controlled by the parameters $N_{\mathrm{O}}$ and 
$N_{\mathrm{Z}}$ which, in general, can be varied completely independently. However, in our 
preliminary calculations we found that near-optimal results are obtained for equal values of these 
parameters, i.e. $N_{\mathrm{O}}=N_{\mathrm{Z}}$. Accuracy gains attainable by an independent 
adjustment of $N_{\mathrm{O}}$ and $N_{\mathrm{Z}}$ are not worth the corresponding increase of 
the complexity. Therefore, we set $N_{\mathrm{O}}=N_{\mathrm{Z}}$ (or $m=m'$) from this point 
onward.

\begin{table}[t!]
\caption{\label{tab:oz-trunc1}
Statistical measures of relative errors (in percent) in the RR-CCSD/cc-pVDZ correlation energy 
(for $\neig=\nmo$) resulting from truncation of the expansions (\ref{osvd2}) and (\ref{zsvd2}) at 
length $N_{\mathrm{O}}=N_{\mathrm{Z}}=mO$, where $O$ is the number of occupied orbitals in the 
system and the value of the parameter $m$ is given in the first column. The statistics comes from 
RR-CCSD/cc-pVDZ calculations for 70 molecules contained in the Adler-Werner benchmark set~\cite{adler11}.
}
\begin{ruledtabular}
\begin{tabular}{lcccc}
 $m$ & mean  & mean abs. & standard  & max. abs. \\
     & error & error     & deviation & error     \\
\hline
$1$ & $-0.167$ & $0.168$ & $0.043$ & $0.340$ \\
$2$ & $-0.069$ & $0.077$ & $0.042$ & $0.165$ \\
$3$ & $-0.032$ & $0.037$ & $0.022$ & $0.066$ \\
$4$ & $-0.015$ & $0.016$ & $0.009$ & $0.032$ \\
\end{tabular}
\end{ruledtabular}
\end{table}

\begin{table}[t!]
\caption{\label{tab:oz-trunc2}
The same data as in Table \ref{tab:oz-trunc1}, but for $\neig=2\cdot\nmo$.
}
\begin{ruledtabular}
\begin{tabular}{lcccc}
 $m$ & mean  & mean abs. & standard  & max. abs. \\
     & error & error     & deviation & error     \\
\hline
$1$ & $-0.169$ & $0.169$ & $0.046$ & $0.321$ \\
$2$ & $-0.074$ & $0.079$ & $0.035$ & $0.152$ \\
$3$ & $-0.034$ & $0.042$ & $0.028$ & $0.097$ \\
$4$ & $-0.016$ & $0.017$ & $0.010$ & $0.033$ \\
\end{tabular}
\end{ruledtabular}
\end{table}

\begin{figure}[ht!]
\includegraphics[scale=0.70]{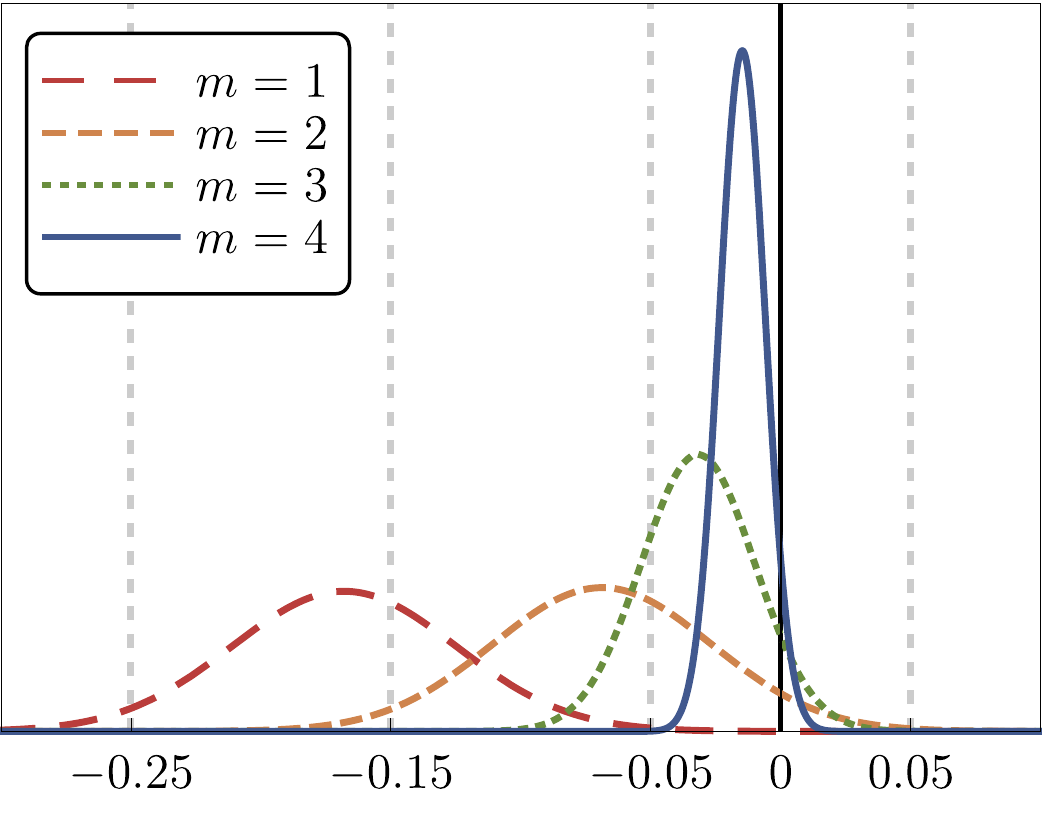}
\\\vspace{0.5cm}
\includegraphics[scale=0.70]{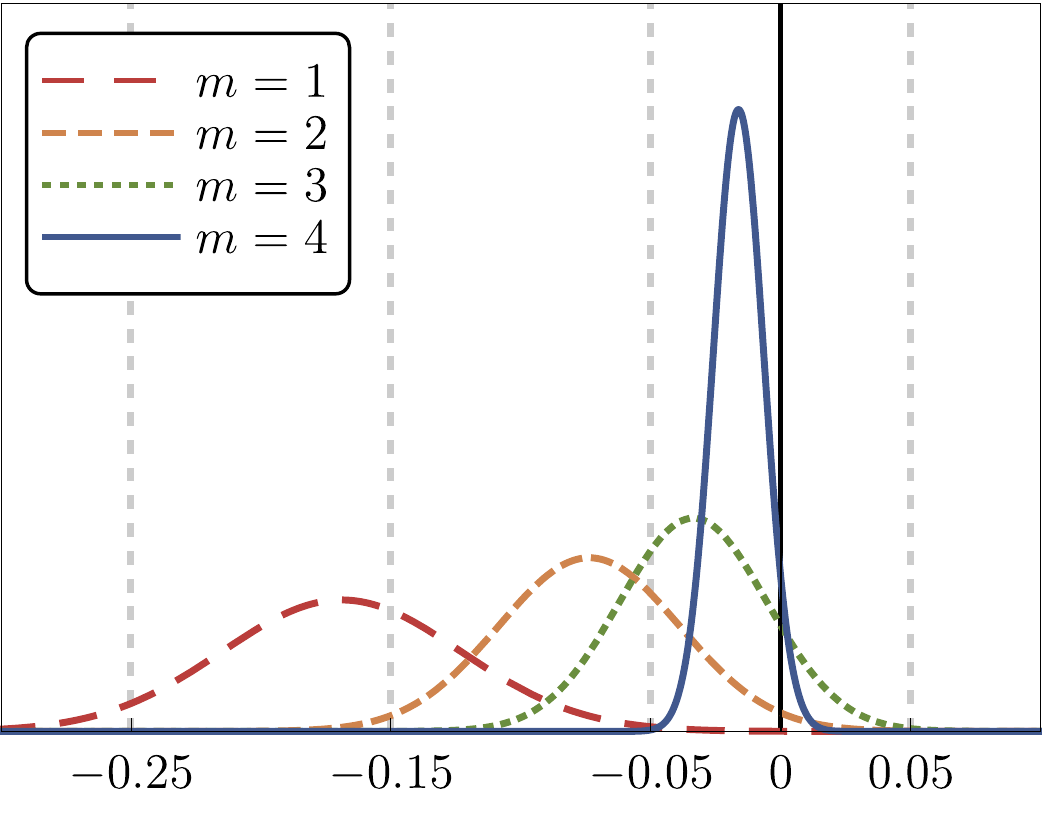}
\caption{\label{fig:oz-accuracy} Distribution of relative error (in percent) resulting from 
the truncation of the expansions (\ref{osvd2}) and (\ref{zsvd2}) at 
length $N_{\mathrm{O}}=N_{\mathrm{Z}}=mO$, where $O$ is the number of occupied orbitals in the 
system and the value of the parameter $m$ is given in the legend. The top and bottom panels 
correspond to $\neig=N_{\mathrm{MO}}$ and $\neig=2\cdot N_{\mathrm{MO}}$, respectively. The 
statistics comes from RR-CCSD/cc-pVDZ calculations for 70 molecules contained in the Adler-Werner 
benchmark set~\cite{adler11}.
}
\end{figure}

The calculations for the Alder-Werner benchmark set were performed for two representative examples 
of $\neig=N_{\mathrm{MO}}$ and $\neig=2\cdot N_{\mathrm{MO}}$, where $N_{\mathrm{MO}}$ is the total 
number of active orbitals in a given system (occupied plus virtual, neglecting the frozen-core orbitals). The expansion 
vectors $U_{ia}^X$ 
come from diagonalization of the MP2 amplitudes, but nearly the same results are obtained with the 
MP3 amplitudes. All data are given for $N_{\mathrm{O}}=N_{\mathrm{Z}}=mO$ with $m=1$, $2$, $3$, 
$4$. As the size of Alder-Werner benchmark set is substantial and comparison of individual results is 
cumbersome, we provide statistical error measures to access the quality of the results for each 
value of the control parameters $N_{\mathrm{O}}=N_{\mathrm{Z}}$. In Tables \ref{tab:oz-trunc1} and 
\ref{tab:oz-trunc2} we report such measures for \emph{relative} errors in the RR-CCSD correlation 
energies: the mean relative error, mean absolute relative error, standard deviation of 
the relative error and maximum absolute relative error. Since the molecules included in the 
Alder-Werner set vary considerably in size, relative errors are preferred due to their 
size-intensive character. We found that for each value of the parameters $m$ and $\neig$ the 
distributions of the (signed) relative errors are well approximated by the normal (Gaussian) 
distribution with the mean and standard deviation indicated in Tables \ref{tab:oz-trunc1} and 
\ref{tab:oz-trunc2}. In Fig. \ref{fig:oz-accuracy} we represent these distributions graphically to 
simplify the analysis of the results. 

In general, the approximate treatment of the $O^{ij}_{kl}$ and $Z_{ij}^{ab}$ intermediates leads to minor errors in the 
RR-CCSD energy. Even with the smallest expansion length considered here 
($N_{\mathrm{O}}=N_{\mathrm{Z}}=O$) more than 99.8\% of the correlation energy is recovered. Beyond 
this point the error vanishes with increasing $N_{\mathrm{O}}$ and $N_{\mathrm{Z}}$ at a rate close 
to exponential; with $N_{\mathrm{O}}=N_{\mathrm{Z}}=4O$ the relative error decreases below 0.02\%. 
Moreover, the error distributions for $\neig=N_{\mathrm{MO}}$ and $\neig=2\cdot N_{\mathrm{MO}}$ 
are remarkably similar, indicating that the truncation error of Eqs. (\ref{osvd2}) and 
(\ref{zsvd2}) is practically independent of the dimension of the double excitation subspace.
It is also noteworthy that the approximate treatment of the $O^{ij}_{kl}$ and $Z_{ij}^{ab}$ intermediates 
systematically underestimates the correlation energy. In summary, the results obtained with 
$N_{\mathrm{O}}=N_{\mathrm{Z}}=3O$ are, on average, sufficiently accurate for routine applications, 
with the mean error of only about 0.03\%. However, the standard deviation of the error obtained 
with $N_{\mathrm{O}}=N_{\mathrm{Z}}=3O$ is still substantial compared to its mean, and hence the 
error distribution is rather broad. From the practical point of view, this negatively impacts the 
reliability of the method since it is not uncommon to encounter ''outliers`` with 
unexpectedly large errors. Therefore, we recommend that $N_{\mathrm{O}}=N_{\mathrm{Z}}=4O$ is used 
in actual applications where the reference results are not available. As is evident from Fig. 
\ref{fig:oz-accuracy} the error distribution for $N_{\mathrm{O}}=N_{\mathrm{Z}}=4O$ is much narrower 
than for $N_{\mathrm{O}}=N_{\mathrm{Z}}=3O$ which translates into a decreased likelihood of 
encountering the outliers. At the same time, the jump from $N_{\mathrm{O}}=N_{\mathrm{Z}}=3O$ to 
$N_{\mathrm{O}}=N_{\mathrm{Z}}=4O$ leads to only a minor increase of the overall computational 
timings. Therefore, all numerical results reported further in this work were obtained with 
$N_{\mathrm{O}}=N_{\mathrm{Z}}=4O$.

It is also important to study how the augmentation of the basis set with diffuse functions influences the accuracy of the approximations adopted for the $O^{ij}_{kl}$ and $Z_{ij}^{ab}$ intermediates. To address this question, we performed analogous calculations for the Alder-Werner benchmark set as described in the previous paragraph, but employing the aug-cc-pVDZ basis set~\cite{kendall92}. As the results are essentially insensitive to the value of the $\neig$ parameter, in supplementary material we report the data for the representative case of $\neig=N_{\mathrm{MO}}$. In summary, the augmentation of the basis set has a tiny influence on the accuracy of the decomposition applied to the $O^{ij}_{kl}$ and $Z_{ij}^{ab}$ intermediates. As an example, for the recommended expansion length ($N_{\mathrm{O}}=N_{\mathrm{Z}}=4O$) the mean relative error in the correlation energy increases only by about one thousandth of a percent upon the augmentation.
Therefore, the approach proposed in the present work with the recommended expansion length ($N_{\mathrm{O}}=N_{\mathrm{Z}}=4O$) can be safely applied in calculations with diffuse basis set functions.

It is worthwhile to point out that there are two equivalent ways of selecting the dimension of the 
subspace used for the expansion of the $O^{ij}_{kl}$ and $Z_{ij}^{ab}$ intermediates. The first is to specify the 
values of parameters $N_{\mathrm{O}}$ and $N_{\mathrm{Z}}$ by relating it to another quantity that scales linearly 
with the system size, as was done in the calculations above ($N_{\mathrm{O}}=N_{\mathrm{Z}}=mO$). In this way the 
values of $N_{\mathrm{O}}$ and $N_{\mathrm{Z}}$ are known before the calculations are even started. However, an 
alternative idea is to form the subspace by taking all singular vectors with singular values larger than the 
predefined numerical threshold $\epsilon$. In other words, $N_{\mathrm{O}}$ and $N_{\mathrm{Z}}$ are found 
dynamically during the SVD procedure based on the parameter $\epsilon$ provided by the user. The main advantage of 
knowing $N_{\mathrm{O}}$ and $N_{\mathrm{Z}}$ in advance, besides the fact that the computational cost and scaling 
of the method can be judged more easily, is purely technical. In fact, implementation of an SVD procedure that 
dynamically adjusts the expansion length in each iteration is significantly more complicated than with fixed 
$N_{\mathrm{O}}$ and $N_{\mathrm{Z}}$, and can be expected to be also less efficient, especially in parallel 
environment. To address the question whether is worth the effort to develop an algorithm that dynamically 
adjusts the expansion length, we performed calculations for a subset of the Alder-Werner benchmark set. In the 
supplementary material we provide a comparison of the fixed and dynamic approach for one molecule we found 
representative of the whole set. As an example, the dynamic adjustment based on $\epsilon$ reduces 
the expansion length by about 15\% if the relative accuracy of 99.95\% is desired. While this reduction is 
non-negligible, this finding has to be understood in a broader context. In fact, in the next section we provide a 
comparison of timings of various steps of the RR-CCSD calculations. We show that the determination of the 
subspace used for the expansion of the $O^{ij}_{kl}$ and $Z_{ij}^{ab}$ intermediates constitutes less than 5\% of 
the total RR-CCSD timings. Therefore, while the cost of handling the $O^{ij}_{kl}$ and $Z_{ij}^{ab}$ intermediates 
alone may be reduced using the dynamic adjustment of the expansion length, this would lead to only a minuscule 
decrease of the overall cost of the RR-CCSD method.

Finally, let us discuss how the approximations to the $O^{ij}_{kl}$ and $Z_{ij}^{ab}$ intermediates
adopted in the present work affect the size-extensivity of the energy and how the present approach can be extended to calculation of, e.g. molecular properties. Similarly as discussed in Ref.~\onlinecite{parrish19}, there are two necessary conditions that the expansion basis used in Eqs. (\ref{osvd2}) and (\ref{zsvd2}) must fulfill. First, the basis vectors must be obtained using approximate doubly-excited amplitudes coming from a method that is size-extensive itself. In the present work MP2 or MP3 amplitudes are used which both fulfill this requirement. Second, the expansion length in Eqs. (\ref{osvd2}) and (\ref{zsvd2}) must be a size-extensive quantity and hence increase linearly with the system size. We verified numerically using the model systems of linear alkanes and water clusters considered above, that the original~\cite{parrish19} and the modified RR-CCSD variants retain the size-extensive property.

Moving on to the calculation of the RR-CCSD properties, in the original formulation of the RR-CCSD method described in Ref.~\onlinecite{parrish19}, the projectors $U_{ia}^X$ are assumed to be perturbation-independent. Therefore, the Lagrangian formulation introduced in Ref.~\onlinecite{parrish19} enables straightforward calculation of molecular properties, using a similar approach as in the conventional coupled-cluster theory. It is reasonable to adopt the same condition for expansion basis in Eqs. (\ref{osvd2}) and (\ref{zsvd2}) which leaves only the core matrices $o_{FG}$ and $z_{FG}$ as additional perturbation-dependent quantities whose response must be taken into account explicitly. In order to extend the RR-CCSD Lagrangian in this direction one requires to specify the stationary conditions that $o_{FG}$ and $z_{FG}$ fulfill. Taking the former matrix as an example, let us define the following quantity
\begin{align}
 \tau = \sum_{ijkl} \Big[ O^{ij}_{kl} - \sum_{FG}\alpha_{ik}^F\,o_{FG}\,\alpha_{jl}^G \Big].
\end{align}
One can show that at convergence of the RR-CCSD iterations this quantity is stationary with respect to $o_{FG}$ in the sense that $\frac{\partial \tau}{\partial o_{FG}}=0$ as this condition becomes equivalent to Eq. (\ref{ocore}). Therefore, the modified RR-CCSD Lagrangian is defined by adding a term $\zeta_{FG}\,\frac{\partial \tau}{\partial o_{FG}}$, where $\zeta_{FG}$ is a new set of Lagrange multipliers. By minimization of this modified Lagrangian with respect to all perturbation-dependent parameters ($t_{XY}$, $o_{FG}$) one obtains equations that have to be solved to find the multipliers. As a result, the Lagrangian is stationary with respect to all parameters which enables straightforward determination of molecular properties using the extended Hellmann-Feynman theorem.

\subsection{\label{subsec:accud} Accuracy and efficiency of the method}

\begin{table}[t!]
\caption{\label{tab:final-error-mp2}
Statistical measures of relative errors (in percent) in the RR-CCSD correlation energy with 
respect to the exact CCSD results. The dimension of the excitation subspace ($\neig$) is expressed 
as $\neig=x\cdot N_{\mathrm{MO}}$, where $N_{\mathrm{MO}}$ is the total number of orbitals in the 
system. The subspace of double excitations was obtained by diagonalization of MP2 amplitudes. The 
statistics comes from calculations for 70 molecules contained in the Adler-Werner benchmark set~\cite{adler11}.
}
\begin{ruledtabular}
\begin{tabular}{lcccc}
 $x$ & mean  & mean abs. & standard  & max. abs. \\
     & error & error     & deviation & error     \\
\hline
\multicolumn{5}{c}{cc-pVDZ basis set} \\
\hline
$0.5$ & 0.349 & 0.365 & 0.310 & 1.419 \\
$1.0$ & 0.608 & 0.608 & 0.268 & 2.034 \\
$1.5$ & 0.285 & 0.290 & 0.128 & 0.538 \\
$2.0$ & 0.197 & 0.203 & 0.118 & 0.427 \\
$2.5$ & 0.227 & 0.228 & 0.195 & 1.601 \\
\hline
\multicolumn{5}{c}{cc-pVTZ basis set} \\
\hline
$0.5$ & 0.342 & 0.343 & 0.164 & 1.275 \\
$1.0$ & 0.309 & 0.312 & 0.140 & 1.124 \\
$1.5$ & 0.088 & 0.119 & 0.113 & 0.333 \\
$2.0$ & 0.106 & 0.124 & 0.149 & 0.300 \\
$2.5$ & 0.123 & 0.131 & 0.079 & 0.259 \\
\end{tabular}
\end{ruledtabular}
\end{table}

\begin{table}[t!]
\caption{\label{tab:final-error-mp3}
The same data as in Table \ref{tab:final-error-mp2}, but the subspace of double excitations was 
obtained by diagonalization of MP3 amplitudes.
}
\begin{ruledtabular}
\begin{tabular}{lcccc}
 $x$ & mean  & mean abs. & standard  & max. abs. \\
     & error & error     & deviation & error     \\
\hline
\multicolumn{5}{c}{cc-pVDZ basis set} \\
\hline
$0.5$ & 0.100 & 0.286 & 0.371 & 1.128 \\
$1.0$ & 0.468 & 0.468 & 0.216 & 2.053 \\
$1.5$ & 0.294 & 0.294 & 0.081 & 0.487 \\
$2.0$ & 0.158 & 0.158 & 0.052 & 0.298 \\
$2.5$ & 0.068 & 0.069 & 0.034 & 0.223 \\
\hline
\multicolumn{5}{c}{cc-pVTZ basis set} \\
\hline
$0.5$ & $-$0.203           & 0.257 & 0.165 & 0.917 \\
$1.0$ & \phantom{$-$}0.259 & 0.258 & 0.143 & 1.372 \\
$1.5$ & \phantom{$-$}0.121 & 0.121 & 0.044 & 0.422 \\
$2.0$ & \phantom{$-$}0.037 & 0.039 & 0.027 & 0.114 \\
$2.5$ & \phantom{$-$}0.004 & 0.018 & 0.024 & 0.076 \\
\end{tabular}
\end{ruledtabular}
\end{table}

\begin{figure}[ht!]
\includegraphics[scale=1.00]{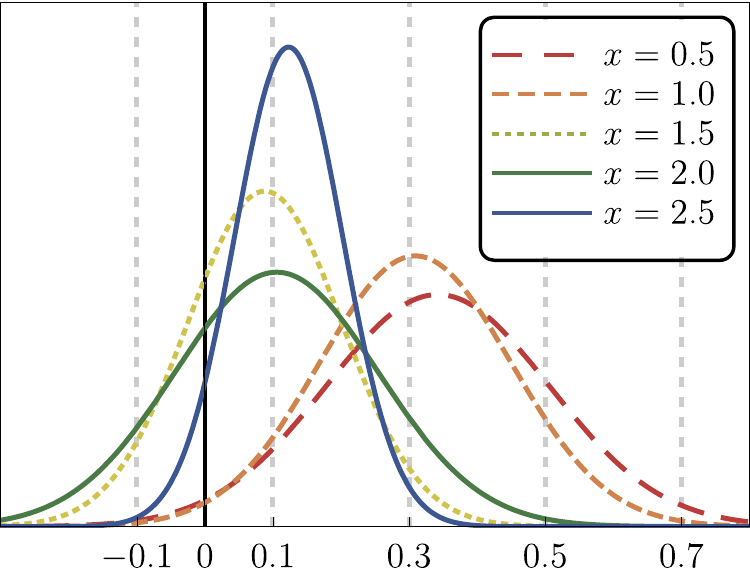}
\\\vspace{0.5cm}\hspace{0.1cm}
\includegraphics[scale=1.02]{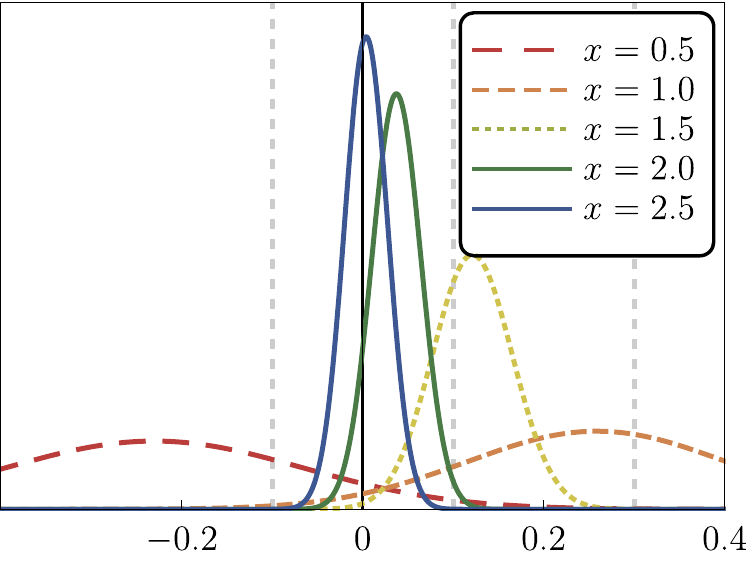}
\caption{\label{fig:final-accuracy-vtz} Distribution of relative error (in percent) in the 
RR-CCSD/cc-pVTZ correlation energy with respect to the exact CCSD/cc-pVTZ results. The dimension of 
the excitation subspace ($\neig$) is expressed as $\neig=x\cdot N_{\mathrm{MO}}$, where 
$N_{\mathrm{MO}}$ is the total number of orbitals in the system. The excitations subspace was 
obtained by diagonalization of MP2 amplitudes (top panel) or MP3 amplitudes (bottom panel). The 
statistics comes from calculations for 70 molecules contained in the Adler-Werner benchmark set~\cite{adler11}. See 
the supplementary material for analogous results obtained within cc-pVDZ basis set.
}
\end{figure}

In this section we study the cumulative error of the RR-CCSD method incurred by the truncation of 
the double excitation subspace (as a function of $\neig$) and the approximate treatment of the 
$O^{ij}_{kl}$ and $Z_{ij}^{ab}$ intermediates. In contrast to Sec. \ref{subsec:scale2}, the exact 
CCSD correlation energies obtained within the same basis set are treated as a reference here. The RR-CCSD calculations 
for the whole 
Alder-Werner benchmark set were performed with $\neig=x\cdot N_{\mathrm{MO}}$, where $x$ is a 
parameter taking values $x=0.50$, $1.00$, $1.50$, $2.00$, $2.50$, and with the recommended 
$N_{\mathrm{O}}=N_{\mathrm{Z}}=4O$.  All calculations were performed within the cc-pVDZ and cc-pVTZ 
basis sets. We 
consider two variants of the method, where the subspace of double excitations is obtained by 
diagonalization of either MP2 or MP3 amplitudes. Statistical measures of relative errors in the 
RR-CCSD correlation energy with respect to the exact CCSD results for both variants 
are given in Tables \ref{tab:final-error-mp2} and \ref{tab:final-error-mp3}. Similarly as in the 
previous section, we found that the corresponding error distributions are approximately normal and
are given in Fig. \ref{fig:final-accuracy-vtz} in the case of the cc-pVTZ basis set. For brevity, plots 
representing analogous results obtained within the cc-pVDZ basis were moved to the supplementary material.

From Tables \ref{tab:final-error-mp2} and \ref{tab:final-error-mp3} one concludes that the MP2 
excitation subspace is not well-suited for highly accurate calculation of the correlation energy. 
While for smaller values of $\neig$ ($x=0.5-1.0$) the MP2 basis performs only marginally worse than MP3, for larger 
$x$ the former method stalls in terms of relative accuracy at the level of $0.2-0.3\%$ in the cc-pVDZ basis 
and $0.1-0.2\%$ in the cc-pVTZ basis. If errors of this magnitude are acceptable, the MP2 basis is 
a reasonable choice due to the marginal cost of its determination. However, it is uneconomical to 
aim at 
the accuracy levels of $0.1\%$ or better with the MP2 basis, as the error decays too slowly as a 
function of $\neig$. As a result, in accurate calculations where relative errors below $0.1\%$ are expected, 
MP3 amplitudes are necessary. The MP3 basis does not suffer from the diminishing 
returns as $\neig$ is increased, and the convergence with respect to $\neig$ is fast even in the 
high-accuracy regime. For $\neig=2\nmo$ the MP3 basis achieves relative accuracy of about $0.1\%$ 
in the cc-pVDZ basis and about $0.04\%$ in the cc-pVTZ basis. As a side note, this demonstrates 
that the amplitudes obtained within the larger basis set are more ''compressible`` and we expect 
this 
phenomenon to prevail as the basis set is increased further. To sum up, we recommend that 
$\neig=2\nmo$ is used to fix the expansion length in Eq. (\ref{t2comp}), both in the case of MP2 
and MP3 excitation bases. This value strikes a balance between the computational cost of the 
RR-CCSD procedure and the accuracy level it offers.

\begin{figure}[t]
 \includegraphics[scale=0.70]{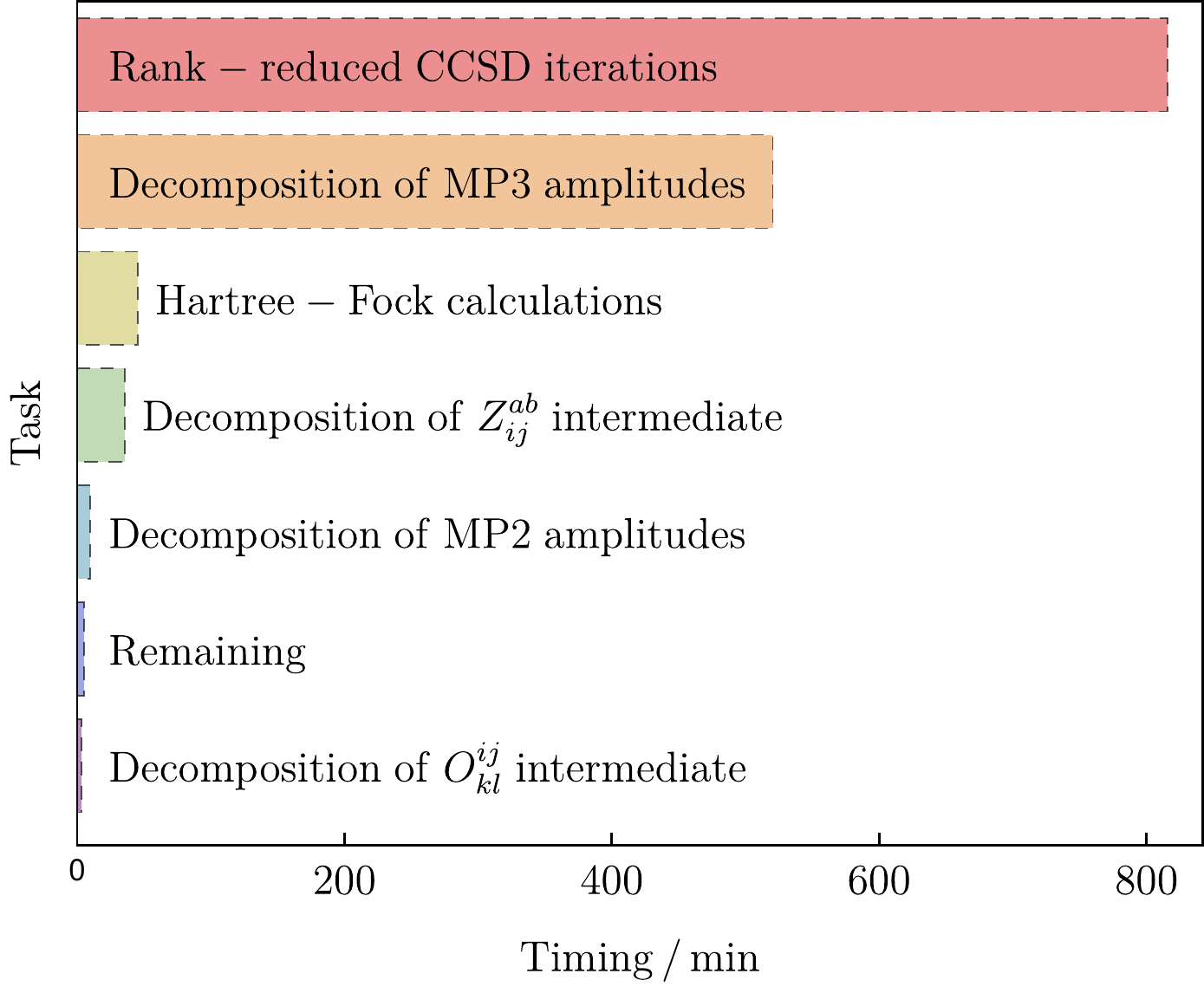}
 \caption{\label{fig:breakdown} Breakdown of the RR-CCSD/cc-pVTZ wall clock timings for the ethylbenzene molecule 
(C$_8$H$_{10}$, 380 molecular orbitals). The timings of the Hartree-Fock calculations are given for comparison (default 
\textsc{Gamess} settings, density matrix convergence threshold $10^{-8}$). The category ''remaining`` includes minor 
tasks such as updating the coupled-cluster amplitudes, evaluating the energy, etc. The calculations were 
performed using a single core of AMD Opteron\texttrademark\, Processor 6174 (no parallelization).}
\end{figure}

To study the computational efficiency of the RR-CCSD method we first analyze which 
steps of the RR-CCSD algorithm bring the dominant contribution to the total timings. To this end, we consider the 
largest molecule in the Adler-Werner set (ethylbenzene, 
C$_8$H$_{10}$) employing the cc-pVTZ basis set (380 molecular orbitals). We employ the recommended 
values of the truncation parameters, namely $\neig=2\nmo$, $N_{\mathrm{O}}=N_{\mathrm{Z}}=4O$.
In Fig. \ref{fig:breakdown} we present a breakdown of the total RR-CCSD wall clock timings into 
individual components of the algorithm. For comparison, timings of the 
Hartree-Fock calculations obtained with the default \textsc{Gamess} settings and the density 
matrix convergence threshold of $10^{-8}$ are also given. From Fig. \ref{fig:breakdown} it is clear 
that two steps of the proposed RR-CCSD algorithm, namely diagonalization of the MP2 amplitudes and 
decomposition of the $O^{ij}_{kl}$ intermediate, are essentially negligible in terms of the 
computational effort. The treatment of $Z_{ij}^{ab}$ intermediate is somewhat more costly, but 
still comparable with the conventional Hartree-Fock calculations. Therefore, while the 
decomposition of 
the $O^{ij}_{kl}$ and $Z_{ij}^{ab}$ intermediates advocated in this work scales formally as $N^5$ 
with the system size, the prefactor of this procedure is small and hence this step does not 
contribute significantly to the overall workload. On the other hand, the diagonalization of the MP3 
amplitudes introduces a considerable overhead. With the current implementation the total costs of 
finding the MP3 excitation subspace and the subsequent RR-CCSD iterations are comparable.

In general, the pilot implementation of the RR-CCSD method reported in this paper is limited to $600-700$ basis set 
functions, but this limitation results from overuse of disk files for storage of intermediate quantities. We are 
currently working on an improved implementation that avoids this problem and hence should vastly exceed the 
capabilities of the conventional CCSD implementations. In fact, shortly after the present manuscript was submitted 
for publication, another article was published~\cite{hohenstein21a} that describes a GPU-accelerated parallel 
implementation of the RR-CCSD method applicable up to ca. 2000 basis set functions. Notably, this was achieved 
without any special treatment of the non-factorizable terms in the RR-CCSD residual, and hence the implementation 
reported in Ref.~\onlinecite{hohenstein21a} scales as $N^6$. Because of that, the non-factorizable terms turned out to 
be the bottleneck of this implementation for large systems. Clearly, an efficient implementation of the RR-CCSD 
method with the treatment of the non-factorizable terms described in the present work should be capable of reaching 
even larger systems. An alternative method of eliminating the non-factorizable terms based on the tensor 
hypercontraction (THC) decomposition of the doubly-excited amplitudes has also been reported 
recently~\cite{hohenstein21b}.

\begin{figure}[t]
 \includegraphics[scale=0.70]{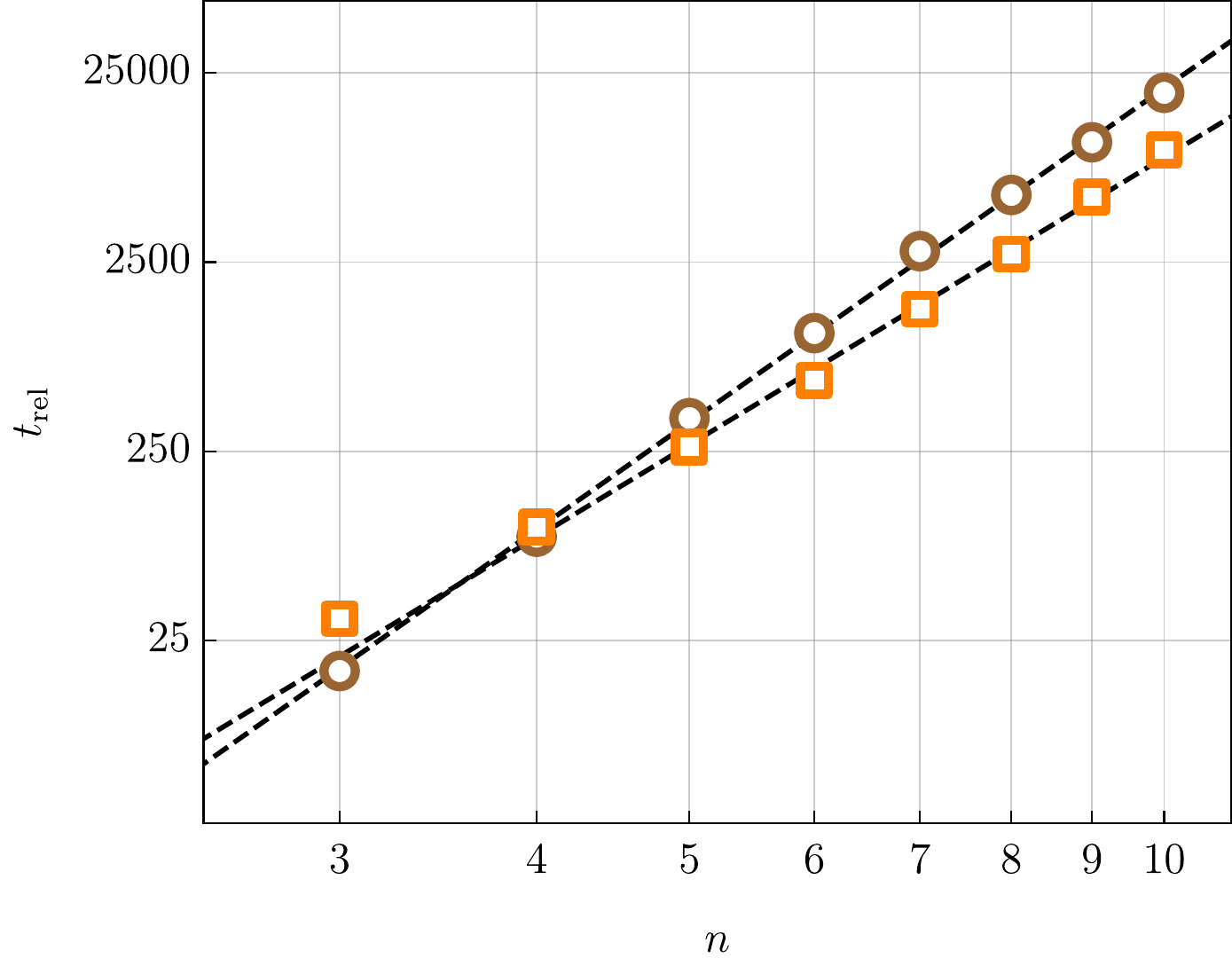}
 \caption{\label{fig:timings} Timings of the RR-CCSD/cc-pVDZ calculations with $\neig=2\nmo$, 
$N_{\mathrm{O}}=N_{\mathrm{Z}}=4O$ (orange squares) and of the exact CCSD/cc-pVDZ calculations (brown circles) 
for linear alkanes C$_n$H$_{2n+2}$ as a function of the chain length, $n$ (logarithmic scale on both axes). The timings 
are given in relation to the CCSD calculations for methane ($t_{\mathrm{rel}}$). The calculations were performed using 
a single core of AMD Opteron\texttrademark\, Processor 6174 (no parallelization). The black dashed lines were obtained 
by least-squares fitting of the data points with the functional form $a\cdot n^b$ (represented by linear functions on 
doubly-logarithmic scale).
}
\end{figure}

Finally, to compare the performance of the RR-CCSD and the exact CCSD methods, we analyze their 
timings for the linear alkanes C$_n$H$_{2n+2}$ previously considered in Sec. \ref{subsec:scale1}. 
In Fig. \ref{fig:timings} we report the total wall clock times of the CCSD and RR-CCSD calculations 
(cc-pVDZ basis set) as a function of the alkanes chain length, $n$. Similarly as above, we employ 
the recommended values of the truncation parameters, namely $\neig=2\nmo$, 
$N_{\mathrm{O}}=N_{\mathrm{Z}}=4O$. The timings for the RR-CCSD method include determination of the 
double-excitation subspace, treatment of the $O^{ij}_{kl}$ and 
$Z_{ij}^{ab}$ intermediates and all other steps discussed in the previous paragraph.
To convert the data into dimensionless units the timings are given in relation to the CCSD 
calculations for methane. 

To confirm numerically that the scaling observed in Fig. \ref{fig:timings} matches the theoretical 
predictions from Sec. \ref{subsec:scale2} we fitted the timings with the functional form $a\cdot 
n^b$ for $n=3-10$. We obtained the exponents $b=5.04$ for the RR-CCSD method and $5.86$ for the 
exact CCSD theory, in a good agreement with the conclusions of Sec. \ref{subsec:scale2}. Another important issue is to 
estimate for how large systems the RR-CCSD method becomes advantageous in terms of computational timings. From 
Fig. \ref{fig:timings} we see that the break-even point for linear alkanes occurs rather early, around 
$n=4$ (butane). Beyond this point the RR-CCSD is less computationally expensive, with the gap 
increasing linearly with the molecular size. For the fixed values of the control parameters 
($\neig=2\nmo$, $N_{\mathrm{O}}$, and $N_{\mathrm{Z}}$) we expect this finding to be approximately 
valid also for other systems, with the break-even point occurring for about thirty or so active 
electrons.

\section{\label{sec:triples} Perturbative triples corrections}

\subsection{\label{subsec:trip-def} Problem formulation}

It is well-known that the conventional coupled-cluster theory with only single and double 
excitations included in the cluster operator is insufficient to obtain chemically-accurate 
predictions. In fact, only after triple excitations are accounted for, levels of accuracy of 1 
kcal/mol or better become routinely accessible\cite{bartlett90,hopkins04,bak00,tajti04,karton06,riley10}. However, the 
computational cost of the 
coupled-cluster theory with full inclusion of triple excitations is prohibitively high for systems 
comprising more than a few non-hydrogen atoms. For this reason, numerous approximate schemes were 
proposed to 
account for the effects of triple excitations in a more affordable way without sacrificing too much 
accuracy. The most widely-used method of this type is the CCSD(T) theory of Raghavachari 
\emph{et al.}\cite{ragha89}, frequently referred to as the ''gold standard`` of quantum chemistry.

The CCSD(T) theory is perturbative in nature and adds a non-iterative correction, denoted 
shortly $E_{\mathrm{(T)}}$ further in the text, on top of the standard CCSD energy. This correction 
is a sum of two terms
\begin{align}
\label{ept}
 E_{\mathrm{(T)}} = E_{\mathrm{T}}^{[4]} + E_{\mathrm{ST}}^{[5]}
\end{align}
defined as
\begin{align}
\label{e4t}
 E_{\mathrm{T}}^{[4]} = \langle T_2^{\mathrm{SD}} | \Big[ W, T_3 \Big] \rangle,
\end{align}
and 
\begin{align}
\label{e5st}
 E_{\mathrm{ST}}^{[5]} = \langle T_1^{\mathrm{SD}} | \Big[ W, T_3 \Big] \rangle,
\end{align}
where $T_1^{\mathrm{SD}}$ and $T_2^{\mathrm{SD}}$ is an abbreviation for cluster 
operators~(\ref{t12}) obtained at the CCSD level of theory. The triple excitation operator present 
in Eqs. (\ref{ept}) and (\ref{e4t}) is given by the standard formula
\begin{align}
 T_3 = \frac{1}{6}\,t_{ijk}^{abc} \,E_{ai}\,E_{bj}\,E_{ck},
\end{align}
where the amplitudes $t_{ijk}^{abc}$ are approximated as
\begin{align}
\label{t32}
&t_{ijk}^{abc} \approx (\epsilon_{ijk}^{abc})^{-1}\,\Gamma_{ijk}^{abc},\\
&\Gamma_{ijk}^{abc} = \langle \,_{ijk}^{abc} | \big[ W, T_2^{\mathrm{SD}} \big]\rangle,
\end{align}
and $\epsilon_{ijk}^{abc} = \epsilon_i^a + \epsilon_j^b + \epsilon_k^c$ is the three-particle 
energy denominator. The evaluation of the corrections $E_{\mathrm{T}}^{[4]}$ and 
$E_{\mathrm{ST}}^{[5]}$ scales as $N^7$ with the system size, if no further 
approximations are introduced.

A natural extension of the RR-CCSD method is to evaluate the $E_{\mathrm{(T)}}$ correction with the 
singly- and doubly-excited amplitudes obtained within the rank-reduced formalism and add it to the 
RR-CCSD correlation 
energy. The resulting method is abbreviated RR-CCSD(T) further in the paper and we expect it to 
faithfully reproduce the exact CCSD(T) results. Unfortunately, the steep $N^7$ scaling of the 
$E_{\mathrm{(T)}}$ correction would constitute a severe bottleneck in applications to larger systems, in 
comparison to the more subdued $N^5$ cost of the RR-CCSD iterations. To the best of our knowledge, 
the $N^7$ scaling cannot be reduced by exploiting solely the rank-reduced form of the $T_2^{\mathrm{SD}}$ amplitudes 
given by Eq. (\ref{t2comp}). Therefore, additional approximations are needed to make the 
RR-CCSD(T) method advantageous which is explored in the subsequent section.

\subsection{\label{subsec:trip-comp} Compression of the triply excited amplitudes}

To decrease the scaling of the $E_{\mathrm{(T)}}$ correction removal of the 
three-particle energy denominator from Eq. (\ref{t32}) is a priority. To this end, 
we employ the same min-max quadrature as in Sec. \ref{subsec:diag} for the MP2 and MP3 amplitudes. 
The Laplace transformation formula now reads
\begin{align}
\label{ltd3}
 (\epsilon_{ijk}^{abc})^{-1} = w_g\,e^{-t_g \big( \epsilon_i^a + \epsilon_j^b + 
 \epsilon_k^c \big)},
\end{align}
where the notation for all quantities in the same as in Sec. \ref{subsec:diag}. As demonstrated in 
the paper of Constans \emph{et al.}\cite{pere00} application of the Laplace transformation of the three-particle 
energy denominator alone is sufficient to reduce the scaling of the $E_{\mathrm{T}}^{[4]}$ and $E_{\mathrm{ST}}^{[5]}$ 
terms to the level of $N^6$. Unfortunately, the subsequent factorization yields numerous terms with a large prefactor 
($OV^5$ scaling) and hence the computational benefits are achieved only for very large systems. To 
avoid this problem, in Ref.~\onlinecite{pere00} the $E_{\mathrm{T}}^{[4]}$ and $E_{\mathrm{ST}}^{[5]}$ corrections 
were rewritten in terms of CCSD natural orbitals, enabling an efficient screening procedure to 
eliminate negligible contributions. Here we propose an alternative approach where the 
triply-excited amplitudes (\ref{t32}) are approximately represented in the Tucker-3 format\cite{tucker66}
\begin{align}
\label{t3comp}
 t_{ijk}^{abc} = t_{ABC}\,V_{ia}^A\,V_{jb}^B\,V_{kc}^C.
\end{align}
In analogy to Eq. (\ref{t2comp}) the basis vectors $V_{ia}^A$ span the subspace of triple 
excitations. The dimension of this subspace, i.e. the length of the summations over the variables 
$A$, $B$, $C$ in Eq. (\ref{t3comp}), is denoted $\ntri$ further in the paper. Note that the Tucker-3 
decomposition has been recently applied to the full CCSDT method\cite{lesiuk20} with the quantities $V_{ia}^A$ 
obtained by higher-order singular value decomposition\cite{delath00,vannie12} (HOSVD) of Eq. 
(\ref{t32}). More importantly, it 
has also been shown that $\ntri$ has to scale only linearly with the system size in order to maintain a constant 
relative accuracy in the correlation energy. As demonstrated further in the text, this allows to calculate the 
$E_{\mathrm{T}}^{[4]}$ and $E_{\mathrm{ST}}^{[5]}$ corrections with the cost proportional to $N^6$ and an acceptable 
prefactor. Unfortunately, the HOSVD method adopted in Ref.~\onlinecite{lesiuk20} is not feasible in the present context 
due to a prohibitive cost. To achieve the decomposition 
(\ref{t3comp}) we thus employ a variant of the higher-order orthogonal iteration (HOOI) 
procedure\cite{delath00b,elden09} which is a particular method of minimizing the least-squares error
\begin{align}
\label{hooi_ls}
 \tau = \sum_{ijk}\sum_{abc} \bigg[ t_{ijk}^{abc} - t_{ABC}\,V_{ia}^A\,V_{jb}^B\,V_{kc}^C \bigg]^2,
\end{align}
subject to the orthonormality condition $V_{ia}^A\,V_{ia}^B=\delta_{AB}$. While HOOI is a 
well-known tool in the mathematics literature, we are aware of only one paper where it is used in the context
of the electronic structure theory\cite{bell10}. 

To apply the HOOI procedure to the $t_{ijk}^{abc}$ 
amplitudes one requires an initial guess of the basis vectors $V_{ia}^A$. In our implementation this 
guess is generated by taking the basis vectors $U_{ia}^X$ (obtained previously for the 
doubly-excited amplitudes) that correspond to the largest absolute eigenvalues. While a more 
sophisticated and effective guess can definitely be proposed, we found this simple and self-contained 
approach to be entirely adequate. The HOOI procedure consists of two basic steps
\begin{itemize}
 \item evaluate the partially contracted quantity
 \begin{align}
  t_{ia,BC}=t_{ijk}^{abc}\,V_{jb}^B\,V_{kc}^C,
 \end{align}
 with current estimation of the factors $V_{ia}^A$;
 \item compute SVD of $t_{ia,BC}$ reshaped as a $OV\times\ntri^2$ matrix and take left-singular 
vectors that correspond to the largest singular values as the next $V_{ia}^A$.
\end{itemize}
These steps are repeated until convergence; the choice of the stopping criteria is discussed 
further in the text. The computational costs of both steps scale as the fifth power of the system 
size. This is straightforward to prove in the case of the second step by noting that we need to 
find only a subset of $\ntri$ singular vectors with the largest singular values. As $\ntri$ is 
proportional to the system size, application of the decomposition algorithm described in Sec. 
\ref{subsec:bidiag} immediately results in $N^5$ cost. The first step of 
the HOOI algorithm can also be accomplished with the same scaling. To show that we recall the 
explicit 
formula for the quantity $\Gamma_{ijk}^{abc}$ defined in Eq. (\ref{t32}):
\begin{align}
\label{gammaijk}
\begin{split}
 \Gamma_{ijk}^{abc} &=
 \Big( 1 + P_{jk}^{bc} \Big) \Big( 1 + P_{ij}^{ab} + P_{ik}^{ac}\Big) \\
 &\times\Big[ t_{ij}^{ad}\,(ck|bd) - t_{il}^{ab}\,(ck|lj) \Big],
\end{split}
\end{align}
By inserting the density-fitting form of the two-electron integrals together with the decomposed 
amplitudes (\ref{t2comp}) and defining the intermediate
\begin{align}
 \bar{D}_{jb}^{QX} = \Big( B_{bd}^Q\,U_{jd}^Y - B_{lj}^Q\,U_{lb}^Y \Big)\,t_{XY},
\end{align}
we bring Eq. (\ref{gammaijk}) into a simpler form
\begin{align}
 \Gamma_{ijk}^{abc} &= \Big( 1 + P_{jk}^{bc} \Big) \Big( 1 + P_{ij}^{ab} + P_{ik}^{ac}\Big)
 U_{ia}^X\,\bar{D}_{jb}^{QX}\,B_{kc}^Q.
\end{align}
This leads to the working expression for the partially contracted quantity $t_{ia,BC}$ required in 
the HOOI algorithm
\begin{align}
\begin{split}
 t_{ia,BC} &= \big(1+P_{BC}\big) w_g\,e^{-t_g \epsilon_i^a} \\
 &\times\Big[ U_{ia}^X\,\Big( \bar{D}_{jb}^{QX}\,V_{jb}^B\,e^{-t_g \epsilon_j^b} \Big)
 \Big( B_{kc}^Q\,V_{kc}^C\,e^{-t_g \epsilon_k^c} \Big) \\
 &+\,\bar{D}_{ia}^{QX}\,\Big( U_{jb}^X\,V_{jb}^B\,e^{-t_g \epsilon_j^b} \Big)
 \Big( B_{kc}^Q\,V_{kc}^C\,e^{-t_g \epsilon_k^c} \Big) \\
 &+\,B_{ia}^Q\,\Big( U_{jb}^X\,V_{jb}^B\,e^{-t_g \epsilon_j^b} \Big)
 \Big( \bar{D}_{kc}^{QX}\,V_{kc}^C\,e^{-t_g \epsilon_k^c} \Big) \Big].
\end{split}
\end{align}
It is straightforward to verify that each elementary contraction in the above formula involves at most 
five indices at the same time, not including the Laplace grid index. As a result, the cost of assembling the quantity 
$t_{ia,BC}$ scales as the fifth power of the system size.

Finally, we discuss the issue of the stopping criteria in the HOOI procedure. The obvious choices are to monitor either 
the least-squares error of Eq. (\ref{t3comp}) or the norm of the difference between $V_{ia}^A$ obtained in two 
subsequent iterations. Unfortunately, both of these ideas are troublesome in practice. The calculation of the 
least-squares error during every HOOI iteration is prohibitively expensive as it involves quantities such as 
$\Big(t_{ijk}^{abc}\Big)^2$. The use of the differences between the expansion vectors $V^A_{ai}$ is not effective due 
to non-uniqueness of the singular vectors which may change from iteration to iteration without affecting 
the error in Eq. (\ref{t3comp}). 

To avoid these problems, we monitor the norm of the core tensor $t_{ABC}$ as a proxy for the convergence of 
the procedure, namely
\begin{align}
\label{tnorm}
 ||t||^2 = \sum_{ABC}t_{ABC}^2 =\sum_{ABC} \Big( \sum_{ia}t_{ia,BC}\,V_{ia}^A\Big) \Big( 
\sum_{jb}t_{jb,BC}\,V_{jb}^A\Big),
\end{align}
where summation symbols were added for clarity. The second equality is a consequence of the orthonormality of the 
$V^A_{ai}$ vectors obtained from the SVD procedure. The HOOI procedure is terminated when the difference in $||t||$ 
between 
two consecutive iterations falls below a predefined threshold. The threshold value 
$10^{-5}$ is sufficient in most applications and has been adopted in the present work. 
The 
cost of computing $||t||$ during every iteration is negligible in comparison with other parts of 
the algorithm as $t_{ia,BC}$ is explicitly available anyway. The reason why this simplified procedure is 
adequate follows from the fact that the HOOI algorithm can be equivalently formulated as a maximization of the 
norm of the core tensor rather than the minimization of the least-squares error as in Eq. (\ref{hooi_ls}), 
see Refs.~\onlinecite{kolda09} and reference therein.

With the triply-excited amplitudes represented in the rank-reduced format (\ref{t3comp}), the 
remaining task is to evaluate the $E_{\mathrm{T}}^{[4]}$ and $E_{\mathrm{ST}}^{[5]}$ corrections. 
Derivation of explicit formulas for these corrections given in terms of basic two-electron 
integrals and 
cluster amplitudes is straightforward, but the resulting expressions are rather lengthy. 
Therefore, they are given in the full form in the supplementary material. However, it is worth 
pointing out that the $E_{\mathrm{ST}}^{[5]}$ correction is expressed as a sum of four distinct 
terms with the computational complexity of $N^5$ or lower. Assuming that $\naux>\ntri$, the most 
expensive of them scales as $O^2\naux\ntri^2$ or $O^2V\naux\ntri$ depending on the ratio of $V$ to 
$\ntri$. Explicit formula for $E_{\mathrm{T}}^{[4]}$ comprises six terms, the most expensive two 
scaling as $O^2V\neig\naux\ntri \propto N^6$. Therefore, evaluation of the $E_{\mathrm{(T)}}$ 
correction in the rank-reduced formalism possesses $N^6$ computational 
complexity, lower than the $O^3V^4\propto N^7$ scaling of the conventional algorithms. A rough 
estimate of the crossover point between two algorithms is obtained by recalling that $\neig\approx 
2\nmo\approx 2V$ is sufficient in practice in the RR-CCSD method, and that $\naux\approx 2-4V$ with 
the standard auxiliary basis sets. Therefore, even in the most computationally demanding scenario 
where 
$\ntri\approx\neig$ is needed to achieve sufficient levels of 
accuracy, the crossover point occurs for relatively small systems with $O\approx 10$ or so.

\subsection{\label{subsec:trip-accu-tot} Accuracy of the RR-CCSD(T) method: total energies}

\begin{table}[t!]
\caption{\label{tab:final-error-et}
Statistical measures of relative errors (in percent) in the $E_{\mathrm{(T)}}$ correction in the 
rank-reduced formulation with respect to the exact results. The dimension of the triple excitation 
subspace ($\ntri$) is expressed as $\ntri=y\cdot \nmo$, where $\nmo$ is the total number of 
orbitals in the system. The statistics comes from calculations for 70 molecules contained in the 
Adler-Werner benchmark set~\cite{adler11}.
}
\begin{ruledtabular}
\begin{tabular}{lcccc}
 $y$ & mean  & mean abs. & standard  & max. abs. \\
     & error & error     & deviation & error     \\
\hline
\multicolumn{5}{c}{cc-pVDZ basis set} \\
\hline
$0.50$ & $-$16.63           & 16.65           & 5.81 & 25.00 \\
$0.75$ & \phantom{0}$-$7.33 & \phantom{0}7.71 & 4.34 & 14.42 \\
$1.00$ & \phantom{0}$-$2.89 & \phantom{0}3.76 & 3.17 & \phantom{0}7.37 \\
$1.25$ & \phantom{0}$-$1.58 & \phantom{0}2.57 & 2.59 & \phantom{0}5.65 \\
$1.50$ & \phantom{0}$-$0.71 & \phantom{0}1.88 & 2.19 & \phantom{0}5.19 \\
\hline
\multicolumn{5}{c}{cc-pVTZ basis set} \\
\hline
$0.50$ & $-$6.86 & 7.38 & 4.17 & 12.67 \\
$0.75$ & $-$1.89 & 2.79 & 2.75 & 6.93 \\
$1.00$ & $-$0.53 & 1.67 & 2.04 & 5.26 \\
$1.25$ & $-$0.31 & 1.21 & 1.49 & 4.64 \\
$1.50$ & $-$0.34 & 0.94 & 1.17 & 4.90 \\
\end{tabular}
\end{ruledtabular}
\end{table}

\begin{table}[t!]
\caption{\label{tab:final-error-esdt}
The same data as in Table \ref{tab:final-error-et}, except relative errors (in percent) in the 
\emph{total} RR-CCSD(T) correlation energies are given.
}
\begin{ruledtabular}
\begin{tabular}{lcccc}
 $y$ & mean  & mean abs. & standard  & max. abs. \\
     & error & error     & deviation & error     \\
\hline
\multicolumn{5}{c}{cc-pVDZ basis set} \\
\hline
$0.50$ & $-$0.397            & 0.411 & 0.240 & 0.900 \\
$0.75$ & $-$0.094            & 0.159 & 0.180 & 0.506 \\
$1.00$ & \phantom{$-$}0.053  & 0.099 & 0.133 & 0.534 \\
$1.25$ & \phantom{$-$}0.096  & 0.114 & 0.119 & 0.580 \\
$1.50$ & \phantom{$-$}0.125  & 0.130 & 0.107 & 0.591 \\
\hline
\multicolumn{5}{c}{cc-pVTZ basis set} \\
\hline
$0.50$ & $-$0.257           & 0.320 & 0.273 & 1.200 \\
$0.75$ & $-$0.041           & 0.136 & 0.210 & 1.269 \\
$1.00$ & \phantom{$-$}0.022 & 0.101 & 0.187 & 1.001 \\
$1.25$ & \phantom{$-$}0.034 & 0.083 & 0.171 & 0.871 \\
$1.50$ & \phantom{$-$}0.033 & 0.070 & 0.166 & 0.784 \\
\end{tabular}
\end{ruledtabular}
\end{table}

\begin{figure}[ht!]
\includegraphics[scale=1.00]{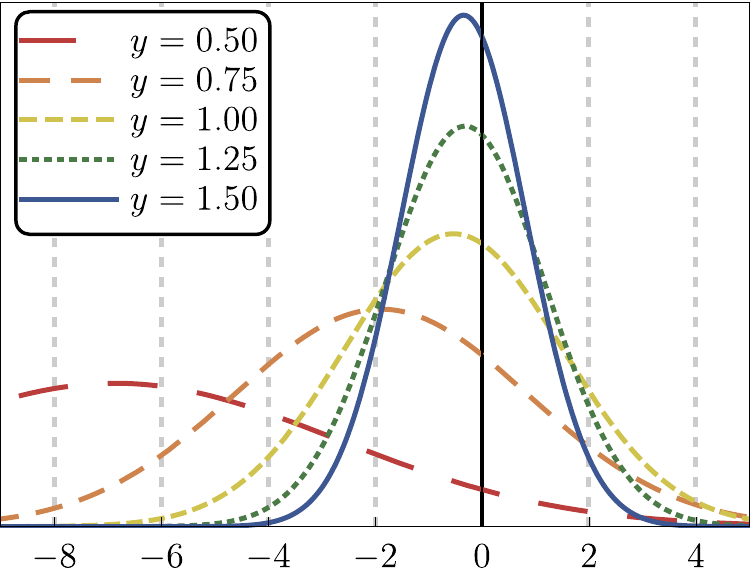}
\caption{\label{fig:trip2-vtz-1} Distribution of relative error (in percent) in the $E_{\mathrm{(T)}}$ correction
with respect to the exact results (cc-pVTZ basis). The dimension of the triples excitation subspace ($\ntri$) is 
expressed as $\ntri=y\cdot \nmo$, where $\nmo$ is the total number of orbitals in the system. The statistics comes from 
calculations for 70 molecules contained in the Adler-Werner benchmark set~\cite{adler11}. See the supplementary material 
for analogous results obtained within the cc-pVDZ basis set.}
\end{figure}
\begin{figure}[ht!]
\includegraphics[scale=1.00]{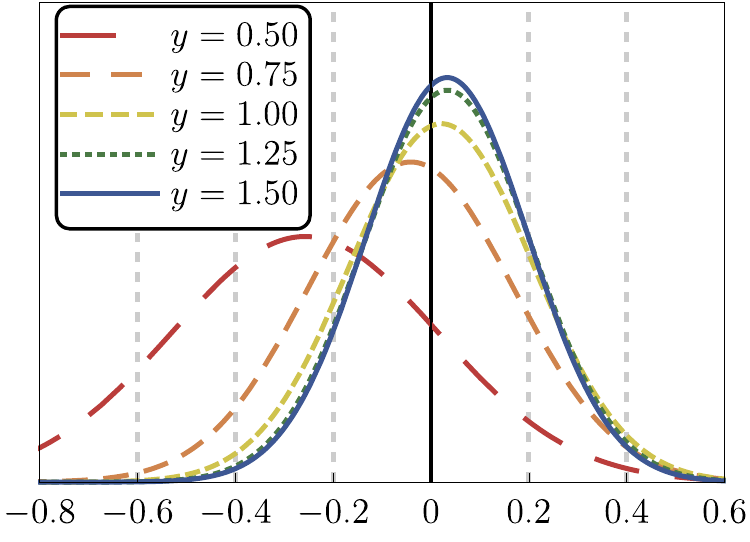}
\caption{\label{fig:trip2-vtz-2} Distribution of relative error (in percent) in the total RR-CCSD(T)/cc-pVTZ correlation energy with respect to the exact CCSD(T)/cc-pVTZ method. The dimension of the triples excitation subspace ($\ntri$) is 
expressed as $\ntri=y\cdot \nmo$, where $\nmo$ is the total number of orbitals in the system. The statistics comes from 
calculations for 70 molecules contained in the Adler-Werner benchmark set~\cite{adler11}. See the supplementary material 
for analogous results obtained within the cc-pVDZ basis set.}
\end{figure}

In order to study the accuracy levels that can realistically be reached with the RR-CCSD(T) method 
and find the value of the parameter $\ntri$ that offers a compromise between accuracy and 
computational costs under typical conditions, we performed RR-CCSD(T) calculations for the 
Alder-Werner benchmark set. The other numerical parameters present in the RR-CCSD(T) method were 
fixed at their recommended values ($\neig=2\nmo$ and $N_{\mathrm{O}}=N_{\mathrm{Z}}=4O$), so that 
the focus is solely on the remaining $\ntri$ parameter. Additionally, in this section we consider 
only the RR-CCSD(T) method with the double excitation subspace obtained by diagonalization of the 
MP3 amplitudes. We found that the $E_{\mathrm{(T)}}$ correction, in contrast to the RR-CCSD energy, 
is rather insensitive to whether MP2 or MP3 amplitudes are used, with relative errors of the 
$E_{\mathrm{(T)}}$ correction differing by just a small fraction of a percent.

For each member of the Alder-Werner set we performed RR-CCSD(T) calculations with $\ntri=y\cdot 
\nmo$, where $y=0.50$, $0.75$, $1.00$, $1.25$, and $1.50$. Note that the $y$ parameter is 
asymptotically independent of the system size and hence we expect it to possess some universal value that 
delivers a decent accuracy in the $E_{\mathrm{(T)}}$ correction for a broad range of systems. 

We aim at relative accuracy level of a few percent in the $E_{\mathrm{(T)}}$ correction. This is a reasonable target 
from the practical point of view, because $E_{\mathrm{(T)}}$ rarely contributes from the 5\% of the total correlation 
energy in well-behaved systems. In Table \ref{tab:final-error-et} we report error statistics for the calculations of 
the $E_{\mathrm{(T)}}$ correction in the rank-reduced formulation. Analogous data are given also in Table 
\ref{tab:final-error-esdt}, but there we consider errors in the \emph{total} RR-CCSD(T) 
correlation energies, i.e. the sum of the RR-CCSD and $E_{\mathrm{(T)}}$ contributions, taking the 
exact CCSD(T) results as a reference. For ease of comparison, the distributions of errors for the 
$E_{\mathrm{(T)}}$ correction alone and for the total RR-CCSD(T) correlation energy (both within the cc-pVTZ basis set) are represented graphically in terms of normal distributions in Figs. \ref{fig:trip2-vtz-1} and \ref{fig:trip2-vtz-2}, respectively. Analogous plots obtained with RR-CCSD(T)/cc-pVDZ method are given in the supplementary material.

The results reported in Fig. \ref{fig:trip2-vtz-1} reveal the overall trend in the accuracy of the 
$E_{\mathrm{(T)}}$ correction as a function of the $y$ parameter. Even with the smallest triple 
excitation subspace dimension considered here ($y=0.50$) a reasonable relative accuracy of several 
percent is obtained. This improves to about 0.5\% when the parameter $y$ is increased to unity.  
Beyond the point $y=1$ the improvement rate slows down considerably. We verified that this 
phenomenon is a consequence of finite accuracy of the 
doubly-excited amplitudes (with the recommended $\neig=2\nmo$) which limit the accuracy of the 
$E_{\mathrm{(T)}}$ correction for $y>1$.

Based on the results reported in Table \ref{tab:final-error-et}, we recommend that for $\neig=2\nmo$ the dimension of 
the triple-excitation subspace is set to $\ntri=\nmo$, corresponding to $y=1$. For this value of the parameter the 
accuracy of the $E_{\mathrm{(T)}}$ correction meets the criteria discussed in the previous 
paragraphs. Moreover, this choice is supported by the observation that 
a further increase of the parameter $y$ leads to minor improvements in the accuracy of the 
\emph{total} RR-CCSD(T) energies, see Table \ref{tab:final-error-esdt}. This is a result of an 
accidental, yet systematic cancellation of errors that occurs for $\neig=2\nmo$ and 
$\ntri=\nmo$ where the RR-CCSD component of the energy is slightly underestimated, while the 
$E_{\mathrm{(T)}}$ correction is overestimated by a comparable amount. However, we verified that 
even in the absence of this fruitful error cancellation, i.e. assuming that both errors are of the 
same sign, the combination of the parameters $\neig=2\nmo$ and $\ntri=\nmo$ would still provide 
accuracy levels better than 0.1\% in the total RR-CCSD(T) energies. Therefore, the choice 
$\neig=2\nmo$ and $\ntri=\nmo$ is both safe and pragmatic, and is adopted further in the paper.

\subsection{\label{subsec:trip-accu-rel} Accuracy of the RR-CCSD(T) method: relative energies}

Finally, we study the accuracy of the RR-CCSD(T) method in reproduction of relative energies and 
compare the results with the reference CCSD(T) data. As the first test we employ the 
benchmark set of 34 isomerization energies of organic molecules 
introduced by Grimme \emph{et al.}\cite{grimme07} (usually abbreviated as ISO34 in the literature). The range of 
isomerization energies included in the ISO34 set spans from a few kJ/mol to a few hundreds kJ/mol.
The RR-CCSD(T) calculations were performed with the recommended settings ($\neig=2\nmo$, 
$N_{\mathrm{O}}=N_{\mathrm{Z}}=4O$, and $\ntri=\nmo$) and are compared with the exact CCSD(T) 
results obtained with \textsc{NWChem} package. Note that in the latter calculations we do not apply 
the density-fitting approximation of the two-electron integrals and hence the error budget of the 
RR-CCSD(T) results formally includes also the density-fitting error. In all calculations we 
employ the cc-pVTZ basis set and the $1s$ core orbitals of the first-row atoms were frozen. Within 
this setup the largest system included in the ISO34 set contains about 500 orbitals and 50 active 
electrons which is near the edge of applicability of the canonical CCSD(T) theory without further 
approximations or a parallelization. 

Raw isomerization energies computed using the RR-CCSD(T) and the exact CCSD(T) methods are listed 
in the supplementary material. To simplify the analysis we consider statistical error measures 
with respect to the reference CCSD(T) method evaluated for the whole ISO34 set. The 
RR-CCSD(T)/cc-pVTZ method exhibits the mean error of $-$0.03 kJ/mol and mean absolute error of 0.32 
kJ/mol. The standard deviation of the error equals to 0.38 kJ/mol. This level of accuracy 
is sufficient for many applications involving polyatomic molecules. Moreover, it is worth pointing out 
that the RR-CCSD(T) method is systematically improvable without a drastic increase of the computational 
costs. Therefore, if accuracy levels of, e.g., 0.1 kJ/mol are needed in a particular application, this 
requirement can be met by increasing the control parameters $\neig$ and $\ntri$ above the values 
recommended currently. 

The maximum absolute deviation among the isomerization energies from the ISO34 set was found for 
reaction 13 (styrene $\rightarrow$ cyclooctatetraene) and amounts to 0.77 kJ/mol. However, it has to 
be pointed out that the total isomerization energy for this reaction is particularly large (152.89 
kJ/mol), so the relative error obtained in this case (about 0.5\%) is still acceptable. At the same 
time, the RR-CCSD(T) method accurately reproduces also small energy differences, indicating a 
systematic error cancellation. For example, consider the smallest two isomerization energies from 
the ISO34 benchmark set equal to 4.66 kJ/mol and 4.73 kJ/mol for reaction 4 (trans-2-butene 
$\rightarrow$ cis-2-butene) and reaction 5 (isobutylene $\rightarrow$ trans-2-butene). The errors of 
the RR-CCSD(T) method for these reactions amount to $-$0.03 kJ/mol and $-$0.06 kJ/mol, respectively. 
This shows that the proposed method is capable of providing uniformly reliable results in a 
chemically-relevant energy range.

\begin{figure}[ht!]
\includegraphics[scale=1.00]{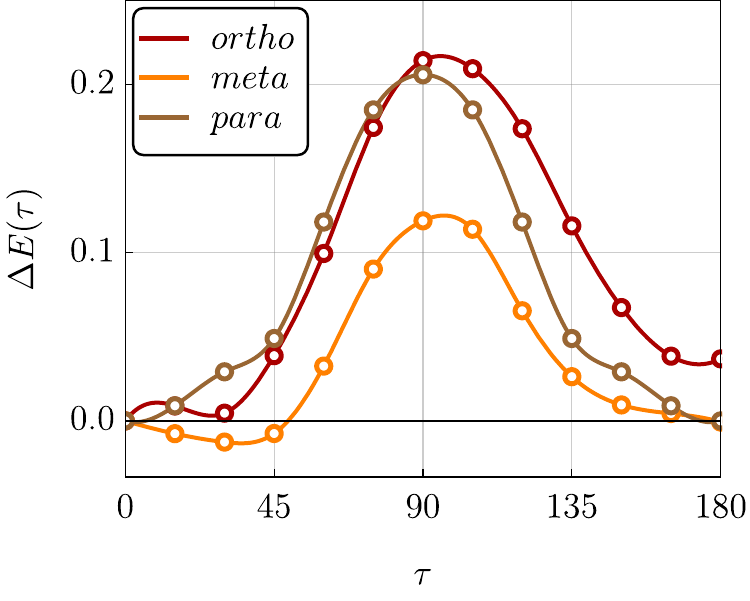}
\includegraphics[scale=1.00]{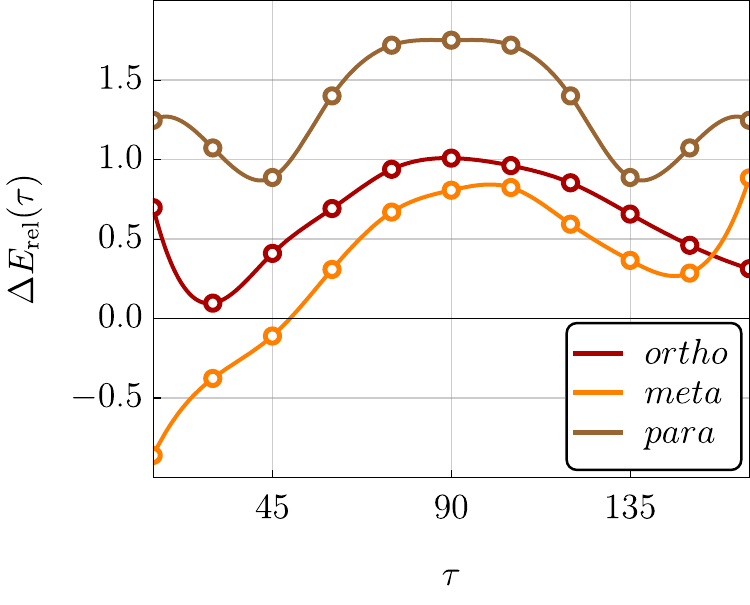}
\caption{\label{fig:phenols} 
Absolute errors (left panel) and percent relative errors (right panel) in the RR-CCSD(T)/cc-pVTZ torsional energy 
curves for the \emph{ortho}-, \emph{meta}- and \emph{para}-fluorophenols. The exact CCSD(T)/cc-pVTZ results are used as 
a reference.}
\end{figure}

The second group of model systems we employ to study the accuracy of the RR-CCSD(T) method in reproduction 
of relative energies are \emph{ortho}-, \emph{meta}- and \emph{para}-fluorophenols. These systems 
have been intensively studied in the literature due to their rich microwave spectrum prototypical 
for hydrogen bond interactions with fluorine\cite{larsen74,larsen86,smeyers87,ratzer03,jaman07,bell17}. Here we consider 
the torsional energy differences 
related to the internal rotation of the hydroxyl moiety in relation to the plane of the aromatic 
ring. The torsional angle, denoted $\tau$ further in the text, is defined by the following sequence 
of four atoms: the hydrogen of the hydroxyl group, the oxygen, the carbon atom closest to the 
oxygen, and the next carbon atom in the ring closest to the fluorine atom (in the case of 
\emph{para}-fluorophenol the last choice is arbitrary). By convention, in the case of 
\emph{ortho} and \emph{meta} isomers the torsional angle $\tau=0$ corresponds to the \emph{trans} 
structure with the maximum distance between the hydrogen of the hydroxyl group and the fluorine 
atom.

\begin{table}[ht!]
\caption{\label{tab:torsional-phenol}
Parameters of the torsional energy curve (see the text for definitions of all quantities) for 
three isomers of fluorophenol computed using the RR-CCSD(T)/cc-pVTZ method (``RR'') and the exact CCSD(T)/cc-pVTZ 
method (``exact''). For each isomer relative energies with respect to its $\tau=0$ conformation are 
given. The angles are given in degrees and the energies~in~kJ/mol.
}
\begin{ruledtabular}
\begin{tabular}{lcccccc}
  & 
 \multicolumn{2}{c}{\emph{ortho}} & 
 \multicolumn{2}{c}{\emph{meta}}  &
 \multicolumn{2}{c}{\emph{para}} \\\cline{2-3}\cline{4-5}\cline{6-7}
 quantity & RR & exact & RR & exact & RR & exact \\
\hline
 $\Delta E_{\mathrm{barrier}}$    & 
 \phantom{0}21.62 & \phantom{0}21.84 &  
 14.63 & 14.75 & 11.55 & 11.76 \\
 $\tau_{\mathrm{barrier}}$       & 
 101.34 & 101.24 &  
 91.18 & 91.26 & 90.00 & 90.00 \\
 $\Delta E_{\mathrm{cis/trans}}$ & 
 \phantom{0}11.36 & \phantom{0}11.40 &
 $-$0.53 & $-$0.53 & 
 \phantom{0}0.00 & \phantom{0}0.00 \\
\end{tabular}
\end{ruledtabular}
\end{table}

We performed an energy scan varying the torsional angle $\tau$ from $0^\circ$ to $180^\circ$ in 
steps of $15^\circ$. The rest of 
the molecular geometry was fully optimized for each $\tau$ at the MP2/cc-pVTZ level of theory. 
Cartesian coordinates of the optimized structures are included in the supplementary material. 
Finally, the RR-CCSD(T)/cc-pVTZ and the exact CCSD(T)/cc-pVTZ calculations are performed on every 
optimized geometry with the same settings as for the ISO34 benchmark set. For each isomer, the 
$\tau=0$ conformation is treated as the zero-energy point and all other energies are given relative 
to it. In Fig. \ref{fig:phenols} we provide errors of the RR-CCSD(T)/cc-pVTZ torsional energies 
with respect to the exact CCSD(T)/cc-pVTZ for each isomer of the fluorophenol. Raw energies used to 
compile this plot are given in the supplementary material. From Fig. \ref{fig:phenols} it is clear 
that the errors in the torsional energies vary smoothly with $\tau$, without major jumps and 
discontinuities. The mean absolute errors (averaged over $\tau=15^\circ,\ldots,180^\circ$) are 
$0.091\,/\,0.038\,/\,0.076$~kJ/mol for the \emph{ortho}/\emph{meta}/\emph{para} isomers, while the 
corresponding standard deviations are $0.079\,/\,0.044\,/\,0.077$~kJ/mol. To further study the performance of 
the RR-CCSD(T) method we calculated three parameters that characterize the potential energy curves 
for each isomer:
\begin{itemize}
 \item the height of the potential energy barrier separating the \emph{trans} ($\tau=0$) and \emph{cis} 
($\tau=180^\circ$) conformations, $\Delta E_{\mathrm{barrier}}$;
 \item the value of the torsional angle corresponding to the maximum of the barrier, 
$\tau_{\mathrm{barrier}}$;
 \item the energy difference between the \emph{trans} and \emph{cis} 
conformations, $\Delta E_{\mathrm{trans/cis}}$.
\end{itemize}
The first two parameters were found with the help of $B$-splines interpolation of the calculated data 
points, followed by application of the Brent algorithm\cite{brent71} to find the minimum of the interpolated curve. The 
numerical 
errors caused by this procedure are essentially negligible. The parameters $\Delta E_{\mathrm{barrier}}$, 
$\tau_{\mathrm{barrier}}$, and $\Delta E_{\mathrm{trans/cis}}$ determined for three conformers are reported in Table 
\ref{tab:torsional-phenol} and compared with the reference CCSD(T) values. The error of determining the barrier 
height is around 0.2~kJ/mol for each conformer, while the error of determining its location is below 0.1~degrees. 
In many applications to polyatomic molecules errors of this magnitude would be negligible in comparison with other 
uncertainties, such as the basis set incompleteness. 

\section{\label{sec:concl} Conclusions}

In this work we have modified and extended the rank-reduced CCSD theory introduced by Parrish and 
collaborators with three major contributions. First, we have shown how a subset of eigenvectors of 
the MP2 and MP3 amplitudes, serving as the the expansion basis for the doubly-excited amplitudes in 
the RR-CCSD method, can be obtained efficiently with $N^5$ scaling. Second, we have eliminated the 
issue of non-factorizable terms from the RR-CCSD residual. We have provided a systematic way to 
approximate these terms using the singular value decomposition and reduced the overall scaling of 
the RR-CCSD iterations down to the level of $N^5$. Finally, we have considered the evaluation of 
the perturbative corrections to the CCSD energies resulting from triply excited configurations. The 
triply-excited amplitudes present in the CCSD(T) method have been decomposed to the Tucker-3 format 
using the higher-order orthogonal iteration (HOOI) procedure. This has enabled to compute the 
energy correction due to triple excitations non-iteratively with $N^6$ cost.

The accuracy of the proposed RR-CCSD(T) method in reproduction of total correlation 
energies has been studied using a diverse set of 70 polyatomic molecules comprising first- and 
second-row atoms. It has been shown that with the recommended values of the control parameters, 
relative accuracy levels better than 99.9\% have been achieved, both in the double- and triple-zeta 
basis sets. Next, we have considered the accuracy of relative energies calculated with the 
RR-CCSD(T) method. Numerical results for isomerization energies of 34 organic molecules and 
conformational energies of substituted phenols have shown that average absolute errors are of the 
order of $0.1-0.3$ kJ/mol. Moreover, the calculated energy surfaces show no discontinuities and are 
suitable for fitting with a properly chosen functional form, which is usually a necessary step in 
nuclear dynamics simulations, for example. We have also compared efficiency of the reduced scaling 
RR-CCSD implementation with the standard CCSD algorithm. While we have shown that the break-even 
point beyond which the RR formulation becomes advantageous occurs for only $30-40$ active electrons, 
an efficient parallelized code is required to compete with carefully-optimized implementations 
reported recently that scale favorably to thousands of cores.

Finally, we point out possible extensions of the present work which are of particular interest. First, the 
rank-reduction concepts can be applied to the symmetry-adapted perturbation theory 
(SAPT)\cite{jeziorski94,hohenstein12c,szalewicz12,jansen14}. Higher-level variants of SAPT\cite{parker14}, such as 
SAPT2+ and SAPT2+(3), share a structure similar to the CCD theory with the exception that the excitation subspace for 
the supermolecule is formed as a union of excitations localized on the monomers. Another promising idea is to extend 
the rank-reduced formalism to the time-dependent coupled-cluster 
theory\cite{huber11,kvaal12,sato18}, where the 
high cost of the calculations is one of the main stumbling blocks that prevent routine applications to polyatomic 
molecules. Indeed, the computational effort of a single time step is usually comparable to several standard CC 
iterations\cite{pedersen19,kristi20} and tens of thousands of time steps may be needed in simulations in strong laser 
fields. The applicability of the rank-reduced formalism to the time-dependent problems shall be the subject of a future 
study.

\section*{\label{sec:supp} Supplementary Material}

See supplementary material for additional numerical results obtained using the RR-CCSD and RR-CCSD(T) methods, tests of accuracy of the Laplace quadrature, and explicit analytical formulas for the factorizable terms in the RR-CCSD residual and for the $E_{\mathrm{T}}^{[4]}$ and $E_{\mathrm{ST}}^{[5]}$ corrections within the rank-reduced framework.

\begin{acknowledgments}
I would like to thank Dr. A. Tucholska for fruitful discussions, and for reading and commenting on 
the manuscript. I am grateful to Prof.~M.~Reiher and all members of his group for their hospitality 
during my stay at Laboratorium f\"{u}r Physikalische Chemie, ETH Z\"{u}rich. This work was supported by the Foundation 
for Polish Science (FNP) and by the Polish National Agency of Academic Exchange through the Bekker programme No. 
PPN/BEK/2019/1/00315/U/00001. Computations presented in this research were carried out with the support of the 
Interdisciplinary Center for Mathematical and Computational Modeling (ICM) at the University of Warsaw, grant number 
G86-1021.
\end{acknowledgments}

\section*{Data Availability Statement}
The data that support the findings of this study are available
within the article and its supplementary material.

\bibliography{rrccsd}

\begin{thebibliography}{163}%
\makeatletter
\providecommand \@ifxundefined [1]{%
 \@ifx{#1\undefined}
}%
\providecommand \@ifnum [1]{%
 \ifnum #1\expandafter \@firstoftwo
 \else \expandafter \@secondoftwo
 \fi
}%
\providecommand \@ifx [1]{%
 \ifx #1\expandafter \@firstoftwo
 \else \expandafter \@secondoftwo
 \fi
}%
\providecommand \natexlab [1]{#1}%
\providecommand \enquote  [1]{``#1''}%
\providecommand \bibnamefont  [1]{#1}%
\providecommand \bibfnamefont [1]{#1}%
\providecommand \citenamefont [1]{#1}%
\providecommand \href@noop [0]{\@secondoftwo}%
\providecommand \href [0]{\begingroup \@sanitize@url \@href}%
\providecommand \@href[1]{\@@startlink{#1}\@@href}%
\providecommand \@@href[1]{\endgroup#1\@@endlink}%
\providecommand \@sanitize@url [0]{\catcode `\\12\catcode `\$12\catcode
  `\&12\catcode `\#12\catcode `\^12\catcode `\_12\catcode `\%12\relax}%
\providecommand \@@startlink[1]{}%
\providecommand \@@endlink[0]{}%
\providecommand \url  [0]{\begingroup\@sanitize@url \@url }%
\providecommand \@url [1]{\endgroup\@href {#1}{\urlprefix }}%
\providecommand \urlprefix  [0]{URL }%
\providecommand \Eprint [0]{\href }%
\providecommand \doibase [0]{http://dx.doi.org/}%
\providecommand \selectlanguage [0]{\@gobble}%
\providecommand \bibinfo  [0]{\@secondoftwo}%
\providecommand \bibfield  [0]{\@secondoftwo}%
\providecommand \translation [1]{[#1]}%
\providecommand \BibitemOpen [0]{}%
\providecommand \bibitemStop [0]{}%
\providecommand \bibitemNoStop [0]{.\EOS\space}%
\providecommand \EOS [0]{\spacefactor3000\relax}%
\providecommand \BibitemShut  [1]{\csname bibitem#1\endcsname}%
\let\auto@bib@innerbib\@empty
\bibitem [{\citenamefont {Crawford}\ and\ \citenamefont
  {Schaefer~III}(2007)}]{crawford07}%
  \BibitemOpen
  \bibfield  {author} {\bibinfo {author} {\bibfnamefont {T.~D.}\ \bibnamefont
  {Crawford}}\ and\ \bibinfo {author} {\bibfnamefont {H.~F.}\ \bibnamefont
  {Schaefer~III}},\ }\enquote {\bibinfo {title} {An introduction to coupled
  cluster theory for computational chemists},}\ in\ \href
  {https://onlinelibrary.wiley.com/doixx/abs/10.1002/9780470125915.ch2} {\emph
  {\bibinfo {booktitle} {Reviews in Computational Chemistry}}}\ (\bibinfo
  {publisher} {John Wiley \& Sons, Ltd},\ \bibinfo {year} {2007})\ pp.\
  \bibinfo {pages} {33--136}\BibitemShut {NoStop}%
\bibitem [{\citenamefont {Bartlett}\ and\ \citenamefont
  {Musia\l{}}(2007)}]{bartlett07}%
  \BibitemOpen
  \bibfield  {author} {\bibinfo {author} {\bibfnamefont {R.~J.}\ \bibnamefont
  {Bartlett}}\ and\ \bibinfo {author} {\bibfnamefont {M.}~\bibnamefont
  {Musia\l{}}},\ }\href {https://link.aps.org/doix/10.1103/RevModPhys.79.291}
  {\bibfield  {journal} {\bibinfo  {journal} {Rev. Mod. Phys.}\ }\textbf
  {\bibinfo {volume} {79}},\ \bibinfo {pages} {291} (\bibinfo {year}
  {2007})}\BibitemShut {NoStop}%
\bibitem [{\citenamefont {Raghavachari}\ \emph {et~al.}(1989)\citenamefont
  {Raghavachari}, \citenamefont {Trucks}, \citenamefont {Pople},\ and\
  \citenamefont {Head-Gordon}}]{ragha89}%
  \BibitemOpen
  \bibfield  {author} {\bibinfo {author} {\bibfnamefont {K.}~\bibnamefont
  {Raghavachari}}, \bibinfo {author} {\bibfnamefont {G.~W.}\ \bibnamefont
  {Trucks}}, \bibinfo {author} {\bibfnamefont {J.~A.}\ \bibnamefont {Pople}}, \
  and\ \bibinfo {author} {\bibfnamefont {M.}~\bibnamefont {Head-Gordon}},\
  }\href {http://www.sciencedirect.com/science/article/pii/S0009261489873956}
  {\bibfield  {journal} {\bibinfo  {journal} {Chem. Phys. Lett.}\ }\textbf
  {\bibinfo {volume} {157}},\ \bibinfo {pages} {479 } (\bibinfo {year}
  {1989})}\BibitemShut {NoStop}%
\bibitem [{\citenamefont {Kobayashi}\ and\ \citenamefont
  {Rendell}(1997)}]{kobayashi97}%
  \BibitemOpen
  \bibfield  {author} {\bibinfo {author} {\bibfnamefont {R.}~\bibnamefont
  {Kobayashi}}\ and\ \bibinfo {author} {\bibfnamefont {A.~P.}\ \bibnamefont
  {Rendell}},\ }\href
  {https://www.sciencedirect.com/science/article/pii/S0009261496013875}
  {\bibfield  {journal} {\bibinfo  {journal} {Chem. Phys. Lett.}\ }\textbf
  {\bibinfo {volume} {265}},\ \bibinfo {pages} {1} (\bibinfo {year}
  {1997})}\BibitemShut {NoStop}%
\bibitem [{\citenamefont {Hirata}(2003)}]{hirata03}%
  \BibitemOpen
  \bibfield  {author} {\bibinfo {author} {\bibfnamefont {S.}~\bibnamefont
  {Hirata}},\ }\href {\doibase 10.1021/jp034596z} {\bibfield  {journal}
  {\bibinfo  {journal} {The Journal of Physical Chemistry A}\ }\textbf
  {\bibinfo {volume} {107}},\ \bibinfo {pages} {9887} (\bibinfo {year}
  {2003})},\ \Eprint {http://arxiv.org/abs/https://doi.org/10.1021/jp034596z}
  {https://doi.org/10.1021/jp034596z} \BibitemShut {NoStop}%
\bibitem [{\citenamefont {Auer}\ \emph {et~al.}(2006)\citenamefont {Auer},
  \citenamefont {Baumgartner}, \citenamefont {Bernholdt}, \citenamefont
  {Bibireata}, \citenamefont {Choppella}, \citenamefont {Cociorva},
  \citenamefont {Gao}, \citenamefont {Harrison}, \citenamefont
  {Krishnamoorthy}, \citenamefont {Krishnan}, \citenamefont {Lam},
  \citenamefont {Lu}, \citenamefont {Nooijen}, \citenamefont {Pitzer},
  \citenamefont {Ramanujam}, \citenamefont {Sadayappan},\ and\ \citenamefont
  {Sibiryakov}}]{auer06}%
  \BibitemOpen
  \bibfield  {author} {\bibinfo {author} {\bibfnamefont {A.~A.}\ \bibnamefont
  {Auer}}, \bibinfo {author} {\bibfnamefont {G.}~\bibnamefont {Baumgartner}},
  \bibinfo {author} {\bibfnamefont {D.~E.}\ \bibnamefont {Bernholdt}}, \bibinfo
  {author} {\bibfnamefont {A.}~\bibnamefont {Bibireata}}, \bibinfo {author}
  {\bibfnamefont {V.}~\bibnamefont {Choppella}}, \bibinfo {author}
  {\bibfnamefont {D.}~\bibnamefont {Cociorva}}, \bibinfo {author}
  {\bibfnamefont {X.}~\bibnamefont {Gao}}, \bibinfo {author} {\bibfnamefont
  {R.}~\bibnamefont {Harrison}}, \bibinfo {author} {\bibfnamefont
  {S.}~\bibnamefont {Krishnamoorthy}}, \bibinfo {author} {\bibfnamefont
  {S.}~\bibnamefont {Krishnan}}, \bibinfo {author} {\bibfnamefont {C.-C.}\
  \bibnamefont {Lam}}, \bibinfo {author} {\bibfnamefont {Q.}~\bibnamefont
  {Lu}}, \bibinfo {author} {\bibfnamefont {M.}~\bibnamefont {Nooijen}},
  \bibinfo {author} {\bibfnamefont {R.}~\bibnamefont {Pitzer}}, \bibinfo
  {author} {\bibfnamefont {J.}~\bibnamefont {Ramanujam}}, \bibinfo {author}
  {\bibfnamefont {P.}~\bibnamefont {Sadayappan}}, \ and\ \bibinfo {author}
  {\bibfnamefont {A.}~\bibnamefont {Sibiryakov}},\ }\href {\doibase
  10.1080/00268970500275780} {\bibfield  {journal} {\bibinfo  {journal}
  {Molecular Physics}\ }\textbf {\bibinfo {volume} {104}},\ \bibinfo {pages}
  {211} (\bibinfo {year} {2006})},\ \Eprint
  {http://arxiv.org/abs/https://doi.org/10.1080/00268970500275780}
  {https://doi.org/10.1080/00268970500275780} \BibitemShut {NoStop}%
\bibitem [{\citenamefont {Olson}\ \emph {et~al.}(2007)\citenamefont {Olson},
  \citenamefont {Bentz}, \citenamefont {Kendall}, \citenamefont {Schmidt},\
  and\ \citenamefont {Gordon}}]{olson07}%
  \BibitemOpen
  \bibfield  {author} {\bibinfo {author} {\bibfnamefont {R.~M.}\ \bibnamefont
  {Olson}}, \bibinfo {author} {\bibfnamefont {J.~L.}\ \bibnamefont {Bentz}},
  \bibinfo {author} {\bibfnamefont {R.~A.}\ \bibnamefont {Kendall}}, \bibinfo
  {author} {\bibfnamefont {M.~W.}\ \bibnamefont {Schmidt}}, \ and\ \bibinfo
  {author} {\bibfnamefont {M.~S.}\ \bibnamefont {Gordon}},\ }\href@noop {}
  {\bibfield  {journal} {\bibinfo  {journal} {J. Chem. Theory Comp.}\ }\textbf
  {\bibinfo {volume} {3}},\ \bibinfo {pages} {1312} (\bibinfo {year}
  {2007})}\BibitemShut {NoStop}%
\bibitem [{\citenamefont {Janowski}, \citenamefont {Ford},\ and\ \citenamefont
  {Pulay}(2007)}]{janowski07}%
  \BibitemOpen
  \bibfield  {author} {\bibinfo {author} {\bibfnamefont {T.}~\bibnamefont
  {Janowski}}, \bibinfo {author} {\bibfnamefont {A.~R.}\ \bibnamefont {Ford}},
  \ and\ \bibinfo {author} {\bibfnamefont {P.}~\bibnamefont {Pulay}},\
  }\href@noop {} {\bibfield  {journal} {\bibinfo  {journal} {J. Chem. Theory
  Comp.}\ }\textbf {\bibinfo {volume} {3}},\ \bibinfo {pages} {1368} (\bibinfo
  {year} {2007})}\BibitemShut {NoStop}%
\bibitem [{\citenamefont {Janowski}\ and\ \citenamefont
  {Pulay}(2008)}]{janowski08}%
  \BibitemOpen
  \bibfield  {author} {\bibinfo {author} {\bibfnamefont {T.}~\bibnamefont
  {Janowski}}\ and\ \bibinfo {author} {\bibfnamefont {P.}~\bibnamefont
  {Pulay}},\ }\href@noop {} {\bibfield  {journal} {\bibinfo  {journal} {J.
  Chem. Theory Comp.}\ }\textbf {\bibinfo {volume} {4}},\ \bibinfo {pages}
  {1585} (\bibinfo {year} {2008})}\BibitemShut {NoStop}%
\bibitem [{\citenamefont {van Dam}\ \emph {et~al.}(2011)\citenamefont {van
  Dam}, \citenamefont {de~Jong}, \citenamefont {Bylaska}, \citenamefont
  {Govind}, \citenamefont {Kowalski}, \citenamefont {Straatsma},\ and\
  \citenamefont {Valiev}}]{vandam11}%
  \BibitemOpen
  \bibfield  {author} {\bibinfo {author} {\bibfnamefont {H.}~\bibnamefont {van
  Dam}}, \bibinfo {author} {\bibfnamefont {W.}~\bibnamefont {de~Jong}},
  \bibinfo {author} {\bibfnamefont {E.}~\bibnamefont {Bylaska}}, \bibinfo
  {author} {\bibfnamefont {N.}~\bibnamefont {Govind}}, \bibinfo {author}
  {\bibfnamefont {K.}~\bibnamefont {Kowalski}}, \bibinfo {author}
  {\bibfnamefont {T.}~\bibnamefont {Straatsma}}, \ and\ \bibinfo {author}
  {\bibfnamefont {M.}~\bibnamefont {Valiev}},\ }\href {\doibase
  https://doi.org/10.1002/wcms.62} {\bibfield  {journal} {\bibinfo  {journal}
  {WIREs Computational Molecular Science}\ }\textbf {\bibinfo {volume} {1}},\
  \bibinfo {pages} {888} (\bibinfo {year} {2011})},\ \Eprint
  {http://arxiv.org/abs/https://wires.onlinelibrary.wiley.com/doi/pdf/10.1002/wcms.62}
  {https://wires.onlinelibrary.wiley.com/doi/pdf/10.1002/wcms.62} \BibitemShut
  {NoStop}%
\bibitem [{\citenamefont {Deumens}\ \emph {et~al.}(2011)\citenamefont
  {Deumens}, \citenamefont {Lotrich}, \citenamefont {Perera}, \citenamefont
  {Ponton}, \citenamefont {Sanders},\ and\ \citenamefont
  {Bartlett}}]{deumens11}%
  \BibitemOpen
  \bibfield  {author} {\bibinfo {author} {\bibfnamefont {E.}~\bibnamefont
  {Deumens}}, \bibinfo {author} {\bibfnamefont {V.~F.}\ \bibnamefont
  {Lotrich}}, \bibinfo {author} {\bibfnamefont {A.}~\bibnamefont {Perera}},
  \bibinfo {author} {\bibfnamefont {M.~J.}\ \bibnamefont {Ponton}}, \bibinfo
  {author} {\bibfnamefont {B.~A.}\ \bibnamefont {Sanders}}, \ and\ \bibinfo
  {author} {\bibfnamefont {R.~J.}\ \bibnamefont {Bartlett}},\ }\href
  {https://wires.onlinelibrary.wiley.com/doix/abs/10.1002/wcms.77} {\bibfield
  {journal} {\bibinfo  {journal} {WIREs Comput. Mol. Sci.}\ }\textbf {\bibinfo
  {volume} {1}},\ \bibinfo {pages} {895} (\bibinfo {year} {2011})}\BibitemShut
  {NoStop}%
\bibitem [{\citenamefont {Anisimov}\ \emph {et~al.}(2014)\citenamefont
  {Anisimov}, \citenamefont {Bauer}, \citenamefont {Chadalavada}, \citenamefont
  {Olson}, \citenamefont {Glenski}, \citenamefont {Kramer}, \citenamefont
  {Aprà},\ and\ \citenamefont {Kowalski}}]{anisimov14}%
  \BibitemOpen
  \bibfield  {author} {\bibinfo {author} {\bibfnamefont {V.~M.}\ \bibnamefont
  {Anisimov}}, \bibinfo {author} {\bibfnamefont {G.~H.}\ \bibnamefont {Bauer}},
  \bibinfo {author} {\bibfnamefont {K.}~\bibnamefont {Chadalavada}}, \bibinfo
  {author} {\bibfnamefont {R.~M.}\ \bibnamefont {Olson}}, \bibinfo {author}
  {\bibfnamefont {J.~W.}\ \bibnamefont {Glenski}}, \bibinfo {author}
  {\bibfnamefont {W.~T.~C.}\ \bibnamefont {Kramer}}, \bibinfo {author}
  {\bibfnamefont {E.}~\bibnamefont {Aprà}}, \ and\ \bibinfo {author}
  {\bibfnamefont {K.}~\bibnamefont {Kowalski}},\ }\href@noop {} {\bibfield
  {journal} {\bibinfo  {journal} {J. Chem. Theory Comp.}\ }\textbf {\bibinfo
  {volume} {10}},\ \bibinfo {pages} {4307} (\bibinfo {year}
  {2014})}\BibitemShut {NoStop}%
\bibitem [{\citenamefont {Solomonik}\ \emph {et~al.}(2014)\citenamefont
  {Solomonik}, \citenamefont {Matthews}, \citenamefont {Hammond}, \citenamefont
  {Stanton},\ and\ \citenamefont {Demmel}}]{solomonik14}%
  \BibitemOpen
  \bibfield  {author} {\bibinfo {author} {\bibfnamefont {E.}~\bibnamefont
  {Solomonik}}, \bibinfo {author} {\bibfnamefont {D.}~\bibnamefont {Matthews}},
  \bibinfo {author} {\bibfnamefont {J.~R.}\ \bibnamefont {Hammond}}, \bibinfo
  {author} {\bibfnamefont {J.~F.}\ \bibnamefont {Stanton}}, \ and\ \bibinfo
  {author} {\bibfnamefont {J.}~\bibnamefont {Demmel}},\ }\href
  {https://www.sciencedirect.com/science/article/pii/S074373151400104X}
  {\bibfield  {journal} {\bibinfo  {journal} {J. Parallel Distr. Comp.}\
  }\textbf {\bibinfo {volume} {74}},\ \bibinfo {pages} {3176} (\bibinfo {year}
  {2014})}\BibitemShut {NoStop}%
\bibitem [{\citenamefont {Calvin}, \citenamefont {Lewis},\ and\ \citenamefont
  {Valeev}(2015)}]{calvin15}%
  \BibitemOpen
  \bibfield  {author} {\bibinfo {author} {\bibfnamefont {J.~A.}\ \bibnamefont
  {Calvin}}, \bibinfo {author} {\bibfnamefont {C.~A.}\ \bibnamefont {Lewis}}, \
  and\ \bibinfo {author} {\bibfnamefont {E.~F.}\ \bibnamefont {Valeev}},\ }in\
  \href {\doibase 10.1145/2833179.2833186} {\emph {\bibinfo {booktitle}
  {Proceedings of the 5th Workshop on Irregular Applications: Architectures and
  Algorithms}}},\ \bibinfo {series and number} {IA3 '15}\ (\bibinfo
  {publisher} {Association for Computing Machinery},\ \bibinfo {address} {New
  York, NY, USA},\ \bibinfo {year} {2015})\BibitemShut {NoStop}%
\bibitem [{\citenamefont {Peng}\ \emph {et~al.}(2016)\citenamefont {Peng},
  \citenamefont {Calvin}, \citenamefont {Pavošević}, \citenamefont {Zhang},\
  and\ \citenamefont {Valeev}}]{peng16}%
  \BibitemOpen
  \bibfield  {author} {\bibinfo {author} {\bibfnamefont {C.}~\bibnamefont
  {Peng}}, \bibinfo {author} {\bibfnamefont {J.~A.}\ \bibnamefont {Calvin}},
  \bibinfo {author} {\bibfnamefont {F.}~\bibnamefont {Pavošević}}, \bibinfo
  {author} {\bibfnamefont {J.}~\bibnamefont {Zhang}}, \ and\ \bibinfo {author}
  {\bibfnamefont {E.~F.}\ \bibnamefont {Valeev}},\ }\href {\doibase
  10.1021/acs.jpca.6b10150} {\bibfield  {journal} {\bibinfo  {journal} {The
  Journal of Physical Chemistry A}\ }\textbf {\bibinfo {volume} {120}},\
  \bibinfo {pages} {10231} (\bibinfo {year} {2016})},\ \bibinfo {note} {pMID:
  27966947},\ \Eprint
  {http://arxiv.org/abs/https://doi.org/10.1021/acs.jpca.6b10150}
  {https://doi.org/10.1021/acs.jpca.6b10150} \BibitemShut {NoStop}%
\bibitem [{\citenamefont {Lyakh}(2019)}]{lyakh19}%
  \BibitemOpen
  \bibfield  {author} {\bibinfo {author} {\bibfnamefont {D.~I.}\ \bibnamefont
  {Lyakh}},\ }\href {\doibase https://doi.org/10.1002/qua.25926} {\bibfield
  {journal} {\bibinfo  {journal} {International Journal of Quantum Chemistry}\
  }\textbf {\bibinfo {volume} {119}},\ \bibinfo {pages} {e25926} (\bibinfo
  {year} {2019})},\ \Eprint
  {http://arxiv.org/abs/https://onlinelibrary.wiley.com/doi/pdf/10.1002/qua.25926}
  {https://onlinelibrary.wiley.com/doi/pdf/10.1002/qua.25926} \BibitemShut
  {NoStop}%
\bibitem [{\citenamefont {Gyevi-Nagy}, \citenamefont {K\'{a}llay},\ and\
  \citenamefont {Nagy}(2020)}]{nagy20}%
  \BibitemOpen
  \bibfield  {author} {\bibinfo {author} {\bibfnamefont {L.}~\bibnamefont
  {Gyevi-Nagy}}, \bibinfo {author} {\bibfnamefont {M.}~\bibnamefont
  {K\'{a}llay}}, \ and\ \bibinfo {author} {\bibfnamefont {P.~R.}\ \bibnamefont
  {Nagy}},\ }\href@noop {} {\bibfield  {journal} {\bibinfo  {journal} {J. Chem.
  Theory Comp.}\ }\textbf {\bibinfo {volume} {16}},\ \bibinfo {pages} {366}
  (\bibinfo {year} {2020})}\BibitemShut {NoStop}%
\bibitem [{\citenamefont {Peng}\ \emph {et~al.}(2020)\citenamefont {Peng},
  \citenamefont {Lewis}, \citenamefont {Wang}, \citenamefont {Clement},
  \citenamefont {Pierce}, \citenamefont {Rishi}, \citenamefont {Pavošević},
  \citenamefont {Slattery}, \citenamefont {Zhang}, \citenamefont {Teke},
  \citenamefont {Kumar}, \citenamefont {Masteran}, \citenamefont {Asadchev},
  \citenamefont {Calvin},\ and\ \citenamefont {Valeev}}]{peng2020}%
  \BibitemOpen
  \bibfield  {author} {\bibinfo {author} {\bibfnamefont {C.}~\bibnamefont
  {Peng}}, \bibinfo {author} {\bibfnamefont {C.~A.}\ \bibnamefont {Lewis}},
  \bibinfo {author} {\bibfnamefont {X.}~\bibnamefont {Wang}}, \bibinfo {author}
  {\bibfnamefont {M.~C.}\ \bibnamefont {Clement}}, \bibinfo {author}
  {\bibfnamefont {K.}~\bibnamefont {Pierce}}, \bibinfo {author} {\bibfnamefont
  {V.}~\bibnamefont {Rishi}}, \bibinfo {author} {\bibfnamefont
  {F.}~\bibnamefont {Pavošević}}, \bibinfo {author} {\bibfnamefont
  {S.}~\bibnamefont {Slattery}}, \bibinfo {author} {\bibfnamefont
  {J.}~\bibnamefont {Zhang}}, \bibinfo {author} {\bibfnamefont
  {N.}~\bibnamefont {Teke}}, \bibinfo {author} {\bibfnamefont {A.}~\bibnamefont
  {Kumar}}, \bibinfo {author} {\bibfnamefont {C.}~\bibnamefont {Masteran}},
  \bibinfo {author} {\bibfnamefont {A.}~\bibnamefont {Asadchev}}, \bibinfo
  {author} {\bibfnamefont {J.~A.}\ \bibnamefont {Calvin}}, \ and\ \bibinfo
  {author} {\bibfnamefont {E.~F.}\ \bibnamefont {Valeev}},\ }\href {\doibase
  10.1063/5.0005889} {\bibfield  {journal} {\bibinfo  {journal} {The Journal of
  Chemical Physics}\ }\textbf {\bibinfo {volume} {153}},\ \bibinfo {pages}
  {044120} (\bibinfo {year} {2020})},\ \Eprint
  {http://arxiv.org/abs/https://doi.org/10.1063/5.0005889}
  {https://doi.org/10.1063/5.0005889} \BibitemShut {NoStop}%
\bibitem [{\citenamefont {Datta}\ and\ \citenamefont {Gordon}(2021)}]{datta21}%
  \BibitemOpen
  \bibfield  {author} {\bibinfo {author} {\bibfnamefont {D.}~\bibnamefont
  {Datta}}\ and\ \bibinfo {author} {\bibfnamefont {M.~S.}\ \bibnamefont
  {Gordon}},\ }\href@noop {} {\bibfield  {journal} {\bibinfo  {journal} {J.
  Chem. Theory Comp.}\ }\textbf {\bibinfo {volume} {17}},\ \bibinfo {pages}
  {4799} (\bibinfo {year} {2021})}\BibitemShut {NoStop}%
\bibitem [{\citenamefont {Gyevi-Nagy}, \citenamefont {K\'{a}llay},\ and\
  \citenamefont {Nagy}(2021)}]{nagy21}%
  \BibitemOpen
  \bibfield  {author} {\bibinfo {author} {\bibfnamefont {L.}~\bibnamefont
  {Gyevi-Nagy}}, \bibinfo {author} {\bibfnamefont {M.}~\bibnamefont
  {K\'{a}llay}}, \ and\ \bibinfo {author} {\bibfnamefont {P.~R.}\ \bibnamefont
  {Nagy}},\ }\href@noop {} {\bibfield  {journal} {\bibinfo  {journal} {J. Chem.
  Theory Comp.}\ }\textbf {\bibinfo {volume} {17}},\ \bibinfo {pages} {860}
  (\bibinfo {year} {2021})}\BibitemShut {NoStop}%
\bibitem [{\citenamefont {Kowalski}\ \emph {et~al.}(2021)\citenamefont
  {Kowalski}, \citenamefont {Bair}, \citenamefont {Bauman}, \citenamefont
  {Boschen}, \citenamefont {Bylaska}, \citenamefont {Daily}, \citenamefont
  {de~Jong}, \citenamefont {Dunning}, \citenamefont {Govind}, \citenamefont
  {Harrison}, \citenamefont {Keçeli}, \citenamefont {Keipert}, \citenamefont
  {Krishnamoorthy}, \citenamefont {Kumar}, \citenamefont {Mutlu}, \citenamefont
  {Palmer}, \citenamefont {Panyala}, \citenamefont {Peng}, \citenamefont
  {Richard}, \citenamefont {Straatsma}, \citenamefont {Sushko}, \citenamefont
  {Valeev}, \citenamefont {Valiev}, \citenamefont {van Dam}, \citenamefont
  {Waldrop}, \citenamefont {Williams-Young}, \citenamefont {Yang},
  \citenamefont {Zalewski},\ and\ \citenamefont {Windus}}]{kowalski21}%
  \BibitemOpen
  \bibfield  {author} {\bibinfo {author} {\bibfnamefont {K.}~\bibnamefont
  {Kowalski}}, \bibinfo {author} {\bibfnamefont {R.}~\bibnamefont {Bair}},
  \bibinfo {author} {\bibfnamefont {N.~P.}\ \bibnamefont {Bauman}}, \bibinfo
  {author} {\bibfnamefont {J.~S.}\ \bibnamefont {Boschen}}, \bibinfo {author}
  {\bibfnamefont {E.~J.}\ \bibnamefont {Bylaska}}, \bibinfo {author}
  {\bibfnamefont {J.}~\bibnamefont {Daily}}, \bibinfo {author} {\bibfnamefont
  {W.~A.}\ \bibnamefont {de~Jong}}, \bibinfo {author} {\bibfnamefont
  {T.}~\bibnamefont {Dunning}}, \bibinfo {author} {\bibfnamefont
  {N.}~\bibnamefont {Govind}}, \bibinfo {author} {\bibfnamefont {R.~J.}\
  \bibnamefont {Harrison}}, \bibinfo {author} {\bibfnamefont {M.}~\bibnamefont
  {Keçeli}}, \bibinfo {author} {\bibfnamefont {K.}~\bibnamefont {Keipert}},
  \bibinfo {author} {\bibfnamefont {S.}~\bibnamefont {Krishnamoorthy}},
  \bibinfo {author} {\bibfnamefont {S.}~\bibnamefont {Kumar}}, \bibinfo
  {author} {\bibfnamefont {E.}~\bibnamefont {Mutlu}}, \bibinfo {author}
  {\bibfnamefont {B.}~\bibnamefont {Palmer}}, \bibinfo {author} {\bibfnamefont
  {A.}~\bibnamefont {Panyala}}, \bibinfo {author} {\bibfnamefont
  {B.}~\bibnamefont {Peng}}, \bibinfo {author} {\bibfnamefont {R.~M.}\
  \bibnamefont {Richard}}, \bibinfo {author} {\bibfnamefont {T.~P.}\
  \bibnamefont {Straatsma}}, \bibinfo {author} {\bibfnamefont {P.}~\bibnamefont
  {Sushko}}, \bibinfo {author} {\bibfnamefont {E.~F.}\ \bibnamefont {Valeev}},
  \bibinfo {author} {\bibfnamefont {M.}~\bibnamefont {Valiev}}, \bibinfo
  {author} {\bibfnamefont {H.~J.~J.}\ \bibnamefont {van Dam}}, \bibinfo
  {author} {\bibfnamefont {J.~M.}\ \bibnamefont {Waldrop}}, \bibinfo {author}
  {\bibfnamefont {D.~B.}\ \bibnamefont {Williams-Young}}, \bibinfo {author}
  {\bibfnamefont {C.}~\bibnamefont {Yang}}, \bibinfo {author} {\bibfnamefont
  {M.}~\bibnamefont {Zalewski}}, \ and\ \bibinfo {author} {\bibfnamefont
  {T.~L.}\ \bibnamefont {Windus}},\ }\href {\doibase
  10.1021/acs.chemrev.0c00998} {\bibfield  {journal} {\bibinfo  {journal}
  {Chemical Reviews}\ }\textbf {\bibinfo {volume} {121}},\ \bibinfo {pages}
  {4962} (\bibinfo {year} {2021})},\ \bibinfo {note} {pMID: 33788546},\ \Eprint
  {http://arxiv.org/abs/https://doi.org/10.1021/acs.chemrev.0c00998}
  {https://doi.org/10.1021/acs.chemrev.0c00998} \BibitemShut {NoStop}%
\bibitem [{\citenamefont {Calvin}\ \emph {et~al.}(2021)\citenamefont {Calvin},
  \citenamefont {Peng}, \citenamefont {Rishi}, \citenamefont {Kumar},\ and\
  \citenamefont {Valeev}}]{calvin21}%
  \BibitemOpen
  \bibfield  {author} {\bibinfo {author} {\bibfnamefont {J.~A.}\ \bibnamefont
  {Calvin}}, \bibinfo {author} {\bibfnamefont {C.}~\bibnamefont {Peng}},
  \bibinfo {author} {\bibfnamefont {V.}~\bibnamefont {Rishi}}, \bibinfo
  {author} {\bibfnamefont {A.}~\bibnamefont {Kumar}}, \ and\ \bibinfo {author}
  {\bibfnamefont {E.~F.}\ \bibnamefont {Valeev}},\ }\href {\doibase
  10.1021/acs.chemrev.0c00006} {\bibfield  {journal} {\bibinfo  {journal}
  {Chemical Reviews}\ }\textbf {\bibinfo {volume} {121}},\ \bibinfo {pages}
  {1203} (\bibinfo {year} {2021})},\ \bibinfo {note} {pMID: 33305957},\ \Eprint
  {http://arxiv.org/abs/https://doi.org/10.1021/acs.chemrev.0c00006}
  {https://doi.org/10.1021/acs.chemrev.0c00006} \BibitemShut {NoStop}%
\bibitem [{\citenamefont {DePrince}\ and\ \citenamefont
  {Hammond}(2011)}]{deprince11}%
  \BibitemOpen
  \bibfield  {author} {\bibinfo {author} {\bibfnamefont {A.~E.}\ \bibnamefont
  {DePrince}}\ and\ \bibinfo {author} {\bibfnamefont {J.~R.}\ \bibnamefont
  {Hammond}},\ }\href@noop {} {\bibfield  {journal} {\bibinfo  {journal} {J.
  Chem. Theory Comp.}\ }\textbf {\bibinfo {volume} {7}},\ \bibinfo {pages}
  {1287} (\bibinfo {year} {2011})}\BibitemShut {NoStop}%
\bibitem [{\citenamefont {Ma}\ \emph {et~al.}(2011)\citenamefont {Ma},
  \citenamefont {Krishnamoorthy}, \citenamefont {Villa},\ and\ \citenamefont
  {Kowalski}}]{ma11}%
  \BibitemOpen
  \bibfield  {author} {\bibinfo {author} {\bibfnamefont {W.}~\bibnamefont
  {Ma}}, \bibinfo {author} {\bibfnamefont {S.}~\bibnamefont {Krishnamoorthy}},
  \bibinfo {author} {\bibfnamefont {O.}~\bibnamefont {Villa}}, \ and\ \bibinfo
  {author} {\bibfnamefont {K.}~\bibnamefont {Kowalski}},\ }\href@noop {}
  {\bibfield  {journal} {\bibinfo  {journal} {J. Chem. Theory Comp.}\ }\textbf
  {\bibinfo {volume} {7}},\ \bibinfo {pages} {1316} (\bibinfo {year}
  {2011})}\BibitemShut {NoStop}%
\bibitem [{\citenamefont {A.~Eugene~DePrince}\ \emph
  {et~al.}(2014)\citenamefont {A.~Eugene~DePrince}, \citenamefont {Kennedy},
  \citenamefont {Sumpter},\ and\ \citenamefont {Sherrill}}]{deprince14}%
  \BibitemOpen
  \bibfield  {author} {\bibinfo {author} {\bibfnamefont {I.}~\bibnamefont
  {A.~Eugene~DePrince}}, \bibinfo {author} {\bibfnamefont {M.~R.}\ \bibnamefont
  {Kennedy}}, \bibinfo {author} {\bibfnamefont {B.~G.}\ \bibnamefont
  {Sumpter}}, \ and\ \bibinfo {author} {\bibfnamefont {C.~D.}\ \bibnamefont
  {Sherrill}},\ }\href@noop {} {\bibfield  {journal} {\bibinfo  {journal} {Mol.
  Phys.}\ }\textbf {\bibinfo {volume} {112}},\ \bibinfo {pages} {844} (\bibinfo
  {year} {2014})}\BibitemShut {NoStop}%
\bibitem [{\citenamefont {Kaliman}\ and\ \citenamefont
  {Krylov}(2017)}]{kaliman17}%
  \BibitemOpen
  \bibfield  {author} {\bibinfo {author} {\bibfnamefont {I.~A.}\ \bibnamefont
  {Kaliman}}\ and\ \bibinfo {author} {\bibfnamefont {A.~I.}\ \bibnamefont
  {Krylov}},\ }\href
  {https://onlinelibrary.wiley.com/doix/abs/10.1002/jcc.24713} {\bibfield
  {journal} {\bibinfo  {journal} {J. Comp. Chem.}\ }\textbf {\bibinfo {volume}
  {38}},\ \bibinfo {pages} {842} (\bibinfo {year} {2017})}\BibitemShut
  {NoStop}%
\bibitem [{\citenamefont {DePrince~III}, \citenamefont {Hammond},\ and\
  \citenamefont {Sherrill}(2016)}]{deprince16}%
  \BibitemOpen
  \bibfield  {author} {\bibinfo {author} {\bibfnamefont {A.~E.}\ \bibnamefont
  {DePrince~III}}, \bibinfo {author} {\bibfnamefont {J.~R.}\ \bibnamefont
  {Hammond}}, \ and\ \bibinfo {author} {\bibfnamefont {C.~D.}\ \bibnamefont
  {Sherrill}},\ }\enquote {\bibinfo {title} {Iterative coupled-cluster methods
  on graphics processing units},}\ in\ \href
  {https://onlinelibrary.wiley.com/doix/abs/10.1002/9781118670712.ch13} {\emph
  {\bibinfo {booktitle} {Electronic Structure Calculations on Graphics
  Processing Units}}}\ (\bibinfo  {publisher} {John Wiley \& Sons, Ltd},\
  \bibinfo {year} {2016})\ Chap.~\bibinfo {chapter} {13}, pp.\ \bibinfo {pages}
  {279--300}\BibitemShut {NoStop}%
\bibitem [{\citenamefont {Peng}, \citenamefont {Calvin},\ and\ \citenamefont
  {Valeev}(2019)}]{peng19}%
  \BibitemOpen
  \bibfield  {author} {\bibinfo {author} {\bibfnamefont {C.}~\bibnamefont
  {Peng}}, \bibinfo {author} {\bibfnamefont {J.~A.}\ \bibnamefont {Calvin}}, \
  and\ \bibinfo {author} {\bibfnamefont {E.~F.}\ \bibnamefont {Valeev}},\
  }\href {https://onlinelibrary.wiley.com/doix/abs/10.1002/qua.25894}
  {\bibfield  {journal} {\bibinfo  {journal} {Int. J. Quantum Chem.}\ }\textbf
  {\bibinfo {volume} {119}},\ \bibinfo {pages} {e25894} (\bibinfo {year}
  {2019})}\BibitemShut {NoStop}%
\bibitem [{\citenamefont {Wang}, \citenamefont {Guo},\ and\ \citenamefont
  {Wang}(2020)}]{wang20}%
  \BibitemOpen
  \bibfield  {author} {\bibinfo {author} {\bibfnamefont {Z.}~\bibnamefont
  {Wang}}, \bibinfo {author} {\bibfnamefont {M.}~\bibnamefont {Guo}}, \ and\
  \bibinfo {author} {\bibfnamefont {F.}~\bibnamefont {Wang}},\ }\href
  {http://dx.doix.org/10.1039/D0CP03800H} {\bibfield  {journal} {\bibinfo
  {journal} {Phys. Chem. Chem. Phys.}\ }\textbf {\bibinfo {volume} {22}},\
  \bibinfo {pages} {25103} (\bibinfo {year} {2020})}\BibitemShut {NoStop}%
\bibitem [{\citenamefont {Seritan}\ \emph {et~al.}(2020)\citenamefont
  {Seritan}, \citenamefont {Bannwarth}, \citenamefont {Fales}, \citenamefont
  {Hohenstein}, \citenamefont {Kokkila-Schumacher}, \citenamefont {Luehr},
  \citenamefont {Snyder}, \citenamefont {Song}, \citenamefont {Titov},
  \citenamefont {Ufimtsev},\ and\ \citenamefont {Mart\'{\i}nez}}]{seritan20}%
  \BibitemOpen
  \bibfield  {author} {\bibinfo {author} {\bibfnamefont {S.}~\bibnamefont
  {Seritan}}, \bibinfo {author} {\bibfnamefont {C.}~\bibnamefont {Bannwarth}},
  \bibinfo {author} {\bibfnamefont {B.~S.}\ \bibnamefont {Fales}}, \bibinfo
  {author} {\bibfnamefont {E.~G.}\ \bibnamefont {Hohenstein}}, \bibinfo
  {author} {\bibfnamefont {S.~I.~L.}\ \bibnamefont {Kokkila-Schumacher}},
  \bibinfo {author} {\bibfnamefont {N.}~\bibnamefont {Luehr}}, \bibinfo
  {author} {\bibfnamefont {J.~W.}\ \bibnamefont {Snyder}}, \bibinfo {author}
  {\bibfnamefont {C.}~\bibnamefont {Song}}, \bibinfo {author} {\bibfnamefont
  {A.~V.}\ \bibnamefont {Titov}}, \bibinfo {author} {\bibfnamefont {I.~S.}\
  \bibnamefont {Ufimtsev}}, \ and\ \bibinfo {author} {\bibfnamefont {T.~J.}\
  \bibnamefont {Mart\'{\i}nez}},\ }\href@noop {} {\bibfield  {journal}
  {\bibinfo  {journal} {J. Chem. Phys.}\ }\textbf {\bibinfo {volume} {152}},\
  \bibinfo {pages} {224110} (\bibinfo {year} {2020})}\BibitemShut {NoStop}%
\bibitem [{\citenamefont {Adamowicz}\ and\ \citenamefont
  {Bartlett}(1987)}]{adam87}%
  \BibitemOpen
  \bibfield  {author} {\bibinfo {author} {\bibfnamefont {L.}~\bibnamefont
  {Adamowicz}}\ and\ \bibinfo {author} {\bibfnamefont {R.~J.}\ \bibnamefont
  {Bartlett}},\ }\href@noop {} {\bibfield  {journal} {\bibinfo  {journal} {J.
  Chem. Phys.}\ }\textbf {\bibinfo {volume} {86}},\ \bibinfo {pages} {6314}
  (\bibinfo {year} {1987})}\BibitemShut {NoStop}%
\bibitem [{\citenamefont {Adamowicz}, \citenamefont {Bartlett},\ and\
  \citenamefont {Sadlej}(1988)}]{adam88}%
  \BibitemOpen
  \bibfield  {author} {\bibinfo {author} {\bibfnamefont {L.}~\bibnamefont
  {Adamowicz}}, \bibinfo {author} {\bibfnamefont {R.~J.}\ \bibnamefont
  {Bartlett}}, \ and\ \bibinfo {author} {\bibfnamefont {A.~J.}\ \bibnamefont
  {Sadlej}},\ }\href@noop {} {\bibfield  {journal} {\bibinfo  {journal} {J.
  Chem. Phys.}\ }\textbf {\bibinfo {volume} {88}},\ \bibinfo {pages} {5749}
  (\bibinfo {year} {1988})}\BibitemShut {NoStop}%
\bibitem [{\citenamefont {Neogr\'{a}dy}, \citenamefont {Pito\v{n}\'{a}k},\ and\
  \citenamefont {Urban}(2005)}]{neo05}%
  \BibitemOpen
  \bibfield  {author} {\bibinfo {author} {\bibfnamefont {P.}~\bibnamefont
  {Neogr\'{a}dy}}, \bibinfo {author} {\bibfnamefont {M.}~\bibnamefont
  {Pito\v{n}\'{a}k}}, \ and\ \bibinfo {author} {\bibfnamefont {M.}~\bibnamefont
  {Urban}},\ }\href@noop {} {\bibfield  {journal} {\bibinfo  {journal} {Mol.
  Phys.}\ }\textbf {\bibinfo {volume} {103}},\ \bibinfo {pages} {2141}
  (\bibinfo {year} {2005})}\BibitemShut {NoStop}%
\bibitem [{\citenamefont {Pito\v{n}\'{a}k}\ \emph {et~al.}(2006)\citenamefont
  {Pito\v{n}\'{a}k}, \citenamefont {Holka}, \citenamefont {Neogr\'{a}dy},\ and\
  \citenamefont {Urban}}]{pitoniak06}%
  \BibitemOpen
  \bibfield  {author} {\bibinfo {author} {\bibfnamefont {M.}~\bibnamefont
  {Pito\v{n}\'{a}k}}, \bibinfo {author} {\bibfnamefont {F.}~\bibnamefont
  {Holka}}, \bibinfo {author} {\bibfnamefont {P.}~\bibnamefont {Neogr\'{a}dy}},
  \ and\ \bibinfo {author} {\bibfnamefont {M.}~\bibnamefont {Urban}},\ }\href
  {http://www.sciencedirect.com/science/article/pii/S0166128006002612}
  {\bibfield  {journal} {\bibinfo  {journal} {J. Mol. Struct.}\ }\textbf
  {\bibinfo {volume} {768}},\ \bibinfo {pages} {79 } (\bibinfo {year}
  {2006})}\BibitemShut {NoStop}%
\bibitem [{\citenamefont {Kumar}\ and\ \citenamefont
  {Crawford}(2017)}]{kumar17}%
  \BibitemOpen
  \bibfield  {author} {\bibinfo {author} {\bibfnamefont {A.}~\bibnamefont
  {Kumar}}\ and\ \bibinfo {author} {\bibfnamefont {T.~D.}\ \bibnamefont
  {Crawford}},\ }\href {\doibase 10.1021/acs.jpca.6b11410} {\bibfield
  {journal} {\bibinfo  {journal} {The Journal of Physical Chemistry A}\
  }\textbf {\bibinfo {volume} {121}},\ \bibinfo {pages} {708} (\bibinfo {year}
  {2017})},\ \bibinfo {note} {pMID: 28045265},\ \Eprint
  {http://arxiv.org/abs/https://doi.org/10.1021/acs.jpca.6b11410}
  {https://doi.org/10.1021/acs.jpca.6b11410} \BibitemShut {NoStop}%
\bibitem [{\citenamefont {Yang}\ \emph {et~al.}(2011)\citenamefont {Yang},
  \citenamefont {Kurashige}, \citenamefont {Manby},\ and\ \citenamefont
  {Chan}}]{yang11}%
  \BibitemOpen
  \bibfield  {author} {\bibinfo {author} {\bibfnamefont {J.}~\bibnamefont
  {Yang}}, \bibinfo {author} {\bibfnamefont {Y.}~\bibnamefont {Kurashige}},
  \bibinfo {author} {\bibfnamefont {F.~R.}\ \bibnamefont {Manby}}, \ and\
  \bibinfo {author} {\bibfnamefont {G.~K.~L.}\ \bibnamefont {Chan}},\
  }\href@noop {} {\bibfield  {journal} {\bibinfo  {journal} {J. Chem. Phys.}\
  }\textbf {\bibinfo {volume} {134}},\ \bibinfo {pages} {044123} (\bibinfo
  {year} {2011})}\BibitemShut {NoStop}%
\bibitem [{\citenamefont {Kurashige}\ \emph {et~al.}(2012)\citenamefont
  {Kurashige}, \citenamefont {Yang}, \citenamefont {Chan},\ and\ \citenamefont
  {Manby}}]{kura12}%
  \BibitemOpen
  \bibfield  {author} {\bibinfo {author} {\bibfnamefont {Y.}~\bibnamefont
  {Kurashige}}, \bibinfo {author} {\bibfnamefont {J.}~\bibnamefont {Yang}},
  \bibinfo {author} {\bibfnamefont {G.~K.-L.}\ \bibnamefont {Chan}}, \ and\
  \bibinfo {author} {\bibfnamefont {F.~R.}\ \bibnamefont {Manby}},\ }\href@noop
  {} {\bibfield  {journal} {\bibinfo  {journal} {J. Chem. Phys.}\ }\textbf
  {\bibinfo {volume} {136}},\ \bibinfo {pages} {124106} (\bibinfo {year}
  {2012})}\BibitemShut {NoStop}%
\bibitem [{\citenamefont {Yang}\ \emph {et~al.}(2012)\citenamefont {Yang},
  \citenamefont {Chan}, \citenamefont {Manby}, \citenamefont {Sch\"{u}tz},\
  and\ \citenamefont {Werner}}]{yang12}%
  \BibitemOpen
  \bibfield  {author} {\bibinfo {author} {\bibfnamefont {J.}~\bibnamefont
  {Yang}}, \bibinfo {author} {\bibfnamefont {G.~K.-L.}\ \bibnamefont {Chan}},
  \bibinfo {author} {\bibfnamefont {F.~R.}\ \bibnamefont {Manby}}, \bibinfo
  {author} {\bibfnamefont {M.}~\bibnamefont {Sch\"{u}tz}}, \ and\ \bibinfo
  {author} {\bibfnamefont {H.-J.}\ \bibnamefont {Werner}},\ }\href@noop {}
  {\bibfield  {journal} {\bibinfo  {journal} {J. Chem. Phys.}\ }\textbf
  {\bibinfo {volume} {136}},\ \bibinfo {pages} {144105} (\bibinfo {year}
  {2012})}\BibitemShut {NoStop}%
\bibitem [{\citenamefont {Sch\"{u}tz}\ \emph {et~al.}(2013)\citenamefont
  {Sch\"{u}tz}, \citenamefont {Yang}, \citenamefont {Chan}, \citenamefont
  {Manby},\ and\ \citenamefont {Werner}}]{schutz13}%
  \BibitemOpen
  \bibfield  {author} {\bibinfo {author} {\bibfnamefont {M.}~\bibnamefont
  {Sch\"{u}tz}}, \bibinfo {author} {\bibfnamefont {J.}~\bibnamefont {Yang}},
  \bibinfo {author} {\bibfnamefont {G.~K.-L.}\ \bibnamefont {Chan}}, \bibinfo
  {author} {\bibfnamefont {F.~R.}\ \bibnamefont {Manby}}, \ and\ \bibinfo
  {author} {\bibfnamefont {H.-J.}\ \bibnamefont {Werner}},\ }\href@noop {}
  {\bibfield  {journal} {\bibinfo  {journal} {J. Chem. Phys.}\ }\textbf
  {\bibinfo {volume} {138}},\ \bibinfo {pages} {054109} (\bibinfo {year}
  {2013})}\BibitemShut {NoStop}%
\bibitem [{\citenamefont {Li}, \citenamefont {Ma},\ and\ \citenamefont
  {Jiang}(2002)}]{li02}%
  \BibitemOpen
  \bibfield  {author} {\bibinfo {author} {\bibfnamefont {S.}~\bibnamefont
  {Li}}, \bibinfo {author} {\bibfnamefont {J.}~\bibnamefont {Ma}}, \ and\
  \bibinfo {author} {\bibfnamefont {Y.}~\bibnamefont {Jiang}},\ }\href
  {\doibase https://doi.org/10.1002/jcc.10003} {\bibfield  {journal} {\bibinfo
  {journal} {Journal of Computational Chemistry}\ }\textbf {\bibinfo {volume}
  {23}},\ \bibinfo {pages} {237} (\bibinfo {year} {2002})},\ \Eprint
  {http://arxiv.org/abs/https://onlinelibrary.wiley.com/doi/pdf/10.1002/jcc.10003}
  {https://onlinelibrary.wiley.com/doi/pdf/10.1002/jcc.10003} \BibitemShut
  {NoStop}%
\bibitem [{\citenamefont {Li}\ \emph {et~al.}(2006)\citenamefont {Li},
  \citenamefont {Shen}, \citenamefont {Li},\ and\ \citenamefont
  {Jiang}}]{li06}%
  \BibitemOpen
  \bibfield  {author} {\bibinfo {author} {\bibfnamefont {S.}~\bibnamefont
  {Li}}, \bibinfo {author} {\bibfnamefont {J.}~\bibnamefont {Shen}}, \bibinfo
  {author} {\bibfnamefont {W.}~\bibnamefont {Li}}, \ and\ \bibinfo {author}
  {\bibfnamefont {Y.}~\bibnamefont {Jiang}},\ }\href {\doibase
  10.1063/1.2244566} {\bibfield  {journal} {\bibinfo  {journal} {The Journal of
  Chemical Physics}\ }\textbf {\bibinfo {volume} {125}},\ \bibinfo {pages}
  {074109} (\bibinfo {year} {2006})},\ \Eprint
  {http://arxiv.org/abs/https://doi.org/10.1063/1.2244566}
  {https://doi.org/10.1063/1.2244566} \BibitemShut {NoStop}%
\bibitem [{\citenamefont {Li}\ \emph {et~al.}(2009)\citenamefont {Li},
  \citenamefont {Piecuch}, \citenamefont {Gour},\ and\ \citenamefont
  {Li}}]{li09}%
  \BibitemOpen
  \bibfield  {author} {\bibinfo {author} {\bibfnamefont {W.}~\bibnamefont
  {Li}}, \bibinfo {author} {\bibfnamefont {P.}~\bibnamefont {Piecuch}},
  \bibinfo {author} {\bibfnamefont {J.~R.}\ \bibnamefont {Gour}}, \ and\
  \bibinfo {author} {\bibfnamefont {S.}~\bibnamefont {Li}},\ }\href {\doibase
  10.1063/1.3218842} {\bibfield  {journal} {\bibinfo  {journal} {The Journal of
  Chemical Physics}\ }\textbf {\bibinfo {volume} {131}},\ \bibinfo {pages}
  {114109} (\bibinfo {year} {2009})},\ \Eprint
  {http://arxiv.org/abs/https://doi.org/10.1063/1.3218842}
  {https://doi.org/10.1063/1.3218842} \BibitemShut {NoStop}%
\bibitem [{\citenamefont {Neese}, \citenamefont {Wennmohs},\ and\ \citenamefont
  {Hansen}(2009)}]{neese09b}%
  \BibitemOpen
  \bibfield  {author} {\bibinfo {author} {\bibfnamefont {F.}~\bibnamefont
  {Neese}}, \bibinfo {author} {\bibfnamefont {F.}~\bibnamefont {Wennmohs}}, \
  and\ \bibinfo {author} {\bibfnamefont {A.}~\bibnamefont {Hansen}},\
  }\href@noop {} {\bibfield  {journal} {\bibinfo  {journal} {J. Chem. Phys.}\
  }\textbf {\bibinfo {volume} {130}},\ \bibinfo {pages} {114108} (\bibinfo
  {year} {2009})}\BibitemShut {NoStop}%
\bibitem [{\citenamefont {Li}\ and\ \citenamefont
  {Piecuch}(2010{\natexlab{a}})}]{li10a}%
  \BibitemOpen
  \bibfield  {author} {\bibinfo {author} {\bibfnamefont {W.}~\bibnamefont
  {Li}}\ and\ \bibinfo {author} {\bibfnamefont {P.}~\bibnamefont {Piecuch}},\
  }\href {\doibase 10.1021/jp1038738} {\bibfield  {journal} {\bibinfo
  {journal} {The Journal of Physical Chemistry A}\ }\textbf {\bibinfo {volume}
  {114}},\ \bibinfo {pages} {6721} (\bibinfo {year} {2010}{\natexlab{a}})},\
  \bibinfo {note} {pMID: 20496942},\ \Eprint
  {http://arxiv.org/abs/https://doi.org/10.1021/jp1038738}
  {https://doi.org/10.1021/jp1038738} \BibitemShut {NoStop}%
\bibitem [{\citenamefont {Li}\ and\ \citenamefont
  {Piecuch}(2010{\natexlab{b}})}]{li10b}%
  \BibitemOpen
  \bibfield  {author} {\bibinfo {author} {\bibfnamefont {W.}~\bibnamefont
  {Li}}\ and\ \bibinfo {author} {\bibfnamefont {P.}~\bibnamefont {Piecuch}},\
  }\href {\doibase 10.1021/jp100782u} {\bibfield  {journal} {\bibinfo
  {journal} {The Journal of Physical Chemistry A}\ }\textbf {\bibinfo {volume}
  {114}},\ \bibinfo {pages} {8644} (\bibinfo {year} {2010}{\natexlab{b}})},\
  \bibinfo {note} {pMID: 20373794},\ \Eprint
  {http://arxiv.org/abs/https://doi.org/10.1021/jp100782u}
  {https://doi.org/10.1021/jp100782u} \BibitemShut {NoStop}%
\bibitem [{\citenamefont {Rolik}\ and\ \citenamefont
  {Kállay}(2011)}]{rolik11}%
  \BibitemOpen
  \bibfield  {author} {\bibinfo {author} {\bibfnamefont {Z.}~\bibnamefont
  {Rolik}}\ and\ \bibinfo {author} {\bibfnamefont {M.}~\bibnamefont
  {Kállay}},\ }\href {\doibase 10.1063/1.3632085} {\bibfield  {journal}
  {\bibinfo  {journal} {The Journal of Chemical Physics}\ }\textbf {\bibinfo
  {volume} {135}},\ \bibinfo {pages} {104111} (\bibinfo {year} {2011})},\
  \Eprint {http://arxiv.org/abs/https://doi.org/10.1063/1.3632085}
  {https://doi.org/10.1063/1.3632085} \BibitemShut {NoStop}%
\bibitem [{\citenamefont {Rolik}\ \emph {et~al.}(2013)\citenamefont {Rolik},
  \citenamefont {Szegedy}, \citenamefont {Ladjánszki}, \citenamefont
  {Ladóczki},\ and\ \citenamefont {Kállay}}]{rolik13}%
  \BibitemOpen
  \bibfield  {author} {\bibinfo {author} {\bibfnamefont {Z.}~\bibnamefont
  {Rolik}}, \bibinfo {author} {\bibfnamefont {L.}~\bibnamefont {Szegedy}},
  \bibinfo {author} {\bibfnamefont {I.}~\bibnamefont {Ladjánszki}}, \bibinfo
  {author} {\bibfnamefont {B.}~\bibnamefont {Ladóczki}}, \ and\ \bibinfo
  {author} {\bibfnamefont {M.}~\bibnamefont {Kállay}},\ }\href {\doibase
  10.1063/1.4819401} {\bibfield  {journal} {\bibinfo  {journal} {The Journal of
  Chemical Physics}\ }\textbf {\bibinfo {volume} {139}},\ \bibinfo {pages}
  {094105} (\bibinfo {year} {2013})},\ \Eprint
  {http://arxiv.org/abs/https://doi.org/10.1063/1.4819401}
  {https://doi.org/10.1063/1.4819401} \BibitemShut {NoStop}%
\bibitem [{\citenamefont {Riplinger}\ and\ \citenamefont
  {Neese}(2013)}]{riplinger13a}%
  \BibitemOpen
  \bibfield  {author} {\bibinfo {author} {\bibfnamefont {C.}~\bibnamefont
  {Riplinger}}\ and\ \bibinfo {author} {\bibfnamefont {F.}~\bibnamefont
  {Neese}},\ }\href@noop {} {\bibfield  {journal} {\bibinfo  {journal} {J.
  Chem. Phys.}\ }\textbf {\bibinfo {volume} {138}},\ \bibinfo {pages} {034106}
  (\bibinfo {year} {2013})}\BibitemShut {NoStop}%
\bibitem [{\citenamefont {Riplinger}\ \emph {et~al.}(2013)\citenamefont
  {Riplinger}, \citenamefont {Sandhoefer}, \citenamefont {Hansen},\ and\
  \citenamefont {Neese}}]{riplinger13b}%
  \BibitemOpen
  \bibfield  {author} {\bibinfo {author} {\bibfnamefont {C.}~\bibnamefont
  {Riplinger}}, \bibinfo {author} {\bibfnamefont {B.}~\bibnamefont
  {Sandhoefer}}, \bibinfo {author} {\bibfnamefont {A.}~\bibnamefont {Hansen}},
  \ and\ \bibinfo {author} {\bibfnamefont {F.}~\bibnamefont {Neese}},\
  }\href@noop {} {\bibfield  {journal} {\bibinfo  {journal} {J. Chem. Phys.}\
  }\textbf {\bibinfo {volume} {139}},\ \bibinfo {pages} {134101} (\bibinfo
  {year} {2013})}\BibitemShut {NoStop}%
\bibitem [{\citenamefont {Liakos}\ \emph {et~al.}(2015)\citenamefont {Liakos},
  \citenamefont {Sparta}, \citenamefont {Kesharwani}, \citenamefont {Martin},\
  and\ \citenamefont {Neese}}]{liakos15}%
  \BibitemOpen
  \bibfield  {author} {\bibinfo {author} {\bibfnamefont {D.~G.}\ \bibnamefont
  {Liakos}}, \bibinfo {author} {\bibfnamefont {M.}~\bibnamefont {Sparta}},
  \bibinfo {author} {\bibfnamefont {M.~K.}\ \bibnamefont {Kesharwani}},
  \bibinfo {author} {\bibfnamefont {J.~M.~L.}\ \bibnamefont {Martin}}, \ and\
  \bibinfo {author} {\bibfnamefont {F.}~\bibnamefont {Neese}},\ }\href@noop {}
  {\bibfield  {journal} {\bibinfo  {journal} {J. Chem. Theory Comput.}\
  }\textbf {\bibinfo {volume} {11}},\ \bibinfo {pages} {1525} (\bibinfo {year}
  {2015})}\BibitemShut {NoStop}%
\bibitem [{\citenamefont {Schwilk}\ \emph {et~al.}(2017)\citenamefont
  {Schwilk}, \citenamefont {Ma}, \citenamefont {K\"{o}ppl},\ and\ \citenamefont
  {Werner}}]{schwilk17}%
  \BibitemOpen
  \bibfield  {author} {\bibinfo {author} {\bibfnamefont {M.}~\bibnamefont
  {Schwilk}}, \bibinfo {author} {\bibfnamefont {Q.}~\bibnamefont {Ma}},
  \bibinfo {author} {\bibfnamefont {C.}~\bibnamefont {K\"{o}ppl}}, \ and\
  \bibinfo {author} {\bibfnamefont {H.-J.}\ \bibnamefont {Werner}},\
  }\href@noop {} {\bibfield  {journal} {\bibinfo  {journal} {J. Chem. Theory
  Comput.}\ }\textbf {\bibinfo {volume} {13}},\ \bibinfo {pages} {3650}
  (\bibinfo {year} {2017})}\BibitemShut {NoStop}%
\bibitem [{\citenamefont {Kolda}\ and\ \citenamefont {Bader}(2009)}]{kolda09}%
  \BibitemOpen
  \bibfield  {author} {\bibinfo {author} {\bibfnamefont {T.~G.}\ \bibnamefont
  {Kolda}}\ and\ \bibinfo {author} {\bibfnamefont {B.~W.}\ \bibnamefont
  {Bader}},\ }\href@noop {} {\bibfield  {journal} {\bibinfo  {journal} {SIAM
  Rev.}\ }\textbf {\bibinfo {volume} {51}},\ \bibinfo {pages} {455} (\bibinfo
  {year} {2009})}\BibitemShut {NoStop}%
\bibitem [{\citenamefont {Whitten}(1973)}]{whitten73}%
  \BibitemOpen
  \bibfield  {author} {\bibinfo {author} {\bibfnamefont {J.~L.}\ \bibnamefont
  {Whitten}},\ }\href@noop {} {\bibfield  {journal} {\bibinfo  {journal} {J.
  Chem. Phys.}\ }\textbf {\bibinfo {volume} {58}},\ \bibinfo {pages} {4496}
  (\bibinfo {year} {1973})}\BibitemShut {NoStop}%
\bibitem [{\citenamefont {Baerends}, \citenamefont {Ellis},\ and\ \citenamefont
  {Ros}(1973)}]{baerends73}%
  \BibitemOpen
  \bibfield  {author} {\bibinfo {author} {\bibfnamefont {E.}~\bibnamefont
  {Baerends}}, \bibinfo {author} {\bibfnamefont {D.}~\bibnamefont {Ellis}}, \
  and\ \bibinfo {author} {\bibfnamefont {P.}~\bibnamefont {Ros}},\ }\href
  {http://www.sciencedirect.com/science/article/pii/030101047380059X}
  {\bibfield  {journal} {\bibinfo  {journal} {Chem. Phys.}\ }\textbf {\bibinfo
  {volume} {2}},\ \bibinfo {pages} {41 } (\bibinfo {year} {1973})}\BibitemShut
  {NoStop}%
\bibitem [{\citenamefont {Dunlap}, \citenamefont {Connolly},\ and\
  \citenamefont {Sabin}(1979)}]{dunlap79}%
  \BibitemOpen
  \bibfield  {author} {\bibinfo {author} {\bibfnamefont {B.~I.}\ \bibnamefont
  {Dunlap}}, \bibinfo {author} {\bibfnamefont {J.~W.~D.}\ \bibnamefont
  {Connolly}}, \ and\ \bibinfo {author} {\bibfnamefont {J.~R.}\ \bibnamefont
  {Sabin}},\ }\href@noop {} {\bibfield  {journal} {\bibinfo  {journal} {J.
  Chem. Phys.}\ }\textbf {\bibinfo {volume} {71}},\ \bibinfo {pages} {3396}
  (\bibinfo {year} {1979})}\BibitemShut {NoStop}%
\bibitem [{\citenamefont {Van~Alsenoy}(1988)}]{alsenoy88}%
  \BibitemOpen
  \bibfield  {author} {\bibinfo {author} {\bibfnamefont {C.}~\bibnamefont
  {Van~Alsenoy}},\ }\href
  {https://onlinelibrary.wiley.com/doixx/abs/10.1002/jcc.540090607} {\bibfield
  {journal} {\bibinfo  {journal} {J. Comp. Chem.}\ }\textbf {\bibinfo {volume}
  {9}},\ \bibinfo {pages} {620} (\bibinfo {year} {1988})}\BibitemShut {NoStop}%
\bibitem [{\citenamefont {Vahtras}, \citenamefont {Alml{\"{o}}f},\ and\
  \citenamefont {Feyereisen}(1993)}]{vahtras93}%
  \BibitemOpen
  \bibfield  {author} {\bibinfo {author} {\bibfnamefont {O.}~\bibnamefont
  {Vahtras}}, \bibinfo {author} {\bibfnamefont {J.}~\bibnamefont
  {Alml{\"{o}}f}}, \ and\ \bibinfo {author} {\bibfnamefont {M.}~\bibnamefont
  {Feyereisen}},\ }\href@noop {} {\bibfield  {journal} {\bibinfo  {journal}
  {Chem. Phys. Lett.}\ }\textbf {\bibinfo {volume} {213}},\ \bibinfo {pages}
  {514 } (\bibinfo {year} {1993})}\BibitemShut {NoStop}%
\bibitem [{\citenamefont {Beebe}\ and\ \citenamefont
  {Linderberg}(1997)}]{beebe77}%
  \BibitemOpen
  \bibfield  {author} {\bibinfo {author} {\bibfnamefont {N.~H.~F.}\
  \bibnamefont {Beebe}}\ and\ \bibinfo {author} {\bibfnamefont
  {J.}~\bibnamefont {Linderberg}},\ }\href@noop {} {\bibfield  {journal}
  {\bibinfo  {journal} {Int. J. Quantum Chem.}\ }\textbf {\bibinfo {volume}
  {12}},\ \bibinfo {pages} {683} (\bibinfo {year} {1997})}\BibitemShut
  {NoStop}%
\bibitem [{\citenamefont {Koch}, \citenamefont {S\'{a}nchez~de Mer\'{a}s},\
  and\ \citenamefont {Pedersen}(2003)}]{koch03}%
  \BibitemOpen
  \bibfield  {author} {\bibinfo {author} {\bibfnamefont {H.}~\bibnamefont
  {Koch}}, \bibinfo {author} {\bibfnamefont {A.}~\bibnamefont {S\'{a}nchez~de
  Mer\'{a}s}}, \ and\ \bibinfo {author} {\bibfnamefont {T.~B.}\ \bibnamefont
  {Pedersen}},\ }\href@noop {} {\bibfield  {journal} {\bibinfo  {journal} {J.
  Chem. Phys.}\ }\textbf {\bibinfo {volume} {118}},\ \bibinfo {pages} {9481}
  (\bibinfo {year} {2003})}\BibitemShut {NoStop}%
\bibitem [{\citenamefont {Pedersen}, \citenamefont {S\'{a}nchez~de Mer\'{a}s},\
  and\ \citenamefont {Koch}(2004)}]{pedersen04}%
  \BibitemOpen
  \bibfield  {author} {\bibinfo {author} {\bibfnamefont {T.~B.}\ \bibnamefont
  {Pedersen}}, \bibinfo {author} {\bibfnamefont {A.~M.~J.}\ \bibnamefont
  {S\'{a}nchez~de Mer\'{a}s}}, \ and\ \bibinfo {author} {\bibfnamefont
  {H.}~\bibnamefont {Koch}},\ }\href@noop {} {\bibfield  {journal} {\bibinfo
  {journal} {J. Chem. Phys.}\ }\textbf {\bibinfo {volume} {120}},\ \bibinfo
  {pages} {8887} (\bibinfo {year} {2004})}\BibitemShut {NoStop}%
\bibitem [{\citenamefont {Folkestad}, \citenamefont {Kj\o{}nstad},\ and\
  \citenamefont {Koch}(2019)}]{folkestad19}%
  \BibitemOpen
  \bibfield  {author} {\bibinfo {author} {\bibfnamefont {S.~D.}\ \bibnamefont
  {Folkestad}}, \bibinfo {author} {\bibfnamefont {E.~F.}\ \bibnamefont
  {Kj\o{}nstad}}, \ and\ \bibinfo {author} {\bibfnamefont {H.}~\bibnamefont
  {Koch}},\ }\href@noop {} {\bibfield  {journal} {\bibinfo  {journal} {J. Chem.
  Phys.}\ }\textbf {\bibinfo {volume} {150}},\ \bibinfo {pages} {194112}
  (\bibinfo {year} {2019})}\BibitemShut {NoStop}%
\bibitem [{\citenamefont {Martinez}, \citenamefont {Mehta},\ and\ \citenamefont
  {Carter}(1992)}]{martinez92}%
  \BibitemOpen
  \bibfield  {author} {\bibinfo {author} {\bibfnamefont {T.~J.}\ \bibnamefont
  {Martinez}}, \bibinfo {author} {\bibfnamefont {A.}~\bibnamefont {Mehta}}, \
  and\ \bibinfo {author} {\bibfnamefont {E.~A.}\ \bibnamefont {Carter}},\
  }\href {\doibase 10.1063/1.463176} {\bibfield  {journal} {\bibinfo  {journal}
  {The Journal of Chemical Physics}\ }\textbf {\bibinfo {volume} {97}},\
  \bibinfo {pages} {1876} (\bibinfo {year} {1992})},\ \Eprint
  {http://arxiv.org/abs/https://doi.org/10.1063/1.463176}
  {https://doi.org/10.1063/1.463176} \BibitemShut {NoStop}%
\bibitem [{\citenamefont {Martinez}\ and\ \citenamefont
  {Carter}(1993)}]{martinez93}%
  \BibitemOpen
  \bibfield  {author} {\bibinfo {author} {\bibfnamefont {T.~J.}\ \bibnamefont
  {Martinez}}\ and\ \bibinfo {author} {\bibfnamefont {E.~A.}\ \bibnamefont
  {Carter}},\ }\href {\doibase 10.1063/1.464751} {\bibfield  {journal}
  {\bibinfo  {journal} {The Journal of Chemical Physics}\ }\textbf {\bibinfo
  {volume} {98}},\ \bibinfo {pages} {7081} (\bibinfo {year} {1993})},\ \Eprint
  {http://arxiv.org/abs/https://doi.org/10.1063/1.464751}
  {https://doi.org/10.1063/1.464751} \BibitemShut {NoStop}%
\bibitem [{\citenamefont {Martinez}\ and\ \citenamefont
  {Carter}(1994)}]{martinez94}%
  \BibitemOpen
  \bibfield  {author} {\bibinfo {author} {\bibfnamefont {T.~J.}\ \bibnamefont
  {Martinez}}\ and\ \bibinfo {author} {\bibfnamefont {E.~A.}\ \bibnamefont
  {Carter}},\ }\href {\doibase 10.1063/1.466350} {\bibfield  {journal}
  {\bibinfo  {journal} {The Journal of Chemical Physics}\ }\textbf {\bibinfo
  {volume} {100}},\ \bibinfo {pages} {3631} (\bibinfo {year} {1994})},\ \Eprint
  {http://arxiv.org/abs/https://doi.org/10.1063/1.466350}
  {https://doi.org/10.1063/1.466350} \BibitemShut {NoStop}%
\bibitem [{\citenamefont {Martinez}\ and\ \citenamefont
  {Carter}(1995)}]{martinez95}%
  \BibitemOpen
  \bibfield  {author} {\bibinfo {author} {\bibfnamefont {T.~J.}\ \bibnamefont
  {Martinez}}\ and\ \bibinfo {author} {\bibfnamefont {E.~A.}\ \bibnamefont
  {Carter}},\ }\href {\doibase 10.1063/1.469088} {\bibfield  {journal}
  {\bibinfo  {journal} {The Journal of Chemical Physics}\ }\textbf {\bibinfo
  {volume} {102}},\ \bibinfo {pages} {7564} (\bibinfo {year} {1995})},\ \Eprint
  {http://arxiv.org/abs/https://doi.org/10.1063/1.469088}
  {https://doi.org/10.1063/1.469088} \BibitemShut {NoStop}%
\bibitem [{\citenamefont {Reynolds}, \citenamefont {Martinez},\ and\
  \citenamefont {Carter}(1996)}]{martinez96}%
  \BibitemOpen
  \bibfield  {author} {\bibinfo {author} {\bibfnamefont {G.}~\bibnamefont
  {Reynolds}}, \bibinfo {author} {\bibfnamefont {T.~J.}\ \bibnamefont
  {Martinez}}, \ and\ \bibinfo {author} {\bibfnamefont {E.~A.}\ \bibnamefont
  {Carter}},\ }\href {\doibase 10.1063/1.472495} {\bibfield  {journal}
  {\bibinfo  {journal} {The Journal of Chemical Physics}\ }\textbf {\bibinfo
  {volume} {105}},\ \bibinfo {pages} {6455} (\bibinfo {year} {1996})},\ \Eprint
  {http://arxiv.org/abs/https://doi.org/10.1063/1.472495}
  {https://doi.org/10.1063/1.472495} \BibitemShut {NoStop}%
\bibitem [{\citenamefont {Neese}\ \emph {et~al.}(2009)\citenamefont {Neese},
  \citenamefont {Wennmohs}, \citenamefont {Hansen},\ and\ \citenamefont
  {Becker}}]{neese09}%
  \BibitemOpen
  \bibfield  {author} {\bibinfo {author} {\bibfnamefont {F.}~\bibnamefont
  {Neese}}, \bibinfo {author} {\bibfnamefont {F.}~\bibnamefont {Wennmohs}},
  \bibinfo {author} {\bibfnamefont {A.}~\bibnamefont {Hansen}}, \ and\ \bibinfo
  {author} {\bibfnamefont {U.}~\bibnamefont {Becker}},\ }\href
  {http://www.sciencedirect.com/science/article/pii/S0301010408005089}
  {\bibfield  {journal} {\bibinfo  {journal} {Chem. Phys.}\ }\textbf {\bibinfo
  {volume} {356}},\ \bibinfo {pages} {98 } (\bibinfo {year}
  {2009})}\BibitemShut {NoStop}%
\bibitem [{\citenamefont {Kossmann}\ and\ \citenamefont
  {Neese}(2010)}]{kossmann10}%
  \BibitemOpen
  \bibfield  {author} {\bibinfo {author} {\bibfnamefont {S.}~\bibnamefont
  {Kossmann}}\ and\ \bibinfo {author} {\bibfnamefont {F.}~\bibnamefont
  {Neese}},\ }\href@noop {} {\bibfield  {journal} {\bibinfo  {journal} {J.
  Chem. Theory Comput.}\ }\textbf {\bibinfo {volume} {6}},\ \bibinfo {pages}
  {2325} (\bibinfo {year} {2010})}\BibitemShut {NoStop}%
\bibitem [{\citenamefont {Izs\'{a}k}\ and\ \citenamefont
  {Neese}(2011)}]{izsak11}%
  \BibitemOpen
  \bibfield  {author} {\bibinfo {author} {\bibfnamefont {R.}~\bibnamefont
  {Izs\'{a}k}}\ and\ \bibinfo {author} {\bibfnamefont {F.}~\bibnamefont
  {Neese}},\ }\href@noop {} {\bibfield  {journal} {\bibinfo  {journal} {J.
  Chem. Phys.}\ }\textbf {\bibinfo {volume} {135}},\ \bibinfo {pages} {144105}
  (\bibinfo {year} {2011})}\BibitemShut {NoStop}%
\bibitem [{\citenamefont {Petrenko}, \citenamefont {Kossmann},\ and\
  \citenamefont {Neese}(2011)}]{taras11}%
  \BibitemOpen
  \bibfield  {author} {\bibinfo {author} {\bibfnamefont {T.}~\bibnamefont
  {Petrenko}}, \bibinfo {author} {\bibfnamefont {S.}~\bibnamefont {Kossmann}},
  \ and\ \bibinfo {author} {\bibfnamefont {F.}~\bibnamefont {Neese}},\
  }\href@noop {} {\bibfield  {journal} {\bibinfo  {journal} {J. Chem. Phys.}\
  }\textbf {\bibinfo {volume} {134}},\ \bibinfo {pages} {054116} (\bibinfo
  {year} {2011})}\BibitemShut {NoStop}%
\bibitem [{\citenamefont {Izs\'{a}k}, \citenamefont {Hansen},\ and\
  \citenamefont {Neese}(2012)}]{izsak12}%
  \BibitemOpen
  \bibfield  {author} {\bibinfo {author} {\bibfnamefont {R.}~\bibnamefont
  {Izs\'{a}k}}, \bibinfo {author} {\bibfnamefont {A.}~\bibnamefont {Hansen}}, \
  and\ \bibinfo {author} {\bibfnamefont {F.}~\bibnamefont {Neese}},\
  }\href@noop {} {\bibfield  {journal} {\bibinfo  {journal} {Mol. Phys.}\
  }\textbf {\bibinfo {volume} {110}},\ \bibinfo {pages} {2413} (\bibinfo {year}
  {2012})}\BibitemShut {NoStop}%
\bibitem [{\citenamefont {Izs\'{a}k}\ and\ \citenamefont
  {Neese}(2013)}]{izsak13}%
  \BibitemOpen
  \bibfield  {author} {\bibinfo {author} {\bibfnamefont {R.}~\bibnamefont
  {Izs\'{a}k}}\ and\ \bibinfo {author} {\bibfnamefont {F.}~\bibnamefont
  {Neese}},\ }\href@noop {} {\bibfield  {journal} {\bibinfo  {journal} {Mol.
  Phys.}\ }\textbf {\bibinfo {volume} {111}},\ \bibinfo {pages} {1190}
  (\bibinfo {year} {2013})}\BibitemShut {NoStop}%
\bibitem [{\citenamefont {Dutta}, \citenamefont {Neese},\ and\ \citenamefont
  {Izs\'{a}k}(2016)}]{dutta16}%
  \BibitemOpen
  \bibfield  {author} {\bibinfo {author} {\bibfnamefont {A.~K.}\ \bibnamefont
  {Dutta}}, \bibinfo {author} {\bibfnamefont {F.}~\bibnamefont {Neese}}, \ and\
  \bibinfo {author} {\bibfnamefont {R.}~\bibnamefont {Izs\'{a}k}},\ }\href@noop
  {} {\bibfield  {journal} {\bibinfo  {journal} {J. Chem. Phys.}\ }\textbf
  {\bibinfo {volume} {144}},\ \bibinfo {pages} {034102} (\bibinfo {year}
  {2016})}\BibitemShut {NoStop}%
\bibitem [{\citenamefont {Hohenstein}, \citenamefont {Parrish},\ and\
  \citenamefont {Martínez}(2012)}]{hohenstein12}%
  \BibitemOpen
  \bibfield  {author} {\bibinfo {author} {\bibfnamefont {E.~G.}\ \bibnamefont
  {Hohenstein}}, \bibinfo {author} {\bibfnamefont {R.~M.}\ \bibnamefont
  {Parrish}}, \ and\ \bibinfo {author} {\bibfnamefont {T.~J.}\ \bibnamefont
  {Martínez}},\ }\href {\doibase 10.1063/1.4732310} {\bibfield  {journal}
  {\bibinfo  {journal} {The Journal of Chemical Physics}\ }\textbf {\bibinfo
  {volume} {137}},\ \bibinfo {pages} {044103} (\bibinfo {year} {2012})},\
  \Eprint {http://arxiv.org/abs/https://doi.org/10.1063/1.4732310}
  {https://doi.org/10.1063/1.4732310} \BibitemShut {NoStop}%
\bibitem [{\citenamefont {Parrish}\ \emph {et~al.}(2012)\citenamefont
  {Parrish}, \citenamefont {Hohenstein}, \citenamefont {Mart\'{\i}nez},\ and\
  \citenamefont {Sherrill}}]{parrish12}%
  \BibitemOpen
  \bibfield  {author} {\bibinfo {author} {\bibfnamefont {R.~M.}\ \bibnamefont
  {Parrish}}, \bibinfo {author} {\bibfnamefont {E.~G.}\ \bibnamefont
  {Hohenstein}}, \bibinfo {author} {\bibfnamefont {T.~J.}\ \bibnamefont
  {Mart\'{\i}nez}}, \ and\ \bibinfo {author} {\bibfnamefont {C.~D.}\
  \bibnamefont {Sherrill}},\ }\href@noop {} {\bibfield  {journal} {\bibinfo
  {journal} {J. Chem. Phys.}\ }\textbf {\bibinfo {volume} {137}},\ \bibinfo
  {pages} {224106} (\bibinfo {year} {2012})}\BibitemShut {NoStop}%
\bibitem [{\citenamefont {Parrish}\ \emph
  {et~al.}(2013{\natexlab{a}})\citenamefont {Parrish}, \citenamefont
  {Hohenstein}, \citenamefont {Schunck}, \citenamefont {Sherrill},\ and\
  \citenamefont {Mart\'{\i}nez}}]{parrish13a}%
  \BibitemOpen
  \bibfield  {author} {\bibinfo {author} {\bibfnamefont {R.~M.}\ \bibnamefont
  {Parrish}}, \bibinfo {author} {\bibfnamefont {E.~G.}\ \bibnamefont
  {Hohenstein}}, \bibinfo {author} {\bibfnamefont {N.~F.}\ \bibnamefont
  {Schunck}}, \bibinfo {author} {\bibfnamefont {C.~D.}\ \bibnamefont
  {Sherrill}}, \ and\ \bibinfo {author} {\bibfnamefont {T.~J.}\ \bibnamefont
  {Mart\'{\i}nez}},\ }\href
  {https://link.aps.org/doix/10.1103/PhysRevLett.111.132505} {\bibfield
  {journal} {\bibinfo  {journal} {Phys. Rev. Lett.}\ }\textbf {\bibinfo
  {volume} {111}},\ \bibinfo {pages} {132505} (\bibinfo {year}
  {2013}{\natexlab{a}})}\BibitemShut {NoStop}%
\bibitem [{\citenamefont {Parrish}\ \emph
  {et~al.}(2013{\natexlab{b}})\citenamefont {Parrish}, \citenamefont
  {Hohenstein}, \citenamefont {Mart\'{\i}nez},\ and\ \citenamefont
  {Sherrill}}]{parrish13b}%
  \BibitemOpen
  \bibfield  {author} {\bibinfo {author} {\bibfnamefont {R.~M.}\ \bibnamefont
  {Parrish}}, \bibinfo {author} {\bibfnamefont {E.~G.}\ \bibnamefont
  {Hohenstein}}, \bibinfo {author} {\bibfnamefont {T.~J.}\ \bibnamefont
  {Mart\'{\i}nez}}, \ and\ \bibinfo {author} {\bibfnamefont {C.~D.}\
  \bibnamefont {Sherrill}},\ }\href@noop {} {\bibfield  {journal} {\bibinfo
  {journal} {J. Chem. Phys.}\ }\textbf {\bibinfo {volume} {138}},\ \bibinfo
  {pages} {194107} (\bibinfo {year} {2013}{\natexlab{b}})}\BibitemShut
  {NoStop}%
\bibitem [{\citenamefont {Benedikt}\ \emph {et~al.}(2011)\citenamefont
  {Benedikt}, \citenamefont {Auer}, \citenamefont {Espig},\ and\ \citenamefont
  {Hackbusch}}]{benedikt11}%
  \BibitemOpen
  \bibfield  {author} {\bibinfo {author} {\bibfnamefont {U.}~\bibnamefont
  {Benedikt}}, \bibinfo {author} {\bibfnamefont {A.~A.}\ \bibnamefont {Auer}},
  \bibinfo {author} {\bibfnamefont {M.}~\bibnamefont {Espig}}, \ and\ \bibinfo
  {author} {\bibfnamefont {W.}~\bibnamefont {Hackbusch}},\ }\href@noop {}
  {\bibfield  {journal} {\bibinfo  {journal} {J. Chem. Phys.}\ }\textbf
  {\bibinfo {volume} {134}},\ \bibinfo {pages} {054118} (\bibinfo {year}
  {2011})}\BibitemShut {NoStop}%
\bibitem [{\citenamefont {Benedikt}, \citenamefont {B\"{o}hm},\ and\
  \citenamefont {Auer}(2013)}]{benedikt13}%
  \BibitemOpen
  \bibfield  {author} {\bibinfo {author} {\bibfnamefont {U.}~\bibnamefont
  {Benedikt}}, \bibinfo {author} {\bibfnamefont {K.-H.}\ \bibnamefont
  {B\"{o}hm}}, \ and\ \bibinfo {author} {\bibfnamefont {A.~A.}\ \bibnamefont
  {Auer}},\ }\href@noop {} {\bibfield  {journal} {\bibinfo  {journal} {J. Chem.
  Phys.}\ }\textbf {\bibinfo {volume} {139}},\ \bibinfo {pages} {224101}
  (\bibinfo {year} {2013})}\BibitemShut {NoStop}%
\bibitem [{\citenamefont {Hohenstein}\ \emph
  {et~al.}(2013{\natexlab{a}})\citenamefont {Hohenstein}, \citenamefont
  {Kokkila}, \citenamefont {Parrish},\ and\ \citenamefont
  {Mart\'{\i}nez}}]{hohenstein13a}%
  \BibitemOpen
  \bibfield  {author} {\bibinfo {author} {\bibfnamefont {E.~G.}\ \bibnamefont
  {Hohenstein}}, \bibinfo {author} {\bibfnamefont {S.~I.~L.}\ \bibnamefont
  {Kokkila}}, \bibinfo {author} {\bibfnamefont {R.~M.}\ \bibnamefont
  {Parrish}}, \ and\ \bibinfo {author} {\bibfnamefont {T.~J.}\ \bibnamefont
  {Mart\'{\i}nez}},\ }\href@noop {} {\bibfield  {journal} {\bibinfo  {journal}
  {J. Chem. Phys.}\ }\textbf {\bibinfo {volume} {138}},\ \bibinfo {pages}
  {124111} (\bibinfo {year} {2013}{\natexlab{a}})}\BibitemShut {NoStop}%
\bibitem [{\citenamefont {Kokkila~Schumacher}\ \emph
  {et~al.}(2015)\citenamefont {Kokkila~Schumacher}, \citenamefont {Hohenstein},
  \citenamefont {Parrish}, \citenamefont {Wang},\ and\ \citenamefont
  {Mart\'{\i}nez}}]{schumacher15}%
  \BibitemOpen
  \bibfield  {author} {\bibinfo {author} {\bibfnamefont {S.~I.~L.}\
  \bibnamefont {Kokkila~Schumacher}}, \bibinfo {author} {\bibfnamefont {E.~G.}\
  \bibnamefont {Hohenstein}}, \bibinfo {author} {\bibfnamefont {R.~M.}\
  \bibnamefont {Parrish}}, \bibinfo {author} {\bibfnamefont {L.-P.}\
  \bibnamefont {Wang}}, \ and\ \bibinfo {author} {\bibfnamefont {T.~J.}\
  \bibnamefont {Mart\'{\i}nez}},\ }\href@noop {} {\bibfield  {journal}
  {\bibinfo  {journal} {J. Chem. Theory Comput.}\ }\textbf {\bibinfo {volume}
  {11}},\ \bibinfo {pages} {3042} (\bibinfo {year} {2015})}\BibitemShut
  {NoStop}%
\bibitem [{\citenamefont {Lee}, \citenamefont {Lin},\ and\ \citenamefont
  {Head-Gordon}(2020)}]{lee20}%
  \BibitemOpen
  \bibfield  {author} {\bibinfo {author} {\bibfnamefont {J.}~\bibnamefont
  {Lee}}, \bibinfo {author} {\bibfnamefont {L.}~\bibnamefont {Lin}}, \ and\
  \bibinfo {author} {\bibfnamefont {M.}~\bibnamefont {Head-Gordon}},\
  }\href@noop {} {\bibfield  {journal} {\bibinfo  {journal} {J. Chem. Theory
  Comput.}\ }\textbf {\bibinfo {volume} {16}},\ \bibinfo {pages} {243}
  (\bibinfo {year} {2020})}\BibitemShut {NoStop}%
\bibitem [{\citenamefont {Matthews}(2021)}]{matthews21}%
  \BibitemOpen
  \bibfield  {author} {\bibinfo {author} {\bibfnamefont {D.~A.}\ \bibnamefont
  {Matthews}},\ }\href@noop {} {\bibfield  {journal} {\bibinfo  {journal} {J.
  Chem. Phys.}\ }\textbf {\bibinfo {volume} {154}},\ \bibinfo {pages} {134102}
  (\bibinfo {year} {2021})}\BibitemShut {NoStop}%
\bibitem [{\citenamefont {Bell}, \citenamefont {Lambrecht},\ and\ \citenamefont
  {Head-Gordon}(2010)}]{bell10}%
  \BibitemOpen
  \bibfield  {author} {\bibinfo {author} {\bibfnamefont {F.}~\bibnamefont
  {Bell}}, \bibinfo {author} {\bibfnamefont {D.}~\bibnamefont {Lambrecht}}, \
  and\ \bibinfo {author} {\bibfnamefont {M.}~\bibnamefont {Head-Gordon}},\
  }\href@noop {} {\bibfield  {journal} {\bibinfo  {journal} {Mol. Phys.}\
  }\textbf {\bibinfo {volume} {108}},\ \bibinfo {pages} {2759} (\bibinfo {year}
  {2010})}\BibitemShut {NoStop}%
\bibitem [{\citenamefont {Kinoshita}, \citenamefont {Hino},\ and\ \citenamefont
  {Bartlett}(2003)}]{kinoshita03}%
  \BibitemOpen
  \bibfield  {author} {\bibinfo {author} {\bibfnamefont {T.}~\bibnamefont
  {Kinoshita}}, \bibinfo {author} {\bibfnamefont {O.}~\bibnamefont {Hino}}, \
  and\ \bibinfo {author} {\bibfnamefont {R.~J.}\ \bibnamefont {Bartlett}},\
  }\href@noop {} {\bibfield  {journal} {\bibinfo  {journal} {J. Chem. Phys.}\
  }\textbf {\bibinfo {volume} {119}},\ \bibinfo {pages} {7756} (\bibinfo {year}
  {2003})}\BibitemShut {NoStop}%
\bibitem [{\citenamefont {Hino}, \citenamefont {Kinoshita},\ and\ \citenamefont
  {Bartlett}(2004)}]{hino04}%
  \BibitemOpen
  \bibfield  {author} {\bibinfo {author} {\bibfnamefont {O.}~\bibnamefont
  {Hino}}, \bibinfo {author} {\bibfnamefont {T.}~\bibnamefont {Kinoshita}}, \
  and\ \bibinfo {author} {\bibfnamefont {R.~J.}\ \bibnamefont {Bartlett}},\
  }\href@noop {} {\bibfield  {journal} {\bibinfo  {journal} {J. Chem. Phys.}\
  }\textbf {\bibinfo {volume} {121}},\ \bibinfo {pages} {1206} (\bibinfo {year}
  {2004})}\BibitemShut {NoStop}%
\bibitem [{\citenamefont {Scuseria}, \citenamefont {Henderson},\ and\
  \citenamefont {Sorensen}(2008)}]{scuseria08}%
  \BibitemOpen
  \bibfield  {author} {\bibinfo {author} {\bibfnamefont {G.~E.}\ \bibnamefont
  {Scuseria}}, \bibinfo {author} {\bibfnamefont {T.~M.}\ \bibnamefont
  {Henderson}}, \ and\ \bibinfo {author} {\bibfnamefont {D.~C.}\ \bibnamefont
  {Sorensen}},\ }\href@noop {} {\bibfield  {journal} {\bibinfo  {journal} {J.
  Chem. Phys.}\ }\textbf {\bibinfo {volume} {129}},\ \bibinfo {pages} {231101}
  (\bibinfo {year} {2008})}\BibitemShut {NoStop}%
\bibitem [{\citenamefont {Schutski}\ \emph {et~al.}(2017)\citenamefont
  {Schutski}, \citenamefont {Zhao}, \citenamefont {Henderson},\ and\
  \citenamefont {Scuseria}}]{schutski17}%
  \BibitemOpen
  \bibfield  {author} {\bibinfo {author} {\bibfnamefont {R.}~\bibnamefont
  {Schutski}}, \bibinfo {author} {\bibfnamefont {J.}~\bibnamefont {Zhao}},
  \bibinfo {author} {\bibfnamefont {T.~M.}\ \bibnamefont {Henderson}}, \ and\
  \bibinfo {author} {\bibfnamefont {G.~E.}\ \bibnamefont {Scuseria}},\ }\href
  {\doibase 10.1063/1.4996988} {\bibfield  {journal} {\bibinfo  {journal} {The
  Journal of Chemical Physics}\ }\textbf {\bibinfo {volume} {147}},\ \bibinfo
  {pages} {184113} (\bibinfo {year} {2017})},\ \Eprint
  {http://arxiv.org/abs/https://doi.org/10.1063/1.4996988}
  {https://doi.org/10.1063/1.4996988} \BibitemShut {NoStop}%
\bibitem [{\citenamefont {Hohenstein}\ \emph {et~al.}(2012)\citenamefont
  {Hohenstein}, \citenamefont {Parrish}, \citenamefont {Sherrill},\ and\
  \citenamefont {Mart\'{\i}nez}}]{hohenstein12b}%
  \BibitemOpen
  \bibfield  {author} {\bibinfo {author} {\bibfnamefont {E.~G.}\ \bibnamefont
  {Hohenstein}}, \bibinfo {author} {\bibfnamefont {R.~M.}\ \bibnamefont
  {Parrish}}, \bibinfo {author} {\bibfnamefont {C.~D.}\ \bibnamefont
  {Sherrill}}, \ and\ \bibinfo {author} {\bibfnamefont {T.~J.}\ \bibnamefont
  {Mart\'{\i}nez}},\ }\href@noop {} {\bibfield  {journal} {\bibinfo  {journal}
  {J. Chem. Phys.}\ }\textbf {\bibinfo {volume} {137}},\ \bibinfo {pages}
  {221101} (\bibinfo {year} {2012})}\BibitemShut {NoStop}%
\bibitem [{\citenamefont {Hohenstein}\ \emph
  {et~al.}(2013{\natexlab{b}})\citenamefont {Hohenstein}, \citenamefont
  {Kokkila}, \citenamefont {Parrish},\ and\ \citenamefont
  {Mart\'{\i}nez}}]{hohenstein13b}%
  \BibitemOpen
  \bibfield  {author} {\bibinfo {author} {\bibfnamefont {E.~G.}\ \bibnamefont
  {Hohenstein}}, \bibinfo {author} {\bibfnamefont {S.~I.~L.}\ \bibnamefont
  {Kokkila}}, \bibinfo {author} {\bibfnamefont {R.~M.}\ \bibnamefont
  {Parrish}}, \ and\ \bibinfo {author} {\bibfnamefont {T.~J.}\ \bibnamefont
  {Mart\'{\i}nez}},\ }\href@noop {} {\bibfield  {journal} {\bibinfo  {journal}
  {J. Phys. Chem. B}\ }\textbf {\bibinfo {volume} {117}},\ \bibinfo {pages}
  {12972} (\bibinfo {year} {2013}{\natexlab{b}})}\BibitemShut {NoStop}%
\bibitem [{\citenamefont {Parrish}\ \emph {et~al.}(2014)\citenamefont
  {Parrish}, \citenamefont {Sherrill}, \citenamefont {Hohenstein},
  \citenamefont {Kokkila},\ and\ \citenamefont {Mart\'{\i}nez}}]{parrish14}%
  \BibitemOpen
  \bibfield  {author} {\bibinfo {author} {\bibfnamefont {R.~M.}\ \bibnamefont
  {Parrish}}, \bibinfo {author} {\bibfnamefont {C.~D.}\ \bibnamefont
  {Sherrill}}, \bibinfo {author} {\bibfnamefont {E.~G.}\ \bibnamefont
  {Hohenstein}}, \bibinfo {author} {\bibfnamefont {S.~I.~L.}\ \bibnamefont
  {Kokkila}}, \ and\ \bibinfo {author} {\bibfnamefont {T.~J.}\ \bibnamefont
  {Mart\'{\i}nez}},\ }\href@noop {} {\bibfield  {journal} {\bibinfo  {journal}
  {J. Chem. Phys.}\ }\textbf {\bibinfo {volume} {140}},\ \bibinfo {pages}
  {181102} (\bibinfo {year} {2014})}\BibitemShut {NoStop}%
\bibitem [{\citenamefont {Lesiuk}(2020{\natexlab{a}})}]{lesiuk20}%
  \BibitemOpen
  \bibfield  {author} {\bibinfo {author} {\bibfnamefont {M.}~\bibnamefont
  {Lesiuk}},\ }\href@noop {} {\bibfield  {journal} {\bibinfo  {journal} {J.
  Chem. Theory Comput.}\ }\textbf {\bibinfo {volume} {16}},\ \bibinfo {pages}
  {453} (\bibinfo {year} {2020}{\natexlab{a}})}\BibitemShut {NoStop}%
\bibitem [{\citenamefont {Parrish}\ \emph {et~al.}(2019)\citenamefont
  {Parrish}, \citenamefont {Zhao}, \citenamefont {Hohenstein},\ and\
  \citenamefont {Mart\'{\i}nez}}]{parrish19}%
  \BibitemOpen
  \bibfield  {author} {\bibinfo {author} {\bibfnamefont {R.~M.}\ \bibnamefont
  {Parrish}}, \bibinfo {author} {\bibfnamefont {Y.}~\bibnamefont {Zhao}},
  \bibinfo {author} {\bibfnamefont {E.~G.}\ \bibnamefont {Hohenstein}}, \ and\
  \bibinfo {author} {\bibfnamefont {T.~J.}\ \bibnamefont {Mart\'{\i}nez}},\
  }\href@noop {} {\bibfield  {journal} {\bibinfo  {journal} {J. Chem. Phys.}\
  }\textbf {\bibinfo {volume} {150}},\ \bibinfo {pages} {164118} (\bibinfo
  {year} {2019})}\BibitemShut {NoStop}%
\bibitem [{\citenamefont {Helgaker}\ \emph {et~al.}(1997)\citenamefont
  {Helgaker}, \citenamefont {Klopper}, \citenamefont {Koch},\ and\
  \citenamefont {Noga}}]{helgaker97}%
  \BibitemOpen
  \bibfield  {author} {\bibinfo {author} {\bibfnamefont {T.}~\bibnamefont
  {Helgaker}}, \bibinfo {author} {\bibfnamefont {W.}~\bibnamefont {Klopper}},
  \bibinfo {author} {\bibfnamefont {H.}~\bibnamefont {Koch}}, \ and\ \bibinfo
  {author} {\bibfnamefont {J.}~\bibnamefont {Noga}},\ }\href@noop {} {\bibfield
   {journal} {\bibinfo  {journal} {J. Chem. Phys.}\ }\textbf {\bibinfo {volume}
  {106}},\ \bibinfo {pages} {9639} (\bibinfo {year} {1997})}\BibitemShut
  {NoStop}%
\bibitem [{\citenamefont {Karton}, \citenamefont {Taylor},\ and\ \citenamefont
  {Martin}(2007)}]{karton07}%
  \BibitemOpen
  \bibfield  {author} {\bibinfo {author} {\bibfnamefont {A.}~\bibnamefont
  {Karton}}, \bibinfo {author} {\bibfnamefont {P.~R.}\ \bibnamefont {Taylor}},
  \ and\ \bibinfo {author} {\bibfnamefont {J.~M.~L.}\ \bibnamefont {Martin}},\
  }\href@noop {} {\bibfield  {journal} {\bibinfo  {journal} {J. Chem. Phys.}\
  }\textbf {\bibinfo {volume} {127}},\ \bibinfo {pages} {064104} (\bibinfo
  {year} {2007})}\BibitemShut {NoStop}%
\bibitem [{\citenamefont {Martin}\ and\ \citenamefont
  {de~Oliveira}(1999)}]{martin99}%
  \BibitemOpen
  \bibfield  {author} {\bibinfo {author} {\bibfnamefont {J.~M.~L.}\
  \bibnamefont {Martin}}\ and\ \bibinfo {author} {\bibfnamefont
  {G.}~\bibnamefont {de~Oliveira}},\ }\href@noop {} {\bibfield  {journal}
  {\bibinfo  {journal} {J. Chem. Phys.}\ }\textbf {\bibinfo {volume} {111}},\
  \bibinfo {pages} {1843} (\bibinfo {year} {1999})}\BibitemShut {NoStop}%
\bibitem [{\citenamefont {Tucker}(1966)}]{tucker66}%
  \BibitemOpen
  \bibfield  {author} {\bibinfo {author} {\bibfnamefont {L.~R.}\ \bibnamefont
  {Tucker}},\ }\href@noop {} {\bibfield  {journal} {\bibinfo  {journal}
  {Psychometrika}\ }\textbf {\bibinfo {volume} {31}},\ \bibinfo {pages} {279}
  (\bibinfo {year} {1966})}\BibitemShut {NoStop}%
\bibitem [{\citenamefont {Lesiuk}(2019)}]{lesiuk19}%
  \BibitemOpen
  \bibfield  {author} {\bibinfo {author} {\bibfnamefont {M.}~\bibnamefont
  {Lesiuk}},\ }\href
  {https://onlinelibrary.wiley.com/doix/abs/10.1002/jcc.25788} {\bibfield
  {journal} {\bibinfo  {journal} {J. Comp. Chem.}\ }\textbf {\bibinfo {volume}
  {40}},\ \bibinfo {pages} {1319} (\bibinfo {year} {2019})}\BibitemShut
  {NoStop}%
\bibitem [{\citenamefont {De~Lathauwer}, \citenamefont {De~Moor},\ and\
  \citenamefont {Vandewalle}(2000{\natexlab{a}})}]{delath00}%
  \BibitemOpen
  \bibfield  {author} {\bibinfo {author} {\bibfnamefont {L.}~\bibnamefont
  {De~Lathauwer}}, \bibinfo {author} {\bibfnamefont {B.}~\bibnamefont
  {De~Moor}}, \ and\ \bibinfo {author} {\bibfnamefont {J.}~\bibnamefont
  {Vandewalle}},\ }\href@noop {} {\bibfield  {journal} {\bibinfo  {journal}
  {SIAM J. Matrix Anal. Appl.}\ }\textbf {\bibinfo {volume} {21}},\ \bibinfo
  {pages} {1253} (\bibinfo {year} {2000}{\natexlab{a}})}\BibitemShut {NoStop}%
\bibitem [{\citenamefont {Vannieuwenhoven}, \citenamefont {Vandebril},\ and\
  \citenamefont {Meerbergen}(2012)}]{vannie12}%
  \BibitemOpen
  \bibfield  {author} {\bibinfo {author} {\bibfnamefont {N.}~\bibnamefont
  {Vannieuwenhoven}}, \bibinfo {author} {\bibfnamefont {R.}~\bibnamefont
  {Vandebril}}, \ and\ \bibinfo {author} {\bibfnamefont {K.}~\bibnamefont
  {Meerbergen}},\ }\href@noop {} {\bibfield  {journal} {\bibinfo  {journal}
  {SIAM J. Sci. Comput.}\ }\textbf {\bibinfo {volume} {34}},\ \bibinfo {pages}
  {A1027} (\bibinfo {year} {2012})}\BibitemShut {NoStop}%
\bibitem [{\citenamefont {De~Lathauwer}, \citenamefont {De~Moor},\ and\
  \citenamefont {Vandewalle}(2000{\natexlab{b}})}]{delath00b}%
  \BibitemOpen
  \bibfield  {author} {\bibinfo {author} {\bibfnamefont {L.}~\bibnamefont
  {De~Lathauwer}}, \bibinfo {author} {\bibfnamefont {B.}~\bibnamefont
  {De~Moor}}, \ and\ \bibinfo {author} {\bibfnamefont {J.}~\bibnamefont
  {Vandewalle}},\ }\href@noop {} {\bibfield  {journal} {\bibinfo  {journal}
  {SIAM J. Matrix Anal. Appl.}\ }\textbf {\bibinfo {volume} {21}},\ \bibinfo
  {pages} {1324} (\bibinfo {year} {2000}{\natexlab{b}})}\BibitemShut {NoStop}%
\bibitem [{\citenamefont {Eld\'{e}n}\ and\ \citenamefont
  {Savas}(2009)}]{elden09}%
  \BibitemOpen
  \bibfield  {author} {\bibinfo {author} {\bibfnamefont {L.}~\bibnamefont
  {Eld\'{e}n}}\ and\ \bibinfo {author} {\bibfnamefont {B.}~\bibnamefont
  {Savas}},\ }\href@noop {} {\bibfield  {journal} {\bibinfo  {journal} {SIAM J.
  Matrix Anal. Appl.}\ }\textbf {\bibinfo {volume} {31}},\ \bibinfo {pages}
  {248} (\bibinfo {year} {2009})}\BibitemShut {NoStop}%
\bibitem [{\citenamefont {Cichocki}\ \emph {et~al.}(2015)\citenamefont
  {Cichocki}, \citenamefont {Mandic}, \citenamefont {De~Lathauwer},
  \citenamefont {Zhou}, \citenamefont {Zhao}, \citenamefont {Caiafa},\ and\
  \citenamefont {PHAN}}]{cichocki15}%
  \BibitemOpen
  \bibfield  {author} {\bibinfo {author} {\bibfnamefont {A.}~\bibnamefont
  {Cichocki}}, \bibinfo {author} {\bibfnamefont {D.}~\bibnamefont {Mandic}},
  \bibinfo {author} {\bibfnamefont {L.}~\bibnamefont {De~Lathauwer}}, \bibinfo
  {author} {\bibfnamefont {G.}~\bibnamefont {Zhou}}, \bibinfo {author}
  {\bibfnamefont {Q.}~\bibnamefont {Zhao}}, \bibinfo {author} {\bibfnamefont
  {C.}~\bibnamefont {Caiafa}}, \ and\ \bibinfo {author} {\bibfnamefont {H.~A.}\
  \bibnamefont {PHAN}},\ }\href@noop {} {\bibfield  {journal} {\bibinfo
  {journal} {IEEE Signal Process. Mag.}\ }\textbf {\bibinfo {volume} {32}},\
  \bibinfo {pages} {145} (\bibinfo {year} {2015})}\BibitemShut {NoStop}%
\bibitem [{\citenamefont {Liu}\ \emph {et~al.}(2014)\citenamefont {Liu},
  \citenamefont {Shang}, \citenamefont {Fan}, \citenamefont {Cheng},\ and\
  \citenamefont {Cheng}}]{liu14}%
  \BibitemOpen
  \bibfield  {author} {\bibinfo {author} {\bibfnamefont {Y.}~\bibnamefont
  {Liu}}, \bibinfo {author} {\bibfnamefont {F.}~\bibnamefont {Shang}}, \bibinfo
  {author} {\bibfnamefont {W.}~\bibnamefont {Fan}}, \bibinfo {author}
  {\bibfnamefont {J.}~\bibnamefont {Cheng}}, \ and\ \bibinfo {author}
  {\bibfnamefont {H.}~\bibnamefont {Cheng}},\ }in\ \href
  {https://proceedings.neurips.cc/paper/2014/file/8d6dc35e506fc23349dd10ee68dabb64-Paper.pdf}
  {\emph {\bibinfo {booktitle} {Advances in Neural Information Processing
  Systems}}},\ Vol.~\bibinfo {volume} {27},\ \bibinfo {editor} {edited by\
  \bibinfo {editor} {\bibfnamefont {Z.}~\bibnamefont {Ghahramani}}, \bibinfo
  {editor} {\bibfnamefont {M.}~\bibnamefont {Welling}}, \bibinfo {editor}
  {\bibfnamefont {C.}~\bibnamefont {Cortes}}, \bibinfo {editor} {\bibfnamefont
  {N.}~\bibnamefont {Lawrence}}, \ and\ \bibinfo {editor} {\bibfnamefont
  {K.~Q.}\ \bibnamefont {Weinberger}}}\ (\bibinfo  {publisher} {Curran
  Associates, Inc.},\ \bibinfo {year} {2014})\BibitemShut {NoStop}%
\bibitem [{\citenamefont {M\o{}rup}(2011)}]{morup11}%
  \BibitemOpen
  \bibfield  {author} {\bibinfo {author} {\bibfnamefont {M.}~\bibnamefont
  {M\o{}rup}},\ }\href
  {https://wires.onlinelibrary.wiley.com/doix/abs/10.1002/widm.1} {\bibfield
  {journal} {\bibinfo  {journal} {WIREs Data Min. Knowl. Discov.}\ }\textbf
  {\bibinfo {volume} {1}},\ \bibinfo {pages} {24} (\bibinfo {year}
  {2011})}\BibitemShut {NoStop}%
\bibitem [{\citenamefont {Hummel}, \citenamefont {Tsatsoulis},\ and\
  \citenamefont {Grüneis}(2017)}]{hummel17}%
  \BibitemOpen
  \bibfield  {author} {\bibinfo {author} {\bibfnamefont {F.}~\bibnamefont
  {Hummel}}, \bibinfo {author} {\bibfnamefont {T.}~\bibnamefont {Tsatsoulis}},
  \ and\ \bibinfo {author} {\bibfnamefont {A.}~\bibnamefont {Grüneis}},\
  }\href {\doibase 10.1063/1.4977994} {\bibfield  {journal} {\bibinfo
  {journal} {The Journal of Chemical Physics}\ }\textbf {\bibinfo {volume}
  {146}},\ \bibinfo {pages} {124105} (\bibinfo {year} {2017})},\ \Eprint
  {http://arxiv.org/abs/https://doi.org/10.1063/1.4977994}
  {https://doi.org/10.1063/1.4977994} \BibitemShut {NoStop}%
\bibitem [{\citenamefont {Pierce}, \citenamefont {Rishi},\ and\ \citenamefont
  {Valeev}(2021)}]{pierce21}%
  \BibitemOpen
  \bibfield  {author} {\bibinfo {author} {\bibfnamefont {K.}~\bibnamefont
  {Pierce}}, \bibinfo {author} {\bibfnamefont {V.}~\bibnamefont {Rishi}}, \
  and\ \bibinfo {author} {\bibfnamefont {E.~F.}\ \bibnamefont {Valeev}},\
  }\href {\doibase 10.1021/acs.jctc.0c01310} {\bibfield  {journal} {\bibinfo
  {journal} {Journal of Chemical Theory and Computation}\ }\textbf {\bibinfo
  {volume} {17}},\ \bibinfo {pages} {2217} (\bibinfo {year} {2021})},\ \bibinfo
  {note} {pMID: 33780616},\ \Eprint
  {http://arxiv.org/abs/https://doi.org/10.1021/acs.jctc.0c01310}
  {https://doi.org/10.1021/acs.jctc.0c01310} \BibitemShut {NoStop}%
\bibitem [{\citenamefont {Schmidt}\ \emph {et~al.}(1993)\citenamefont
  {Schmidt}, \citenamefont {Baldridge}, \citenamefont {Boatz}, \citenamefont
  {Elbert}, \citenamefont {Gordon}, \citenamefont {Jensen}, \citenamefont
  {Koseki}, \citenamefont {Matsunaga}, \citenamefont {Nguyen}, \citenamefont
  {Su}, \citenamefont {Windus}, \citenamefont {Dupuis},\ and\ \citenamefont
  {Montgomery}}]{gamess1}%
  \BibitemOpen
  \bibfield  {author} {\bibinfo {author} {\bibfnamefont {M.~W.}\ \bibnamefont
  {Schmidt}}, \bibinfo {author} {\bibfnamefont {K.~K.}\ \bibnamefont
  {Baldridge}}, \bibinfo {author} {\bibfnamefont {J.~A.}\ \bibnamefont
  {Boatz}}, \bibinfo {author} {\bibfnamefont {S.~T.}\ \bibnamefont {Elbert}},
  \bibinfo {author} {\bibfnamefont {M.~S.}\ \bibnamefont {Gordon}}, \bibinfo
  {author} {\bibfnamefont {J.~H.}\ \bibnamefont {Jensen}}, \bibinfo {author}
  {\bibfnamefont {S.}~\bibnamefont {Koseki}}, \bibinfo {author} {\bibfnamefont
  {N.}~\bibnamefont {Matsunaga}}, \bibinfo {author} {\bibfnamefont {K.~A.}\
  \bibnamefont {Nguyen}}, \bibinfo {author} {\bibfnamefont {S.}~\bibnamefont
  {Su}}, \bibinfo {author} {\bibfnamefont {T.~L.}\ \bibnamefont {Windus}},
  \bibinfo {author} {\bibfnamefont {M.}~\bibnamefont {Dupuis}}, \ and\ \bibinfo
  {author} {\bibfnamefont {J.~A.}\ \bibnamefont {Montgomery}},\ }\href@noop {}
  {\bibfield  {journal} {\bibinfo  {journal} {J. Comp. Chem.}\ }\textbf
  {\bibinfo {volume} {14}},\ \bibinfo {pages} {1347} (\bibinfo {year}
  {1993})}\BibitemShut {NoStop}%
\bibitem [{\citenamefont {Barca}\ \emph {et~al.}(2020)\citenamefont {Barca},
  \citenamefont {Bertoni}, \citenamefont {Carrington}, \citenamefont {Datta},
  \citenamefont {De~Silva}, \citenamefont {Deustua}, \citenamefont {Fedorov},
  \citenamefont {Gour}, \citenamefont {Gunina}, \citenamefont {Guidez},
  \citenamefont {Harville}, \citenamefont {Irle}, \citenamefont {Ivanic},
  \citenamefont {Kowalski}, \citenamefont {Leang}, \citenamefont {Li},
  \citenamefont {Li}, \citenamefont {Lutz}, \citenamefont {Magoulas},
  \citenamefont {Mato}, \citenamefont {Mironov}, \citenamefont {Nakata},
  \citenamefont {Pham}, \citenamefont {Piecuch}, \citenamefont {Poole},
  \citenamefont {Pruitt}, \citenamefont {Rendell}, \citenamefont {Roskop},
  \citenamefont {Ruedenberg}, \citenamefont {Sattasathuchana}, \citenamefont
  {Schmidt}, \citenamefont {Shen}, \citenamefont {Slipchenko}, \citenamefont
  {Sosonkina}, \citenamefont {Sundriyal}, \citenamefont {Tiwari}, \citenamefont
  {Galvez~Vallejo}, \citenamefont {Westheimer}, \citenamefont {Włoch},
  \citenamefont {Xu}, \citenamefont {Zahariev},\ and\ \citenamefont
  {Gordon}}]{gamess2}%
  \BibitemOpen
  \bibfield  {author} {\bibinfo {author} {\bibfnamefont {G.~M.~J.}\
  \bibnamefont {Barca}}, \bibinfo {author} {\bibfnamefont {C.}~\bibnamefont
  {Bertoni}}, \bibinfo {author} {\bibfnamefont {L.}~\bibnamefont {Carrington}},
  \bibinfo {author} {\bibfnamefont {D.}~\bibnamefont {Datta}}, \bibinfo
  {author} {\bibfnamefont {N.}~\bibnamefont {De~Silva}}, \bibinfo {author}
  {\bibfnamefont {J.~E.}\ \bibnamefont {Deustua}}, \bibinfo {author}
  {\bibfnamefont {D.~G.}\ \bibnamefont {Fedorov}}, \bibinfo {author}
  {\bibfnamefont {J.~R.}\ \bibnamefont {Gour}}, \bibinfo {author}
  {\bibfnamefont {A.~O.}\ \bibnamefont {Gunina}}, \bibinfo {author}
  {\bibfnamefont {E.}~\bibnamefont {Guidez}}, \bibinfo {author} {\bibfnamefont
  {T.}~\bibnamefont {Harville}}, \bibinfo {author} {\bibfnamefont
  {S.}~\bibnamefont {Irle}}, \bibinfo {author} {\bibfnamefont {J.}~\bibnamefont
  {Ivanic}}, \bibinfo {author} {\bibfnamefont {K.}~\bibnamefont {Kowalski}},
  \bibinfo {author} {\bibfnamefont {S.~S.}\ \bibnamefont {Leang}}, \bibinfo
  {author} {\bibfnamefont {H.}~\bibnamefont {Li}}, \bibinfo {author}
  {\bibfnamefont {W.}~\bibnamefont {Li}}, \bibinfo {author} {\bibfnamefont
  {J.~J.}\ \bibnamefont {Lutz}}, \bibinfo {author} {\bibfnamefont
  {I.}~\bibnamefont {Magoulas}}, \bibinfo {author} {\bibfnamefont
  {J.}~\bibnamefont {Mato}}, \bibinfo {author} {\bibfnamefont {V.}~\bibnamefont
  {Mironov}}, \bibinfo {author} {\bibfnamefont {H.}~\bibnamefont {Nakata}},
  \bibinfo {author} {\bibfnamefont {B.~Q.}\ \bibnamefont {Pham}}, \bibinfo
  {author} {\bibfnamefont {P.}~\bibnamefont {Piecuch}}, \bibinfo {author}
  {\bibfnamefont {D.}~\bibnamefont {Poole}}, \bibinfo {author} {\bibfnamefont
  {S.~R.}\ \bibnamefont {Pruitt}}, \bibinfo {author} {\bibfnamefont {A.~P.}\
  \bibnamefont {Rendell}}, \bibinfo {author} {\bibfnamefont {L.~B.}\
  \bibnamefont {Roskop}}, \bibinfo {author} {\bibfnamefont {K.}~\bibnamefont
  {Ruedenberg}}, \bibinfo {author} {\bibfnamefont {T.}~\bibnamefont
  {Sattasathuchana}}, \bibinfo {author} {\bibfnamefont {M.~W.}\ \bibnamefont
  {Schmidt}}, \bibinfo {author} {\bibfnamefont {J.}~\bibnamefont {Shen}},
  \bibinfo {author} {\bibfnamefont {L.}~\bibnamefont {Slipchenko}}, \bibinfo
  {author} {\bibfnamefont {M.}~\bibnamefont {Sosonkina}}, \bibinfo {author}
  {\bibfnamefont {V.}~\bibnamefont {Sundriyal}}, \bibinfo {author}
  {\bibfnamefont {A.}~\bibnamefont {Tiwari}}, \bibinfo {author} {\bibfnamefont
  {J.~L.}\ \bibnamefont {Galvez~Vallejo}}, \bibinfo {author} {\bibfnamefont
  {B.}~\bibnamefont {Westheimer}}, \bibinfo {author} {\bibfnamefont
  {M.}~\bibnamefont {Włoch}}, \bibinfo {author} {\bibfnamefont
  {P.}~\bibnamefont {Xu}}, \bibinfo {author} {\bibfnamefont {F.}~\bibnamefont
  {Zahariev}}, \ and\ \bibinfo {author} {\bibfnamefont {M.~S.}\ \bibnamefont
  {Gordon}},\ }\href@noop {} {\bibfield  {journal} {\bibinfo  {journal} {J.
  Chem. Phys.}\ }\textbf {\bibinfo {volume} {152}},\ \bibinfo {pages} {154102}
  (\bibinfo {year} {2020})}\BibitemShut {NoStop}%
\bibitem [{\citenamefont {Apr\`{a}}\ \emph {et~al.}(2020)\citenamefont
  {Apr\`{a}}, \citenamefont {Bylaska}, \citenamefont {de~Jong}, \citenamefont
  {Govind}, \citenamefont {Kowalski}, \citenamefont {Straatsma}, \citenamefont
  {Valiev}, \citenamefont {van Dam}, \citenamefont {Alexeev}, \citenamefont
  {Anchell}, \citenamefont {Anisimov}, \citenamefont {Aquino}, \citenamefont
  {Atta-Fynn}, \citenamefont {Autschbach}, \citenamefont {Bauman},
  \citenamefont {Becca}, \citenamefont {Bernholdt}, \citenamefont
  {Bhaskaran-Nair}, \citenamefont {Bogatko}, \citenamefont {Borowski},
  \citenamefont {Boschen}, \citenamefont {Brabec}, \citenamefont {Bruner},
  \citenamefont {Cau\"{e}t}, \citenamefont {Chen}, \citenamefont {Chuev},
  \citenamefont {Cramer}, \citenamefont {Daily}, \citenamefont {Deegan},
  \citenamefont {Dunning}, \citenamefont {Dupuis}, \citenamefont {Dyall},
  \citenamefont {Fann}, \citenamefont {Fischer}, \citenamefont {Fonari},
  \citenamefont {Fr\"{u}chtl}, \citenamefont {Gagliardi}, \citenamefont
  {Garza}, \citenamefont {Gawande}, \citenamefont {Ghosh}, \citenamefont
  {Glaesemann}, \citenamefont {G\"{o}tz}, \citenamefont {Hammond},
  \citenamefont {Helms}, \citenamefont {Hermes}, \citenamefont {Hirao},
  \citenamefont {Hirata}, \citenamefont {Jacquelin}, \citenamefont {Jensen},
  \citenamefont {Johnson}, \citenamefont {J\'{o}nsson}, \citenamefont
  {Kendall}, \citenamefont {Klemm}, \citenamefont {Kobayashi}, \citenamefont
  {Konkov}, \citenamefont {Krishnamoorthy}, \citenamefont {Krishnan},
  \citenamefont {Lin}, \citenamefont {Lins}, \citenamefont {Littlefield},
  \citenamefont {Logsdail}, \citenamefont {Lopata}, \citenamefont {Ma},
  \citenamefont {Marenich}, \citenamefont {Martin~del Campo}, \citenamefont
  {Mejia-Rodriguez}, \citenamefont {Moore}, \citenamefont {Mullin},
  \citenamefont {Nakajima}, \citenamefont {Nascimento}, \citenamefont
  {Nichols}, \citenamefont {Nichols}, \citenamefont {Nieplocha}, \citenamefont
  {Otero-de-la Roza}, \citenamefont {Palmer}, \citenamefont {Panyala},
  \citenamefont {Pirojsirikul}, \citenamefont {Peng}, \citenamefont {Peverati},
  \citenamefont {Pittner}, \citenamefont {Pollack}, \citenamefont {Richard},
  \citenamefont {Sadayappan}, \citenamefont {Schatz}, \citenamefont {Shelton},
  \citenamefont {Silverstein}, \citenamefont {Smith}, \citenamefont {Soares},
  \citenamefont {Song}, \citenamefont {Swart}, \citenamefont {Taylor},
  \citenamefont {Thomas}, \citenamefont {Tipparaju}, \citenamefont {Truhlar},
  \citenamefont {Tsemekhman}, \citenamefont {Van~Voorhis}, \citenamefont
  {V\'{a}zquez-Mayagoitia}, \citenamefont {Verma}, \citenamefont {Villa},
  \citenamefont {Vishnu}, \citenamefont {Vogiatzis}, \citenamefont {Wang},
  \citenamefont {Weare}, \citenamefont {Williamson}, \citenamefont {Windus},
  \citenamefont {Woli\'{n}ski}, \citenamefont {Wong}, \citenamefont {Wu},
  \citenamefont {Yang}, \citenamefont {Yu}, \citenamefont {Zacharias},
  \citenamefont {Zhang}, \citenamefont {Zhao},\ and\ \citenamefont
  {Harrison}}]{nwchem20}%
  \BibitemOpen
  \bibfield  {author} {\bibinfo {author} {\bibfnamefont {E.}~\bibnamefont
  {Apr\`{a}}}, \bibinfo {author} {\bibfnamefont {E.~J.}\ \bibnamefont
  {Bylaska}}, \bibinfo {author} {\bibfnamefont {W.~A.}\ \bibnamefont
  {de~Jong}}, \bibinfo {author} {\bibfnamefont {N.}~\bibnamefont {Govind}},
  \bibinfo {author} {\bibfnamefont {K.}~\bibnamefont {Kowalski}}, \bibinfo
  {author} {\bibfnamefont {T.~P.}\ \bibnamefont {Straatsma}}, \bibinfo {author}
  {\bibfnamefont {M.}~\bibnamefont {Valiev}}, \bibinfo {author} {\bibfnamefont
  {H.~J.~J.}\ \bibnamefont {van Dam}}, \bibinfo {author} {\bibfnamefont
  {Y.}~\bibnamefont {Alexeev}}, \bibinfo {author} {\bibfnamefont
  {J.}~\bibnamefont {Anchell}}, \bibinfo {author} {\bibfnamefont
  {V.}~\bibnamefont {Anisimov}}, \bibinfo {author} {\bibfnamefont {F.~W.}\
  \bibnamefont {Aquino}}, \bibinfo {author} {\bibfnamefont {R.}~\bibnamefont
  {Atta-Fynn}}, \bibinfo {author} {\bibfnamefont {J.}~\bibnamefont
  {Autschbach}}, \bibinfo {author} {\bibfnamefont {N.~P.}\ \bibnamefont
  {Bauman}}, \bibinfo {author} {\bibfnamefont {J.~C.}\ \bibnamefont {Becca}},
  \bibinfo {author} {\bibfnamefont {D.~E.}\ \bibnamefont {Bernholdt}}, \bibinfo
  {author} {\bibfnamefont {K.}~\bibnamefont {Bhaskaran-Nair}}, \bibinfo
  {author} {\bibfnamefont {S.}~\bibnamefont {Bogatko}}, \bibinfo {author}
  {\bibfnamefont {P.}~\bibnamefont {Borowski}}, \bibinfo {author}
  {\bibfnamefont {J.}~\bibnamefont {Boschen}}, \bibinfo {author} {\bibfnamefont
  {J.}~\bibnamefont {Brabec}}, \bibinfo {author} {\bibfnamefont
  {A.}~\bibnamefont {Bruner}}, \bibinfo {author} {\bibfnamefont
  {E.}~\bibnamefont {Cau\"{e}t}}, \bibinfo {author} {\bibfnamefont
  {Y.}~\bibnamefont {Chen}}, \bibinfo {author} {\bibfnamefont {G.~N.}\
  \bibnamefont {Chuev}}, \bibinfo {author} {\bibfnamefont {C.~J.}\ \bibnamefont
  {Cramer}}, \bibinfo {author} {\bibfnamefont {J.}~\bibnamefont {Daily}},
  \bibinfo {author} {\bibfnamefont {M.~J.~O.}\ \bibnamefont {Deegan}}, \bibinfo
  {author} {\bibfnamefont {T.~H.}\ \bibnamefont {Dunning}}, \bibinfo {author}
  {\bibfnamefont {M.}~\bibnamefont {Dupuis}}, \bibinfo {author} {\bibfnamefont
  {K.~G.}\ \bibnamefont {Dyall}}, \bibinfo {author} {\bibfnamefont {G.~I.}\
  \bibnamefont {Fann}}, \bibinfo {author} {\bibfnamefont {S.~A.}\ \bibnamefont
  {Fischer}}, \bibinfo {author} {\bibfnamefont {A.}~\bibnamefont {Fonari}},
  \bibinfo {author} {\bibfnamefont {H.}~\bibnamefont {Fr\"{u}chtl}}, \bibinfo
  {author} {\bibfnamefont {L.}~\bibnamefont {Gagliardi}}, \bibinfo {author}
  {\bibfnamefont {J.}~\bibnamefont {Garza}}, \bibinfo {author} {\bibfnamefont
  {N.}~\bibnamefont {Gawande}}, \bibinfo {author} {\bibfnamefont
  {S.}~\bibnamefont {Ghosh}}, \bibinfo {author} {\bibfnamefont
  {K.}~\bibnamefont {Glaesemann}}, \bibinfo {author} {\bibfnamefont {A.~W.}\
  \bibnamefont {G\"{o}tz}}, \bibinfo {author} {\bibfnamefont {J.}~\bibnamefont
  {Hammond}}, \bibinfo {author} {\bibfnamefont {V.}~\bibnamefont {Helms}},
  \bibinfo {author} {\bibfnamefont {E.~D.}\ \bibnamefont {Hermes}}, \bibinfo
  {author} {\bibfnamefont {K.}~\bibnamefont {Hirao}}, \bibinfo {author}
  {\bibfnamefont {S.}~\bibnamefont {Hirata}}, \bibinfo {author} {\bibfnamefont
  {M.}~\bibnamefont {Jacquelin}}, \bibinfo {author} {\bibfnamefont
  {L.}~\bibnamefont {Jensen}}, \bibinfo {author} {\bibfnamefont {B.~G.}\
  \bibnamefont {Johnson}}, \bibinfo {author} {\bibfnamefont {H.}~\bibnamefont
  {J\'{o}nsson}}, \bibinfo {author} {\bibfnamefont {R.~A.}\ \bibnamefont
  {Kendall}}, \bibinfo {author} {\bibfnamefont {M.}~\bibnamefont {Klemm}},
  \bibinfo {author} {\bibfnamefont {R.}~\bibnamefont {Kobayashi}}, \bibinfo
  {author} {\bibfnamefont {V.}~\bibnamefont {Konkov}}, \bibinfo {author}
  {\bibfnamefont {S.}~\bibnamefont {Krishnamoorthy}}, \bibinfo {author}
  {\bibfnamefont {M.}~\bibnamefont {Krishnan}}, \bibinfo {author}
  {\bibfnamefont {Z.}~\bibnamefont {Lin}}, \bibinfo {author} {\bibfnamefont
  {R.~D.}\ \bibnamefont {Lins}}, \bibinfo {author} {\bibfnamefont {R.~J.}\
  \bibnamefont {Littlefield}}, \bibinfo {author} {\bibfnamefont {A.~J.}\
  \bibnamefont {Logsdail}}, \bibinfo {author} {\bibfnamefont {K.}~\bibnamefont
  {Lopata}}, \bibinfo {author} {\bibfnamefont {W.}~\bibnamefont {Ma}}, \bibinfo
  {author} {\bibfnamefont {A.~V.}\ \bibnamefont {Marenich}}, \bibinfo {author}
  {\bibfnamefont {J.}~\bibnamefont {Martin~del Campo}}, \bibinfo {author}
  {\bibfnamefont {D.}~\bibnamefont {Mejia-Rodriguez}}, \bibinfo {author}
  {\bibfnamefont {J.~E.}\ \bibnamefont {Moore}}, \bibinfo {author}
  {\bibfnamefont {J.~M.}\ \bibnamefont {Mullin}}, \bibinfo {author}
  {\bibfnamefont {T.}~\bibnamefont {Nakajima}}, \bibinfo {author}
  {\bibfnamefont {D.~R.}\ \bibnamefont {Nascimento}}, \bibinfo {author}
  {\bibfnamefont {J.~A.}\ \bibnamefont {Nichols}}, \bibinfo {author}
  {\bibfnamefont {P.~J.}\ \bibnamefont {Nichols}}, \bibinfo {author}
  {\bibfnamefont {J.}~\bibnamefont {Nieplocha}}, \bibinfo {author}
  {\bibfnamefont {A.}~\bibnamefont {Otero-de-la Roza}}, \bibinfo {author}
  {\bibfnamefont {B.}~\bibnamefont {Palmer}}, \bibinfo {author} {\bibfnamefont
  {A.}~\bibnamefont {Panyala}}, \bibinfo {author} {\bibfnamefont
  {T.}~\bibnamefont {Pirojsirikul}}, \bibinfo {author} {\bibfnamefont
  {B.}~\bibnamefont {Peng}}, \bibinfo {author} {\bibfnamefont {R.}~\bibnamefont
  {Peverati}}, \bibinfo {author} {\bibfnamefont {J.}~\bibnamefont {Pittner}},
  \bibinfo {author} {\bibfnamefont {L.}~\bibnamefont {Pollack}}, \bibinfo
  {author} {\bibfnamefont {R.~M.}\ \bibnamefont {Richard}}, \bibinfo {author}
  {\bibfnamefont {P.}~\bibnamefont {Sadayappan}}, \bibinfo {author}
  {\bibfnamefont {G.~C.}\ \bibnamefont {Schatz}}, \bibinfo {author}
  {\bibfnamefont {W.~A.}\ \bibnamefont {Shelton}}, \bibinfo {author}
  {\bibfnamefont {D.~W.}\ \bibnamefont {Silverstein}}, \bibinfo {author}
  {\bibfnamefont {D.~M.~A.}\ \bibnamefont {Smith}}, \bibinfo {author}
  {\bibfnamefont {T.~A.}\ \bibnamefont {Soares}}, \bibinfo {author}
  {\bibfnamefont {D.}~\bibnamefont {Song}}, \bibinfo {author} {\bibfnamefont
  {M.}~\bibnamefont {Swart}}, \bibinfo {author} {\bibfnamefont {H.~L.}\
  \bibnamefont {Taylor}}, \bibinfo {author} {\bibfnamefont {G.~S.}\
  \bibnamefont {Thomas}}, \bibinfo {author} {\bibfnamefont {V.}~\bibnamefont
  {Tipparaju}}, \bibinfo {author} {\bibfnamefont {D.~G.}\ \bibnamefont
  {Truhlar}}, \bibinfo {author} {\bibfnamefont {K.}~\bibnamefont {Tsemekhman}},
  \bibinfo {author} {\bibfnamefont {T.}~\bibnamefont {Van~Voorhis}}, \bibinfo
  {author} {\bibfnamefont {A.}~\bibnamefont {V\'{a}zquez-Mayagoitia}}, \bibinfo
  {author} {\bibfnamefont {P.}~\bibnamefont {Verma}}, \bibinfo {author}
  {\bibfnamefont {O.}~\bibnamefont {Villa}}, \bibinfo {author} {\bibfnamefont
  {A.}~\bibnamefont {Vishnu}}, \bibinfo {author} {\bibfnamefont {K.~D.}\
  \bibnamefont {Vogiatzis}}, \bibinfo {author} {\bibfnamefont {D.}~\bibnamefont
  {Wang}}, \bibinfo {author} {\bibfnamefont {J.~H.}\ \bibnamefont {Weare}},
  \bibinfo {author} {\bibfnamefont {M.~J.}\ \bibnamefont {Williamson}},
  \bibinfo {author} {\bibfnamefont {T.~L.}\ \bibnamefont {Windus}}, \bibinfo
  {author} {\bibfnamefont {K.}~\bibnamefont {Woli\'{n}ski}}, \bibinfo {author}
  {\bibfnamefont {A.~T.}\ \bibnamefont {Wong}}, \bibinfo {author}
  {\bibfnamefont {Q.}~\bibnamefont {Wu}}, \bibinfo {author} {\bibfnamefont
  {C.}~\bibnamefont {Yang}}, \bibinfo {author} {\bibfnamefont {Q.}~\bibnamefont
  {Yu}}, \bibinfo {author} {\bibfnamefont {M.}~\bibnamefont {Zacharias}},
  \bibinfo {author} {\bibfnamefont {Z.}~\bibnamefont {Zhang}}, \bibinfo
  {author} {\bibfnamefont {Y.}~\bibnamefont {Zhao}}, \ and\ \bibinfo {author}
  {\bibfnamefont {R.~J.}\ \bibnamefont {Harrison}},\ }\href@noop {} {\bibfield
  {journal} {\bibinfo  {journal} {J. Chem. Phys.}\ }\textbf {\bibinfo {volume}
  {152}},\ \bibinfo {pages} {184102} (\bibinfo {year} {2020})}\BibitemShut
  {NoStop}%
\bibitem [{\citenamefont {Katouda}\ and\ \citenamefont
  {Nagase}(2009)}]{katouda09}%
  \BibitemOpen
  \bibfield  {author} {\bibinfo {author} {\bibfnamefont {M.}~\bibnamefont
  {Katouda}}\ and\ \bibinfo {author} {\bibfnamefont {S.}~\bibnamefont
  {Nagase}},\ }\href
  {https://onlinelibrary.wiley.com/doixx/abs/10.1002/qua.22068} {\bibfield
  {journal} {\bibinfo  {journal} {Int. J. Quantum Chem.}\ }\textbf {\bibinfo
  {volume} {109}},\ \bibinfo {pages} {2121} (\bibinfo {year}
  {2009})}\BibitemShut {NoStop}%
\bibitem [{\citenamefont {Epifanovsky}\ \emph {et~al.}(2013)\citenamefont
  {Epifanovsky}, \citenamefont {Zuev}, \citenamefont {Feng}, \citenamefont
  {Khistyaev}, \citenamefont {Shao},\ and\ \citenamefont
  {Krylov}}]{epifanovsky13}%
  \BibitemOpen
  \bibfield  {author} {\bibinfo {author} {\bibfnamefont {E.}~\bibnamefont
  {Epifanovsky}}, \bibinfo {author} {\bibfnamefont {D.}~\bibnamefont {Zuev}},
  \bibinfo {author} {\bibfnamefont {X.}~\bibnamefont {Feng}}, \bibinfo {author}
  {\bibfnamefont {K.}~\bibnamefont {Khistyaev}}, \bibinfo {author}
  {\bibfnamefont {Y.}~\bibnamefont {Shao}}, \ and\ \bibinfo {author}
  {\bibfnamefont {A.~I.}\ \bibnamefont {Krylov}},\ }\href@noop {} {\bibfield
  {journal} {\bibinfo  {journal} {J. Chem. Phys.}\ }\textbf {\bibinfo {volume}
  {139}},\ \bibinfo {pages} {134105} (\bibinfo {year} {2013})}\BibitemShut
  {NoStop}%
\bibitem [{\citenamefont {DePrince}\ and\ \citenamefont
  {Sherrill}(2013)}]{deprince13}%
  \BibitemOpen
  \bibfield  {author} {\bibinfo {author} {\bibfnamefont {A.~E.}\ \bibnamefont
  {DePrince}}\ and\ \bibinfo {author} {\bibfnamefont {C.~D.}\ \bibnamefont
  {Sherrill}},\ }\href@noop {} {\bibfield  {journal} {\bibinfo  {journal} {J.
  Chem. Theory Comp.}\ }\textbf {\bibinfo {volume} {9}},\ \bibinfo {pages}
  {2687} (\bibinfo {year} {2013})}\BibitemShut {NoStop}%
\bibitem [{\citenamefont {Lesiuk}(2020{\natexlab{b}})}]{lesiuk20b}%
  \BibitemOpen
  \bibfield  {author} {\bibinfo {author} {\bibfnamefont {M.}~\bibnamefont
  {Lesiuk}},\ }\href@noop {} {\bibfield  {journal} {\bibinfo  {journal} {J.
  Chem. Phys.}\ }\textbf {\bibinfo {volume} {152}},\ \bibinfo {pages} {044104}
  (\bibinfo {year} {2020}{\natexlab{b}})}\BibitemShut {NoStop}%
\bibitem [{\citenamefont {Golub}\ and\ \citenamefont {Kahan}(1965)}]{golub65}%
  \BibitemOpen
  \bibfield  {author} {\bibinfo {author} {\bibfnamefont {G.}~\bibnamefont
  {Golub}}\ and\ \bibinfo {author} {\bibfnamefont {W.}~\bibnamefont {Kahan}},\
  }\href@noop {} {\bibfield  {journal} {\bibinfo  {journal} {SIAM J. Numer.
  Anal.}\ }\textbf {\bibinfo {volume} {2}},\ \bibinfo {pages} {205} (\bibinfo
  {year} {1965})}\BibitemShut {NoStop}%
\bibitem [{\citenamefont {Simon}\ and\ \citenamefont {Zha}(2000)}]{simon00}%
  \BibitemOpen
  \bibfield  {author} {\bibinfo {author} {\bibfnamefont {H.}~\bibnamefont
  {Simon}}\ and\ \bibinfo {author} {\bibfnamefont {H.}~\bibnamefont {Zha}},\
  }\href@noop {} {\bibfield  {journal} {\bibinfo  {journal} {SIAM J. Sci.
  Comput.}\ }\textbf {\bibinfo {volume} {21}},\ \bibinfo {pages} {2257}
  (\bibinfo {year} {2000})}\BibitemShut {NoStop}%
\bibitem [{\citenamefont {Baglama}\ and\ \citenamefont
  {Reichel}(2005)}]{baglama05}%
  \BibitemOpen
  \bibfield  {author} {\bibinfo {author} {\bibfnamefont {J.}~\bibnamefont
  {Baglama}}\ and\ \bibinfo {author} {\bibfnamefont {L.}~\bibnamefont
  {Reichel}},\ }\href@noop {} {\bibfield  {journal} {\bibinfo  {journal} {SIAM
  J. Sci. Comput.}\ }\textbf {\bibinfo {volume} {27}},\ \bibinfo {pages} {19}
  (\bibinfo {year} {2005})}\BibitemShut {NoStop}%
\bibitem [{\citenamefont {Davidson}(1975)}]{davidson75}%
  \BibitemOpen
  \bibfield  {author} {\bibinfo {author} {\bibfnamefont {E.~R.}\ \bibnamefont
  {Davidson}},\ }\href
  {https://www.sciencedirect.com/science/article/pii/0021999175900650}
  {\bibfield  {journal} {\bibinfo  {journal} {J. Comp. Phys.}\ }\textbf
  {\bibinfo {volume} {17}},\ \bibinfo {pages} {87} (\bibinfo {year}
  {1975})}\BibitemShut {NoStop}%
\bibitem [{\citenamefont {Paldus}\ and\ \citenamefont
  {Jeziorski}(1988)}]{paldus88}%
  \BibitemOpen
  \bibfield  {author} {\bibinfo {author} {\bibfnamefont {J.}~\bibnamefont
  {Paldus}}\ and\ \bibinfo {author} {\bibfnamefont {B.}~\bibnamefont
  {Jeziorski}},\ }\href@noop {} {\bibfield  {journal} {\bibinfo  {journal}
  {Theor. Chem. Acc.}\ }\textbf {\bibinfo {volume} {73}},\ \bibinfo {pages}
  {81} (\bibinfo {year} {1988})}\BibitemShut {NoStop}%
\bibitem [{\citenamefont {Pulay}(1980)}]{pulay80}%
  \BibitemOpen
  \bibfield  {author} {\bibinfo {author} {\bibfnamefont {P.}~\bibnamefont
  {Pulay}},\ }\href
  {http://www.sciencedirect.com/science/article/pii/0009261480803964}
  {\bibfield  {journal} {\bibinfo  {journal} {Chem. Phys. Lett.}\ }\textbf
  {\bibinfo {volume} {73}},\ \bibinfo {pages} {393 } (\bibinfo {year}
  {1980})}\BibitemShut {NoStop}%
\bibitem [{\citenamefont {Scuseria}, \citenamefont {Lee},\ and\ \citenamefont
  {Schaefer}(1986)}]{scuseria86}%
  \BibitemOpen
  \bibfield  {author} {\bibinfo {author} {\bibfnamefont {G.~E.}\ \bibnamefont
  {Scuseria}}, \bibinfo {author} {\bibfnamefont {T.~J.}\ \bibnamefont {Lee}}, \
  and\ \bibinfo {author} {\bibfnamefont {H.~F.}\ \bibnamefont {Schaefer}},\
  }\href {http://www.sciencedirect.com/science/article/pii/0009261486804614}
  {\bibfield  {journal} {\bibinfo  {journal} {Chem. Phys. Lett.}\ }\textbf
  {\bibinfo {volume} {130}},\ \bibinfo {pages} {236 } (\bibinfo {year}
  {1986})}\BibitemShut {NoStop}%
\bibitem [{\citenamefont {Purvis}\ and\ \citenamefont
  {Bartlett}(1981)}]{purvis81}%
  \BibitemOpen
  \bibfield  {author} {\bibinfo {author} {\bibfnamefont {G.~D.}\ \bibnamefont
  {Purvis}}\ and\ \bibinfo {author} {\bibfnamefont {R.~J.}\ \bibnamefont
  {Bartlett}},\ }\href@noop {} {\bibfield  {journal} {\bibinfo  {journal} {J.
  Chem. Phys.}\ }\textbf {\bibinfo {volume} {75}},\ \bibinfo {pages} {1284}
  (\bibinfo {year} {1981})}\BibitemShut {NoStop}%
\bibitem [{\citenamefont {Zi\'{o}{\l}kowski}\ \emph {et~al.}(2008)\citenamefont
  {Zi\'{o}{\l}kowski}, \citenamefont {Weijo}, \citenamefont {J\o{}rgensen},\
  and\ \citenamefont {Olsen}}]{ziolo08}%
  \BibitemOpen
  \bibfield  {author} {\bibinfo {author} {\bibfnamefont {M.}~\bibnamefont
  {Zi\'{o}{\l}kowski}}, \bibinfo {author} {\bibfnamefont {V.}~\bibnamefont
  {Weijo}}, \bibinfo {author} {\bibfnamefont {P.}~\bibnamefont {J\o{}rgensen}},
  \ and\ \bibinfo {author} {\bibfnamefont {J.}~\bibnamefont {Olsen}},\
  }\href@noop {} {\bibfield  {journal} {\bibinfo  {journal} {J. Chem. Phys.}\
  }\textbf {\bibinfo {volume} {128}},\ \bibinfo {pages} {204105} (\bibinfo
  {year} {2008})}\BibitemShut {NoStop}%
\bibitem [{\citenamefont {Ettenhuber}\ and\ \citenamefont
  {J\o{}rgensen}(2015)}]{ettenhuber15}%
  \BibitemOpen
  \bibfield  {author} {\bibinfo {author} {\bibfnamefont {P.}~\bibnamefont
  {Ettenhuber}}\ and\ \bibinfo {author} {\bibfnamefont {P.}~\bibnamefont
  {J\o{}rgensen}},\ }\href@noop {} {\bibfield  {journal} {\bibinfo  {journal}
  {J. Chem. Theory Comput.}\ }\textbf {\bibinfo {volume} {11}},\ \bibinfo
  {pages} {1518} (\bibinfo {year} {2015})}\BibitemShut {NoStop}%
\bibitem [{\citenamefont {Alml{\"{o}}f}(1991)}]{almlof91}%
  \BibitemOpen
  \bibfield  {author} {\bibinfo {author} {\bibfnamefont {J.}~\bibnamefont
  {Alml{\"{o}}f}},\ }\href
  {http://www.sciencedirect.com/science/article/pii/000926149180078C}
  {\bibfield  {journal} {\bibinfo  {journal} {Chem. Phys. Lett.}\ }\textbf
  {\bibinfo {volume} {181}},\ \bibinfo {pages} {319 } (\bibinfo {year}
  {1991})}\BibitemShut {NoStop}%
\bibitem [{\citenamefont {H\"{a}ser}\ and\ \citenamefont
  {Alml\"{o}f}(1992)}]{haser92}%
  \BibitemOpen
  \bibfield  {author} {\bibinfo {author} {\bibfnamefont {M.}~\bibnamefont
  {H\"{a}ser}}\ and\ \bibinfo {author} {\bibfnamefont {J.}~\bibnamefont
  {Alml\"{o}f}},\ }\href@noop {} {\bibfield  {journal} {\bibinfo  {journal} {J.
  Chem. Phys.}\ }\textbf {\bibinfo {volume} {96}},\ \bibinfo {pages} {489}
  (\bibinfo {year} {1992})}\BibitemShut {NoStop}%
\bibitem [{\citenamefont {Ayala}\ and\ \citenamefont
  {Scuseria}(1999)}]{ayala99}%
  \BibitemOpen
  \bibfield  {author} {\bibinfo {author} {\bibfnamefont {P.~Y.}\ \bibnamefont
  {Ayala}}\ and\ \bibinfo {author} {\bibfnamefont {G.~E.}\ \bibnamefont
  {Scuseria}},\ }\href@noop {} {\bibfield  {journal} {\bibinfo  {journal} {J.
  Chem. Phys.}\ }\textbf {\bibinfo {volume} {110}},\ \bibinfo {pages} {3660}
  (\bibinfo {year} {1999})}\BibitemShut {NoStop}%
\bibitem [{\citenamefont {Lambrecht}, \citenamefont {Doser},\ and\
  \citenamefont {Ochsenfeld}(2005)}]{lambert05}%
  \BibitemOpen
  \bibfield  {author} {\bibinfo {author} {\bibfnamefont {D.~S.}\ \bibnamefont
  {Lambrecht}}, \bibinfo {author} {\bibfnamefont {B.}~\bibnamefont {Doser}}, \
  and\ \bibinfo {author} {\bibfnamefont {C.}~\bibnamefont {Ochsenfeld}},\
  }\href@noop {} {\bibfield  {journal} {\bibinfo  {journal} {J. Chem. Phys.}\
  }\textbf {\bibinfo {volume} {123}},\ \bibinfo {pages} {184102} (\bibinfo
  {year} {2005})}\BibitemShut {NoStop}%
\bibitem [{\citenamefont {Nakajima}\ and\ \citenamefont
  {Hirao}(2006)}]{nakajima06}%
  \BibitemOpen
  \bibfield  {author} {\bibinfo {author} {\bibfnamefont {T.}~\bibnamefont
  {Nakajima}}\ and\ \bibinfo {author} {\bibfnamefont {K.}~\bibnamefont
  {Hirao}},\ }\href
  {http://www.sciencedirect.com/science/article/pii/S0009261406009122}
  {\bibfield  {journal} {\bibinfo  {journal} {Chem. Phys. Lett.}\ }\textbf
  {\bibinfo {volume} {427}},\ \bibinfo {pages} {225 } (\bibinfo {year}
  {2006})}\BibitemShut {NoStop}%
\bibitem [{\citenamefont {Jung}\ \emph {et~al.}(2004)\citenamefont {Jung},
  \citenamefont {Lochan}, \citenamefont {Dutoi},\ and\ \citenamefont
  {Head-Gordon}}]{jung04}%
  \BibitemOpen
  \bibfield  {author} {\bibinfo {author} {\bibfnamefont {Y.}~\bibnamefont
  {Jung}}, \bibinfo {author} {\bibfnamefont {R.~C.}\ \bibnamefont {Lochan}},
  \bibinfo {author} {\bibfnamefont {A.~D.}\ \bibnamefont {Dutoi}}, \ and\
  \bibinfo {author} {\bibfnamefont {M.}~\bibnamefont {Head-Gordon}},\
  }\href@noop {} {\bibfield  {journal} {\bibinfo  {journal} {J. Chem. Phys.}\
  }\textbf {\bibinfo {volume} {121}},\ \bibinfo {pages} {9793} (\bibinfo {year}
  {2004})}\BibitemShut {NoStop}%
\bibitem [{\citenamefont {Kats}, \citenamefont {Usvyat},\ and\ \citenamefont
  {Sch\"{u}tz}(2008)}]{kats08}%
  \BibitemOpen
  \bibfield  {author} {\bibinfo {author} {\bibfnamefont {D.}~\bibnamefont
  {Kats}}, \bibinfo {author} {\bibfnamefont {D.}~\bibnamefont {Usvyat}}, \ and\
  \bibinfo {author} {\bibfnamefont {M.}~\bibnamefont {Sch\"{u}tz}},\ }\href
  {http://dx.doix.org/10.1039/B802993H} {\bibfield  {journal} {\bibinfo
  {journal} {Phys. Chem. Chem. Phys.}\ }\textbf {\bibinfo {volume} {10}},\
  \bibinfo {pages} {3430} (\bibinfo {year} {2008})}\BibitemShut {NoStop}%
\bibitem [{\citenamefont {Takatsuka}, \citenamefont {Ten-no},\ and\
  \citenamefont {Hackbusch}(2008)}]{takatsuka08}%
  \BibitemOpen
  \bibfield  {author} {\bibinfo {author} {\bibfnamefont {A.}~\bibnamefont
  {Takatsuka}}, \bibinfo {author} {\bibfnamefont {S.}~\bibnamefont {Ten-no}}, \
  and\ \bibinfo {author} {\bibfnamefont {W.}~\bibnamefont {Hackbusch}},\
  }\href@noop {} {\bibfield  {journal} {\bibinfo  {journal} {J. Chem. Phys.}\
  }\textbf {\bibinfo {volume} {129}},\ \bibinfo {pages} {044112} (\bibinfo
  {year} {2008})}\BibitemShut {NoStop}%
\bibitem [{\citenamefont {Braess}\ and\ \citenamefont
  {Hackbusch}(2005)}]{braess05}%
  \BibitemOpen
  \bibfield  {author} {\bibinfo {author} {\bibfnamefont {D.}~\bibnamefont
  {Braess}}\ and\ \bibinfo {author} {\bibfnamefont {W.}~\bibnamefont
  {Hackbusch}},\ }\href {https://doix.org/10.1093/imanum/dri015} {\bibfield
  {journal} {\bibinfo  {journal} {IMA J. Numer. Anal.}\ }\textbf {\bibinfo
  {volume} {25}},\ \bibinfo {pages} {685} (\bibinfo {year} {2005})}\BibitemShut
  {NoStop}%
\bibitem [{\citenamefont {Helmich-Paris}\ and\ \citenamefont
  {Visscher}(2016)}]{paris16}%
  \BibitemOpen
  \bibfield  {author} {\bibinfo {author} {\bibfnamefont {B.}~\bibnamefont
  {Helmich-Paris}}\ and\ \bibinfo {author} {\bibfnamefont {L.}~\bibnamefont
  {Visscher}},\ }\href
  {http://www.sciencedirect.com/science/article/pii/S0021999116302364}
  {\bibfield  {journal} {\bibinfo  {journal} {J. Comp. Phys.}\ }\textbf
  {\bibinfo {volume} {321}},\ \bibinfo {pages} {927 } (\bibinfo {year}
  {2016})}\BibitemShut {NoStop}%
\bibitem [{\citenamefont {Dunning}(1989)}]{dunning89}%
  \BibitemOpen
  \bibfield  {author} {\bibinfo {author} {\bibfnamefont {T.~H.}\ \bibnamefont
  {Dunning}},\ }\href@noop {} {\bibfield  {journal} {\bibinfo  {journal} {J.
  Chem. Phys.}\ }\textbf {\bibinfo {volume} {90}},\ \bibinfo {pages} {1007}
  (\bibinfo {year} {1989})}\BibitemShut {NoStop}%
\bibitem [{\citenamefont {Weigend}, \citenamefont {Köhn},\ and\ \citenamefont
  {Hättig}(2002)}]{weigend02b}%
  \BibitemOpen
  \bibfield  {author} {\bibinfo {author} {\bibfnamefont {F.}~\bibnamefont
  {Weigend}}, \bibinfo {author} {\bibfnamefont {A.}~\bibnamefont {Köhn}}, \
  and\ \bibinfo {author} {\bibfnamefont {C.}~\bibnamefont {Hättig}},\
  }\href@noop {} {\bibfield  {journal} {\bibinfo  {journal} {J. Chem. Phys.}\
  }\textbf {\bibinfo {volume} {116}},\ \bibinfo {pages} {3175} (\bibinfo {year}
  {2002})}\BibitemShut {NoStop}%
\bibitem [{\citenamefont {Adler}\ and\ \citenamefont {Werner}(2011)}]{adler11}%
  \BibitemOpen
  \bibfield  {author} {\bibinfo {author} {\bibfnamefont {T.~B.}\ \bibnamefont
  {Adler}}\ and\ \bibinfo {author} {\bibfnamefont {H.-J.}\ \bibnamefont
  {Werner}},\ }\href@noop {} {\bibfield  {journal} {\bibinfo  {journal} {J.
  Chem. Phys.}\ }\textbf {\bibinfo {volume} {135}},\ \bibinfo {pages} {144117}
  (\bibinfo {year} {2011})}\BibitemShut {NoStop}%
\bibitem [{\citenamefont {Kendall}, \citenamefont {Dunning},\ and\
  \citenamefont {Harrison}(1992)}]{kendall92}%
  \BibitemOpen
  \bibfield  {author} {\bibinfo {author} {\bibfnamefont {R.~A.}\ \bibnamefont
  {Kendall}}, \bibinfo {author} {\bibfnamefont {T.~H.}\ \bibnamefont
  {Dunning}}, \ and\ \bibinfo {author} {\bibfnamefont {R.~J.}\ \bibnamefont
  {Harrison}},\ }\href {\doibase 10.1063/1.462569} {\bibfield  {journal}
  {\bibinfo  {journal} {J. Chem. Phys.}\ }\textbf {\bibinfo {volume} {96}},\
  \bibinfo {pages} {6796} (\bibinfo {year} {1992})},\ \Eprint
  {http://arxiv.org/abs/https://doi.org/10.1063/1.462569}
  {https://doi.org/10.1063/1.462569} \BibitemShut {NoStop}%
\bibitem [{\citenamefont {Bartlett}\ \emph {et~al.}(1990)\citenamefont
  {Bartlett}, \citenamefont {Watts}, \citenamefont {Kucharski},\ and\
  \citenamefont {Noga}}]{bartlett90}%
  \BibitemOpen
  \bibfield  {author} {\bibinfo {author} {\bibfnamefont {R.~J.}\ \bibnamefont
  {Bartlett}}, \bibinfo {author} {\bibfnamefont {J.}~\bibnamefont {Watts}},
  \bibinfo {author} {\bibfnamefont {S.}~\bibnamefont {Kucharski}}, \ and\
  \bibinfo {author} {\bibfnamefont {J.}~\bibnamefont {Noga}},\ }\href
  {https://www.sciencedirect.com/science/article/pii/000926149087031L}
  {\bibfield  {journal} {\bibinfo  {journal} {Chem. Phys. Lett.}\ }\textbf
  {\bibinfo {volume} {165}},\ \bibinfo {pages} {513} (\bibinfo {year}
  {1990})}\BibitemShut {NoStop}%
\bibitem [{\citenamefont {Hopkins}\ and\ \citenamefont
  {Tschumper}(2004)}]{hopkins04}%
  \BibitemOpen
  \bibfield  {author} {\bibinfo {author} {\bibfnamefont {B.~W.}\ \bibnamefont
  {Hopkins}}\ and\ \bibinfo {author} {\bibfnamefont {G.~S.}\ \bibnamefont
  {Tschumper}},\ }\href@noop {} {\bibfield  {journal} {\bibinfo  {journal} {J.
  Phys. Chem. A}\ }\textbf {\bibinfo {volume} {108}},\ \bibinfo {pages} {2941}
  (\bibinfo {year} {2004})}\BibitemShut {NoStop}%
\bibitem [{\citenamefont {Bak}\ \emph {et~al.}(2000)\citenamefont {Bak},
  \citenamefont {J\o{}rgensen}, \citenamefont {Olsen}, \citenamefont
  {Helgaker},\ and\ \citenamefont {Klopper}}]{bak00}%
  \BibitemOpen
  \bibfield  {author} {\bibinfo {author} {\bibfnamefont {K.~L.}\ \bibnamefont
  {Bak}}, \bibinfo {author} {\bibfnamefont {P.}~\bibnamefont {J\o{}rgensen}},
  \bibinfo {author} {\bibfnamefont {J.}~\bibnamefont {Olsen}}, \bibinfo
  {author} {\bibfnamefont {T.}~\bibnamefont {Helgaker}}, \ and\ \bibinfo
  {author} {\bibfnamefont {W.}~\bibnamefont {Klopper}},\ }\href@noop {}
  {\bibfield  {journal} {\bibinfo  {journal} {J. Chem. Phys.}\ }\textbf
  {\bibinfo {volume} {112}},\ \bibinfo {pages} {9229} (\bibinfo {year}
  {2000})}\BibitemShut {NoStop}%
\bibitem [{\citenamefont {Tajti}\ \emph {et~al.}(2004)\citenamefont {Tajti},
  \citenamefont {Szalay}, \citenamefont {Cs\'{a}sz\'{a}r}, \citenamefont
  {K\'{a}llay}, \citenamefont {Gauss}, \citenamefont {Valeev}, \citenamefont
  {Flowers}, \citenamefont {V\'{a}zquez},\ and\ \citenamefont
  {Stanton}}]{tajti04}%
  \BibitemOpen
  \bibfield  {author} {\bibinfo {author} {\bibfnamefont {A.}~\bibnamefont
  {Tajti}}, \bibinfo {author} {\bibfnamefont {P.~G.}\ \bibnamefont {Szalay}},
  \bibinfo {author} {\bibfnamefont {A.~G.}\ \bibnamefont {Cs\'{a}sz\'{a}r}},
  \bibinfo {author} {\bibfnamefont {M.}~\bibnamefont {K\'{a}llay}}, \bibinfo
  {author} {\bibfnamefont {J.}~\bibnamefont {Gauss}}, \bibinfo {author}
  {\bibfnamefont {E.~F.}\ \bibnamefont {Valeev}}, \bibinfo {author}
  {\bibfnamefont {B.~A.}\ \bibnamefont {Flowers}}, \bibinfo {author}
  {\bibfnamefont {J.}~\bibnamefont {V\'{a}zquez}}, \ and\ \bibinfo {author}
  {\bibfnamefont {J.~F.}\ \bibnamefont {Stanton}},\ }\href@noop {} {\bibfield
  {journal} {\bibinfo  {journal} {J. Chem. Phys.}\ }\textbf {\bibinfo {volume}
  {121}},\ \bibinfo {pages} {11599} (\bibinfo {year} {2004})}\BibitemShut
  {NoStop}%
\bibitem [{\citenamefont {Karton}\ \emph {et~al.}(2006)\citenamefont {Karton},
  \citenamefont {Rabinovich}, \citenamefont {Martin},\ and\ \citenamefont
  {Ruscic}}]{karton06}%
  \BibitemOpen
  \bibfield  {author} {\bibinfo {author} {\bibfnamefont {A.}~\bibnamefont
  {Karton}}, \bibinfo {author} {\bibfnamefont {E.}~\bibnamefont {Rabinovich}},
  \bibinfo {author} {\bibfnamefont {J.~M.~L.}\ \bibnamefont {Martin}}, \ and\
  \bibinfo {author} {\bibfnamefont {B.}~\bibnamefont {Ruscic}},\ }\href@noop {}
  {\bibfield  {journal} {\bibinfo  {journal} {J. Chem. Phys.}\ }\textbf
  {\bibinfo {volume} {125}},\ \bibinfo {pages} {144108} (\bibinfo {year}
  {2006})}\BibitemShut {NoStop}%
\bibitem [{\citenamefont {Riley}\ \emph {et~al.}(2010)\citenamefont {Riley},
  \citenamefont {Pito\v{n}\'{a}k}, \citenamefont {Jure\u{c}ka},\ and\
  \citenamefont {Hobza}}]{riley10}%
  \BibitemOpen
  \bibfield  {author} {\bibinfo {author} {\bibfnamefont {K.~E.}\ \bibnamefont
  {Riley}}, \bibinfo {author} {\bibfnamefont {M.}~\bibnamefont
  {Pito\v{n}\'{a}k}}, \bibinfo {author} {\bibfnamefont {P.}~\bibnamefont
  {Jure\u{c}ka}}, \ and\ \bibinfo {author} {\bibfnamefont {P.}~\bibnamefont
  {Hobza}},\ }\href@noop {} {\bibfield  {journal} {\bibinfo  {journal} {Chem.
  Rev.}\ }\textbf {\bibinfo {volume} {110}},\ \bibinfo {pages} {5023} (\bibinfo
  {year} {2010})}\BibitemShut {NoStop}%
\bibitem [{\citenamefont {Constans}, \citenamefont {Ayala},\ and\ \citenamefont
  {Scuseria}(2000)}]{pere00}%
  \BibitemOpen
  \bibfield  {author} {\bibinfo {author} {\bibfnamefont {P.}~\bibnamefont
  {Constans}}, \bibinfo {author} {\bibfnamefont {P.~Y.}\ \bibnamefont {Ayala}},
  \ and\ \bibinfo {author} {\bibfnamefont {G.~E.}\ \bibnamefont {Scuseria}},\
  }\href@noop {} {\bibfield  {journal} {\bibinfo  {journal} {J. Chem. Phys.}\
  }\textbf {\bibinfo {volume} {113}},\ \bibinfo {pages} {10451} (\bibinfo
  {year} {2000})}\BibitemShut {NoStop}%
\bibitem [{\citenamefont {Grimme}, \citenamefont {Steinmetz},\ and\
  \citenamefont {Korth}(2007)}]{grimme07}%
  \BibitemOpen
  \bibfield  {author} {\bibinfo {author} {\bibfnamefont {S.}~\bibnamefont
  {Grimme}}, \bibinfo {author} {\bibfnamefont {M.}~\bibnamefont {Steinmetz}}, \
  and\ \bibinfo {author} {\bibfnamefont {M.}~\bibnamefont {Korth}},\
  }\href@noop {} {\bibfield  {journal} {\bibinfo  {journal} {J. Org. Chem.}\
  }\textbf {\bibinfo {volume} {72}},\ \bibinfo {pages} {2118} (\bibinfo {year}
  {2007})}\BibitemShut {NoStop}%
\bibitem [{\citenamefont {Larsen}\ and\ \citenamefont
  {Nicolaisen}(1974)}]{larsen74}%
  \BibitemOpen
  \bibfield  {author} {\bibinfo {author} {\bibfnamefont {N.}~\bibnamefont
  {Larsen}}\ and\ \bibinfo {author} {\bibfnamefont {F.}~\bibnamefont
  {Nicolaisen}},\ }\href
  {https://www.sciencedirect.com/science/article/pii/0022286074800657}
  {\bibfield  {journal} {\bibinfo  {journal} {J. Mol. Struct.}\ }\textbf
  {\bibinfo {volume} {22}},\ \bibinfo {pages} {29} (\bibinfo {year}
  {1974})}\BibitemShut {NoStop}%
\bibitem [{\citenamefont {Larsen}(1986)}]{larsen86}%
  \BibitemOpen
  \bibfield  {author} {\bibinfo {author} {\bibfnamefont {N.}~\bibnamefont
  {Larsen}},\ }\href
  {https://www.sciencedirect.com/science/article/pii/0022286086801697}
  {\bibfield  {journal} {\bibinfo  {journal} {J. Mol. Struct.}\ }\textbf
  {\bibinfo {volume} {144}},\ \bibinfo {pages} {83} (\bibinfo {year}
  {1986})}\BibitemShut {NoStop}%
\bibitem [{\citenamefont {Smeyers}\ and\ \citenamefont
  {Hern\'{a}ndez-Laguna}(1987)}]{smeyers87}%
  \BibitemOpen
  \bibfield  {author} {\bibinfo {author} {\bibfnamefont {Y.~G.}\ \bibnamefont
  {Smeyers}}\ and\ \bibinfo {author} {\bibfnamefont {A.}~\bibnamefont
  {Hern\'{a}ndez-Laguna}},\ }\href
  {https://www.sciencedirect.com/science/article/pii/0166128087800544}
  {\bibfield  {journal} {\bibinfo  {journal} {J. Mol. Struct.}\ }\textbf
  {\bibinfo {volume} {149}},\ \bibinfo {pages} {127} (\bibinfo {year}
  {1987})}\BibitemShut {NoStop}%
\bibitem [{\citenamefont {Ratzer}, \citenamefont {Nispel},\ and\ \citenamefont
  {Schmitt}(2003)}]{ratzer03}%
  \BibitemOpen
  \bibfield  {author} {\bibinfo {author} {\bibfnamefont {C.}~\bibnamefont
  {Ratzer}}, \bibinfo {author} {\bibfnamefont {M.}~\bibnamefont {Nispel}}, \
  and\ \bibinfo {author} {\bibfnamefont {M.}~\bibnamefont {Schmitt}},\ }\href
  {http://dx.doix.org/10.1039/B210188B} {\bibfield  {journal} {\bibinfo
  {journal} {Phys. Chem. Chem. Phys.}\ }\textbf {\bibinfo {volume} {5}},\
  \bibinfo {pages} {812} (\bibinfo {year} {2003})}\BibitemShut {NoStop}%
\bibitem [{\citenamefont {Jaman}(2007)}]{jaman07}%
  \BibitemOpen
  \bibfield  {author} {\bibinfo {author} {\bibfnamefont {A.}~\bibnamefont
  {Jaman}},\ }\href
  {https://www.sciencedirect.com/science/article/pii/S0022285207001701}
  {\bibfield  {journal} {\bibinfo  {journal} {J. Mol. Spectrosc.}\ }\textbf
  {\bibinfo {volume} {245}},\ \bibinfo {pages} {21} (\bibinfo {year}
  {2007})}\BibitemShut {NoStop}%
\bibitem [{\citenamefont {Bell}\ \emph {et~al.}(2017)\citenamefont {Bell},
  \citenamefont {Singer}, \citenamefont {Desmond}, \citenamefont {Mahassneh},\
  and\ \citenamefont {{van Wijngaarden}}}]{bell17}%
  \BibitemOpen
  \bibfield  {author} {\bibinfo {author} {\bibfnamefont {A.}~\bibnamefont
  {Bell}}, \bibinfo {author} {\bibfnamefont {J.}~\bibnamefont {Singer}},
  \bibinfo {author} {\bibfnamefont {D.}~\bibnamefont {Desmond}}, \bibinfo
  {author} {\bibfnamefont {O.}~\bibnamefont {Mahassneh}}, \ and\ \bibinfo
  {author} {\bibfnamefont {J.}~\bibnamefont {{van Wijngaarden}}},\ }\href
  {https://www.sciencedirect.com/science/article/pii/S0022285216302521}
  {\bibfield  {journal} {\bibinfo  {journal} {J. Mol. Spectrosc.}\ }\textbf
  {\bibinfo {volume} {331}},\ \bibinfo {pages} {53} (\bibinfo {year}
  {2017})}\BibitemShut {NoStop}%
\bibitem [{\citenamefont {Brent}(1971)}]{brent71}%
  \BibitemOpen
  \bibfield  {author} {\bibinfo {author} {\bibfnamefont {R.~P.}\ \bibnamefont
  {Brent}},\ }\href {https://doix.org/10.1093/comjnl/14.4.422} {\bibfield
  {journal} {\bibinfo  {journal} {Comput. J.}\ }\textbf {\bibinfo {volume}
  {14}},\ \bibinfo {pages} {422} (\bibinfo {year} {1971})}\BibitemShut
  {NoStop}%
\bibitem [{\citenamefont {Jeziorski}, \citenamefont {Moszynski},\ and\
  \citenamefont {Szalewicz}(1994)}]{jeziorski94}%
  \BibitemOpen
  \bibfield  {author} {\bibinfo {author} {\bibfnamefont {B.}~\bibnamefont
  {Jeziorski}}, \bibinfo {author} {\bibfnamefont {R.}~\bibnamefont
  {Moszynski}}, \ and\ \bibinfo {author} {\bibfnamefont {K.}~\bibnamefont
  {Szalewicz}},\ }\href@noop {} {\bibfield  {journal} {\bibinfo  {journal}
  {Chem. Rev.}\ }\textbf {\bibinfo {volume} {94}},\ \bibinfo {pages} {1887}
  (\bibinfo {year} {1994})}\BibitemShut {NoStop}%
\bibitem [{\citenamefont {Hohenstein}\ and\ \citenamefont
  {Sherrill}()}]{hohenstein12c}%
  \BibitemOpen
  \bibfield  {author} {\bibinfo {author} {\bibfnamefont {E.~G.}\ \bibnamefont
  {Hohenstein}}\ and\ \bibinfo {author} {\bibfnamefont {C.~D.}\ \bibnamefont
  {Sherrill}},\ }\href
  {https://wires.onlinelibrary.wiley.com/doix/abs/10.1002/wcms.84} {\bibfield
  {journal} {\bibinfo  {journal} {WIREs Comput. Mol. Sci.}\ }\textbf {\bibinfo
  {volume} {2}},\ \bibinfo {pages} {304}}\BibitemShut {NoStop}%
\bibitem [{\citenamefont {Szalewicz}(2012)}]{szalewicz12}%
  \BibitemOpen
  \bibfield  {author} {\bibinfo {author} {\bibfnamefont {K.}~\bibnamefont
  {Szalewicz}},\ }\href
  {https://wires.onlinelibrary.wiley.com/doix/abs/10.1002/wcms.86} {\bibfield
  {journal} {\bibinfo  {journal} {WIREs Comput. Mol. Sci.}\ }\textbf {\bibinfo
  {volume} {2}},\ \bibinfo {pages} {254} (\bibinfo {year} {2012})}\BibitemShut
  {NoStop}%
\bibitem [{\citenamefont {Jansen}(2014)}]{jansen14}%
  \BibitemOpen
  \bibfield  {author} {\bibinfo {author} {\bibfnamefont {G.}~\bibnamefont
  {Jansen}},\ }\href
  {https://wires.onlinelibrary.wiley.com/doix/abs/10.1002/wcms.1164} {\bibfield
   {journal} {\bibinfo  {journal} {WIREs Comput. Mol. Sci.}\ }\textbf {\bibinfo
  {volume} {4}},\ \bibinfo {pages} {127} (\bibinfo {year} {2014})}\BibitemShut
  {NoStop}%
\bibitem [{\citenamefont {Parker}\ \emph {et~al.}(2014)\citenamefont {Parker},
  \citenamefont {Burns}, \citenamefont {Parrish}, \citenamefont {Ryno},\ and\
  \citenamefont {Sherrill}}]{parker14}%
  \BibitemOpen
  \bibfield  {author} {\bibinfo {author} {\bibfnamefont {T.~M.}\ \bibnamefont
  {Parker}}, \bibinfo {author} {\bibfnamefont {L.~A.}\ \bibnamefont {Burns}},
  \bibinfo {author} {\bibfnamefont {R.~M.}\ \bibnamefont {Parrish}}, \bibinfo
  {author} {\bibfnamefont {A.~G.}\ \bibnamefont {Ryno}}, \ and\ \bibinfo
  {author} {\bibfnamefont {C.~D.}\ \bibnamefont {Sherrill}},\ }\href@noop {}
  {\bibfield  {journal} {\bibinfo  {journal} {J. Chem. Phys.}\ }\textbf
  {\bibinfo {volume} {140}},\ \bibinfo {pages} {094106} (\bibinfo {year}
  {2014})}\BibitemShut {NoStop}%
\bibitem [{\citenamefont {Huber}\ and\ \citenamefont
  {Klamroth}(2011)}]{huber11}%
  \BibitemOpen
  \bibfield  {author} {\bibinfo {author} {\bibfnamefont {C.}~\bibnamefont
  {Huber}}\ and\ \bibinfo {author} {\bibfnamefont {T.}~\bibnamefont
  {Klamroth}},\ }\href@noop {} {\bibfield  {journal} {\bibinfo  {journal} {J.
  Chem. Phys.}\ }\textbf {\bibinfo {volume} {134}},\ \bibinfo {pages} {054113}
  (\bibinfo {year} {2011})}\BibitemShut {NoStop}%
\bibitem [{\citenamefont {Kvaal}(2012)}]{kvaal12}%
  \BibitemOpen
  \bibfield  {author} {\bibinfo {author} {\bibfnamefont {S.}~\bibnamefont
  {Kvaal}},\ }\href@noop {} {\bibfield  {journal} {\bibinfo  {journal} {J.
  Chem. Phys.}\ }\textbf {\bibinfo {volume} {136}},\ \bibinfo {pages} {194109}
  (\bibinfo {year} {2012})}\BibitemShut {NoStop}%
\bibitem [{\citenamefont {Sato}\ \emph {et~al.}(2018)\citenamefont {Sato},
  \citenamefont {Pathak}, \citenamefont {Orimo},\ and\ \citenamefont
  {Ishikawa}}]{sato18}%
  \BibitemOpen
  \bibfield  {author} {\bibinfo {author} {\bibfnamefont {T.}~\bibnamefont
  {Sato}}, \bibinfo {author} {\bibfnamefont {H.}~\bibnamefont {Pathak}},
  \bibinfo {author} {\bibfnamefont {Y.}~\bibnamefont {Orimo}}, \ and\ \bibinfo
  {author} {\bibfnamefont {K.~L.}\ \bibnamefont {Ishikawa}},\ }\href@noop {}
  {\bibfield  {journal} {\bibinfo  {journal} {J. Chem. Phys.}\ }\textbf
  {\bibinfo {volume} {148}},\ \bibinfo {pages} {051101} (\bibinfo {year}
  {2018})}\BibitemShut {NoStop}%
\bibitem [{\citenamefont {Pedersen}\ and\ \citenamefont
  {Kvaal}(2019)}]{pedersen19}%
  \BibitemOpen
  \bibfield  {author} {\bibinfo {author} {\bibfnamefont {T.~B.}\ \bibnamefont
  {Pedersen}}\ and\ \bibinfo {author} {\bibfnamefont {S.}~\bibnamefont
  {Kvaal}},\ }\href@noop {} {\bibfield  {journal} {\bibinfo  {journal} {J.
  Chem. Phys.}\ }\textbf {\bibinfo {volume} {150}},\ \bibinfo {pages} {144106}
  (\bibinfo {year} {2019})}\BibitemShut {NoStop}%
\bibitem [{\citenamefont {Kristiansen}\ \emph {et~al.}(2020)\citenamefont
  {Kristiansen}, \citenamefont {Sch\o{}yen}, \citenamefont {Kvaal},\ and\
  \citenamefont {Pedersen}}]{kristi20}%
  \BibitemOpen
  \bibfield  {author} {\bibinfo {author} {\bibfnamefont {H.~E.}\ \bibnamefont
  {Kristiansen}}, \bibinfo {author} {\bibfnamefont {O.~S.}\ \bibnamefont
  {Sch\o{}yen}}, \bibinfo {author} {\bibfnamefont {S.}~\bibnamefont {Kvaal}}, \
  and\ \bibinfo {author} {\bibfnamefont {T.~B.}\ \bibnamefont {Pedersen}},\
  }\href@noop {} {\bibfield  {journal} {\bibinfo  {journal} {J. Chem. Phys.}\
  }\textbf {\bibinfo {volume} {152}},\ \bibinfo {pages} {071102} (\bibinfo
  {year} {2020})}\BibitemShut {NoStop}%
\end{thebibliography}%


\begin{thebibliography}{0}%
\makeatletter
\providecommand \@ifxundefined [1]{%
 \@ifx{#1\undefined}
}%
\providecommand \@ifnum [1]{%
 \ifnum #1\expandafter \@firstoftwo
 \else \expandafter \@secondoftwo
 \fi
}%
\providecommand \@ifx [1]{%
 \ifx #1\expandafter \@firstoftwo
 \else \expandafter \@secondoftwo
 \fi
}%
\providecommand \natexlab [1]{#1}%
\providecommand \enquote  [1]{``#1''}%
\providecommand \bibnamefont  [1]{#1}%
\providecommand \bibfnamefont [1]{#1}%
\providecommand \citenamefont [1]{#1}%
\providecommand \href@noop [0]{\@secondoftwo}%
\providecommand \href [0]{\begingroup \@sanitize@url \@href}%
\providecommand \@href[1]{\@@startlink{#1}\@@href}%
\providecommand \@@href[1]{\endgroup#1\@@endlink}%
\providecommand \@sanitize@url [0]{\catcode `\\12\catcode `\$12\catcode
  `\&12\catcode `\#12\catcode `\^12\catcode `\_12\catcode `\%12\relax}%
\providecommand \@@startlink[1]{}%
\providecommand \@@endlink[0]{}%
\providecommand \url  [0]{\begingroup\@sanitize@url \@url }%
\providecommand \@url [1]{\endgroup\@href {#1}{\urlprefix }}%
\providecommand \urlprefix  [0]{URL }%
\providecommand \Eprint [0]{\href }%
\providecommand \doibase [0]{http://dx.doi.org/}%
\providecommand \selectlanguage [0]{\@gobble}%
\providecommand \bibinfo  [0]{\@secondoftwo}%
\providecommand \bibfield  [0]{\@secondoftwo}%
\providecommand \translation [1]{[#1]}%
\providecommand \BibitemOpen [0]{}%
\providecommand \bibitemStop [0]{}%
\providecommand \bibitemNoStop [0]{.\EOS\space}%
\providecommand \EOS [0]{\spacefactor3000\relax}%
\providecommand \BibitemShut  [1]{\csname bibitem#1\endcsname}%
\let\auto@bib@innerbib\@empty
\end{thebibliography}%

\end{document}


\preprint{AIP/123-QED}

\title{Supplemental material for the paper:\\ Quintic-scaling rank-reduced coupled cluster theory 
with single and double excitations}

\author{Micha\l\ Lesiuk}
\email{m.lesiuk@uw.edu.pl}
\affiliation{\sl Faculty of Chemistry, University of Warsaw, Pasteura 1, 02-093 Warsaw, Poland}

\date{\today}

\maketitle
\tableofcontents

\newpage

\section{\label{sec:laplace} Accuracy of the Laplace quadrature}

\begin{table}[h!]
\caption{\label{tab:laplace-error-mp2}
RR-CCSD correlation energies (in mH) obtained with the MP2 excitation basis ($\neig=2\nmo$, 
$N_{\mathrm{O}}=4O$). The number of Laplace quadrature points ($N_g$) used in the diagonalization 
of the MP2 amplitudes is given in the first column. The result obtained with $N_g=20$ are exact to all 
digits given.
}
\begin{ruledtabular}
\begin{tabular}{lcccc}
 $N_g$ & \multicolumn{2}{c}{cc-pVDZ} & \multicolumn{2}{c}{cc-pVTZ} \\\cline{2-3}\cline{4-5}
  & HF$^a$ & CH$_4^b$  & HF$^a$ & CH$_4^b$ \\
\hline
  2 & $-$209.732 & $-$187.501 & $-$287.217 & $-$235.346 \\
  3 & $-$209.929 & $-$187.683 & $-$287.920 & $-$235.660 \\
  4 & $-$209.907 & $-$187.676 & $-$288.002 & $-$235.686 \\
  6 & $-$209.891 & $-$187.672 & $-$288.003 & $-$235.667 \\
 10 & $-$209.889 & $-$187.670 & $-$288.000 & $-$235.665 \\
 20 & $-$209.890 & $-$187.671 & $-$288.000 & $-$235.665 \\
\end{tabular}
\end{ruledtabular}
\vspace{-0.75cm}
\begin{flushleft}
 $^a$ HF bond length: 1.732 a.u. \\\vspace{-0.3cm}
 $^b$ tetrahedral geometry, C--H bond length: 2.048 a.u.
\end{flushleft}
\end{table}

\begin{table}[h!]
\caption{\label{tab:laplace-error-mp3}
RR-CCSD correlation energies (in mH) obtained with the MP3 excitation basis ($\neig=2\nmo$, 
$N_{\mathrm{O}}=4O$). Ten quadrature points are used in the first part of Eq. (15) from the main text, while $N_g$ 
points are used in the second part. The result obtained with $N_g=10$ are
exact to all digits given.
}
\begin{ruledtabular}
\begin{tabular}{lcccc}
 $N_g$ & \multicolumn{2}{c}{cc-pVDZ} & \multicolumn{2}{c}{cc-pVTZ} \\\cline{2-3}\cline{4-5}
  & HF$^a$ & CH$_4^b$  & HF$^a$ & CH$_4^b$ \\
\hline
  2 & $-$209.125 & $-$187.417 & $-$287.334 & $-$235.453 \\
  3 & $-$209.113 & $-$187.453 & $-$287.282 & $-$235.484 \\
  4 & $-$209.109 & $-$187.459 & $-$287.288 & $-$235.479 \\
  6 & $-$209.111 & $-$187.457 & $-$287.291 & $-$235.479 \\
 10 & $-$209.111 & $-$187.457 & $-$287.291 & $-$235.479 \\
\end{tabular}
\end{ruledtabular}
\end{table}

\newpage

\section{\label{sec:fact-rxy} Factorizable terms in the RR-CCSD residual}

All formulas provided here correspond to the $t_1$-dressed two-electron integrals, i.e. 
$(pq\widetilde{|}rs)=\widetilde{B}_{pq}^Q\,\widetilde{B}_{rs}^Q$. The necessary intermediates read
\begin{align}
 T_{ia}^X = U_{ia}^X\,t_{XY},\;\;\;
 S_{ia}^X = \Big( T_{ib}^Z\,U_{jb}^X \Big) U_{ja}^Z,\;\;\;
 \bar{T}_{ia}^X = 2T_{ia}^X - S_{ia}^X,
\end{align}
and
\begin{align}
 B_{ia}^{QX} = \widetilde{B}_{ji}^Q\,U_{ja}^X, \;\;\;
 B_{ai}^{QX} = \widetilde{B}_{ab}^Q\,U_{ib}^X, \;\;\;
 B_{ij}^{QX} = \widetilde{B}_{ia}^Q\,U_{ja}^X, \;\;\;
 A_X^Q = B_{ia}^Q\,U_{ia}^X,\;\;\;
 \bar{A}_X^Q = B_{ai}^Q\,U_{ia}^X,
\end{align}
and
\begin{align}
 &X_{ab} = -\widetilde{F}_{ab} + 2\Big( A_Y^Q\,B_{jb}^Q \Big) T_{ja}^Y 
         - \Big( B_{ij}^{QX}\,B_{jb}^Q \Big) T_{ia}^X, \\
 &X_{ji} = +\widetilde{F}_{ij} + 2\Big( A_X^Q\,B_{ia}^Q \Big) T_{ja}^X 
         - \Big( B_{ik}^{QX}\,B_{kb}^Q \Big) T_{jb}^X,
\end{align}
and
\begin{align}
 &W_{kc}^Y = B_{kc}^Q\,\bar{A}_Y^Q + \half B_{kc}^Q\Big( B_{ld}^Q\,\bar{T}_{ld}^Y \Big), \\
 &Y_{ia}^Y = B_{ij}^Q \Big( B_{ba}^Q\,U_{jb}^Y \Big) 
           - \Big( B_{ic}^Q\,S_{kc}^Y \Big) B_{ka}^Q
           + \Big( B_{ic}^Q\,T_{kc}^Y \Big) B_{ka}^Q.
\end{align}
The RR-CCSD residual reads
\begin{align}
\begin{split}
 r_{XY} = P_{XY}&\bigg[ \half\,\bar{A}_X^Q\,\bar{A}_Y^Q 
        + \half\,\Big[U_{ia}^X\Big( B_{ia}^{QZ} - B_{ai}^{QZ} \Big)\Big] t_{ZW}
          \Big[\Big( B_{jb}^{QW} - B_{bj}^{QW} \Big) U_{jb}^Y\Big] \\
        &+ U_{ia}^X\,\Big( X_{ac}\,T_{ic}^Y + X_{il}\,T_{la}^Y \Big)
        - \bar{T}_{kc}^X\,W_{kc}^Y + T_{kc}^X\,Y_{kc}^Y \bigg]
        + \mbox{\;non-factorizable terms.}
\end{split}
\end{align}
\newpage

\section{\label{sec:oz-vtz} Effective rank of $O^{ij}_{kl}$ and $Z_{ij}^{ab}$ intermediates in the 
double-zeta basis set}
\vspace{-0.5cm}

\begin{figure}[ht]
\includegraphics[scale=0.60]{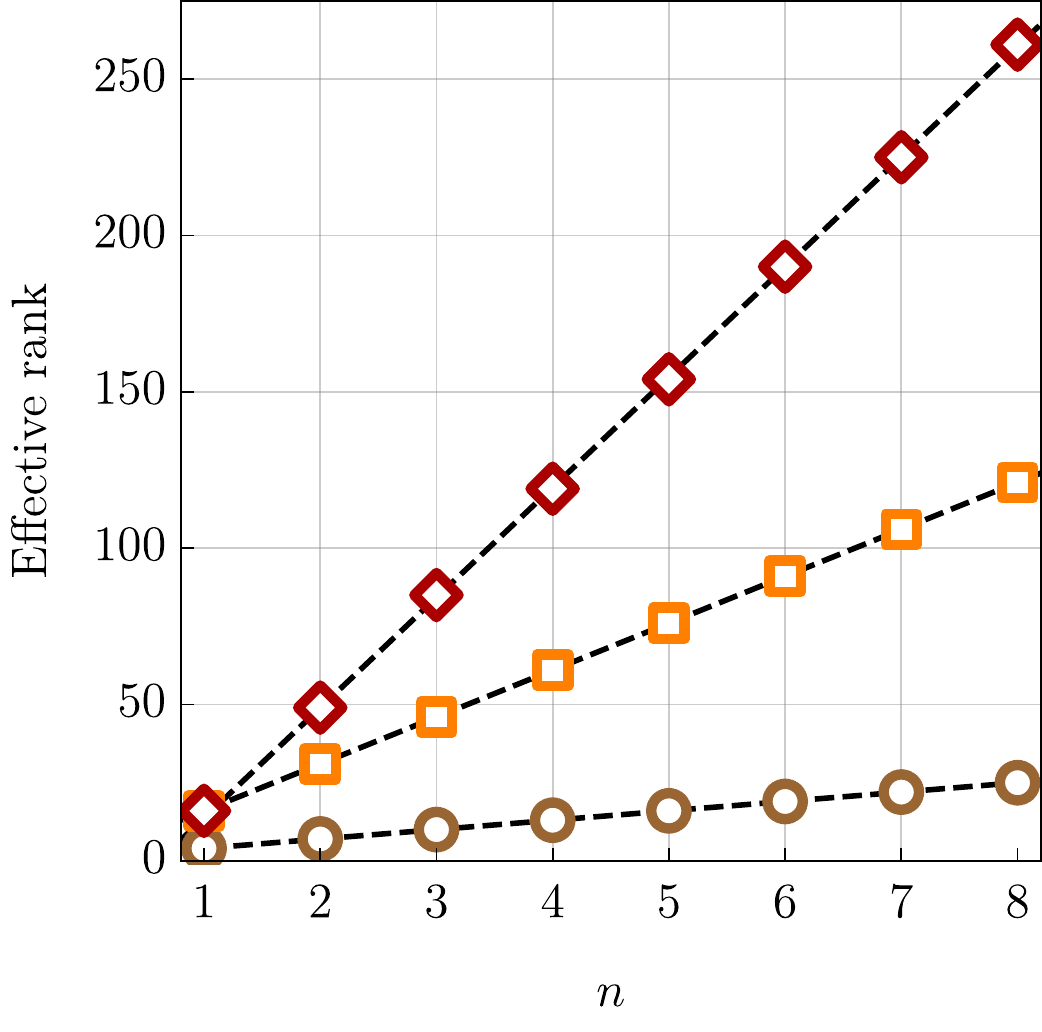} \hspace{0.5cm}
\includegraphics[scale=0.60]{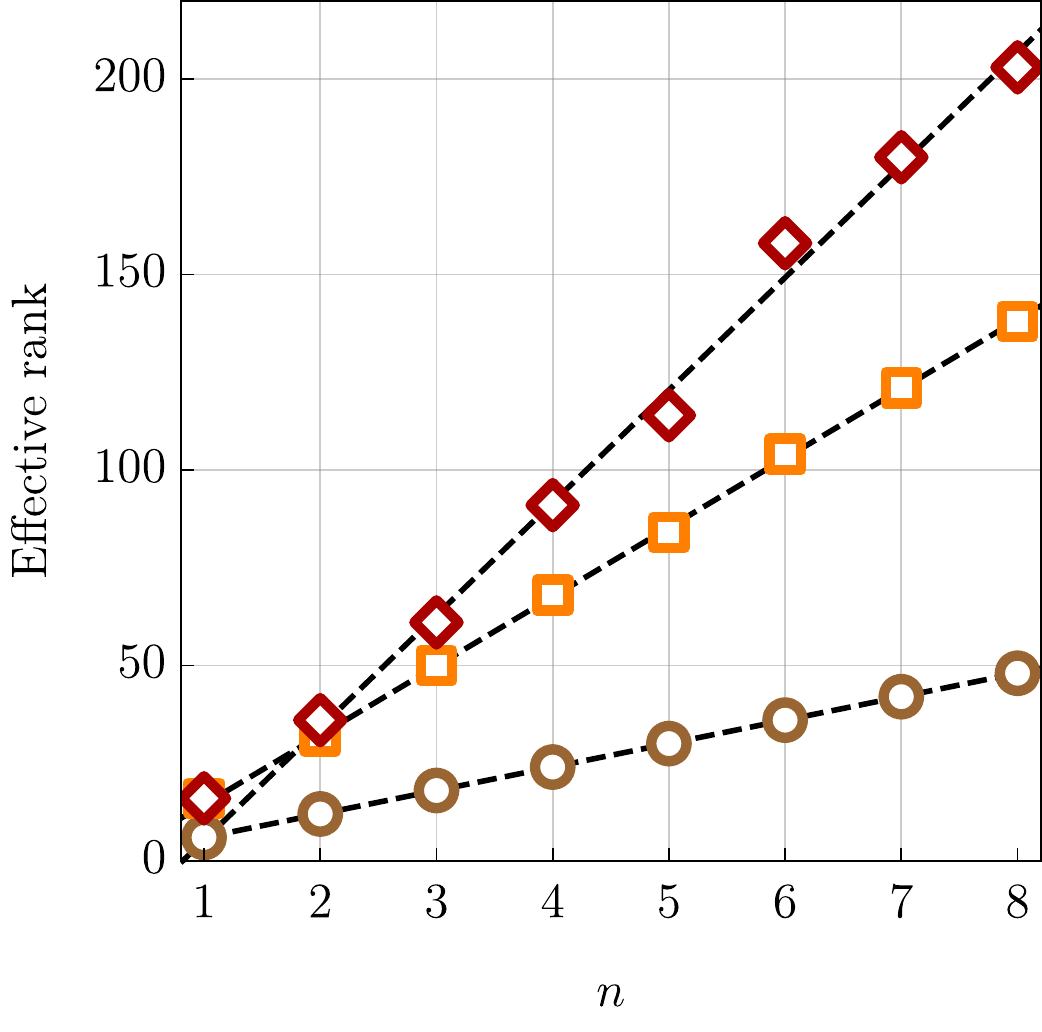} \vspace{-0.8cm}
\caption{\label{fig:o-scaling} Effective rank of the $O^{ij}_{kl}$ intermediate for the linear 
alkanes 
C$_n$H$_{2n+2}$ (left panel) and water clusters $\big($H$_2$O$\big)_n$ (right panel) extracted 
from the CCSD/cc-pVDZ calculations. The brown circles, orange squares and red diamonds indicate the 
effective rank obtained with the thresholds $\varepsilon=10^{-2}$, $10^{-3}$, and $10^{-4}$, 
respectively. The black dashed lines were obtained by least-squares fitting to the corresponding 
data points.}
\end{figure}
\begin{figure}[ht]
\includegraphics[scale=0.60]{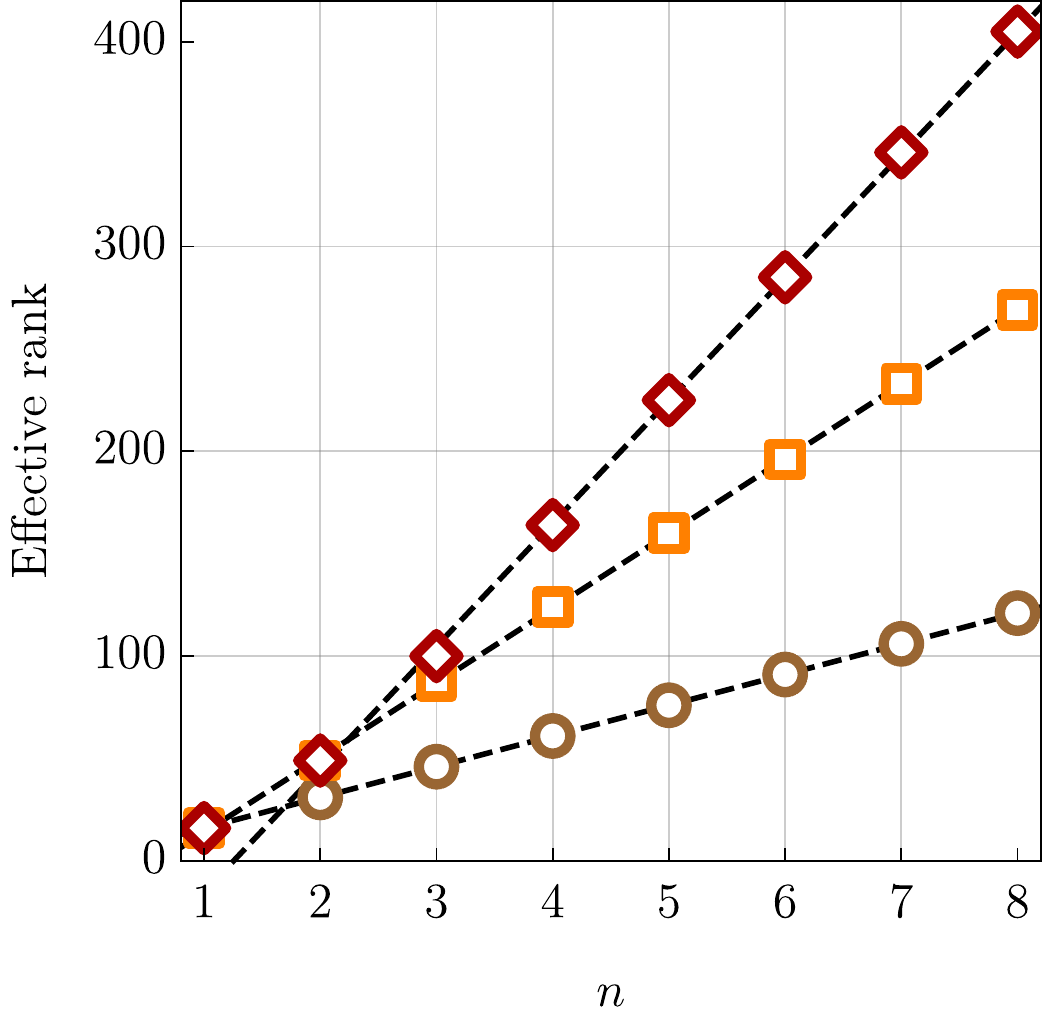} \hspace{0.5cm}
\includegraphics[scale=0.60]{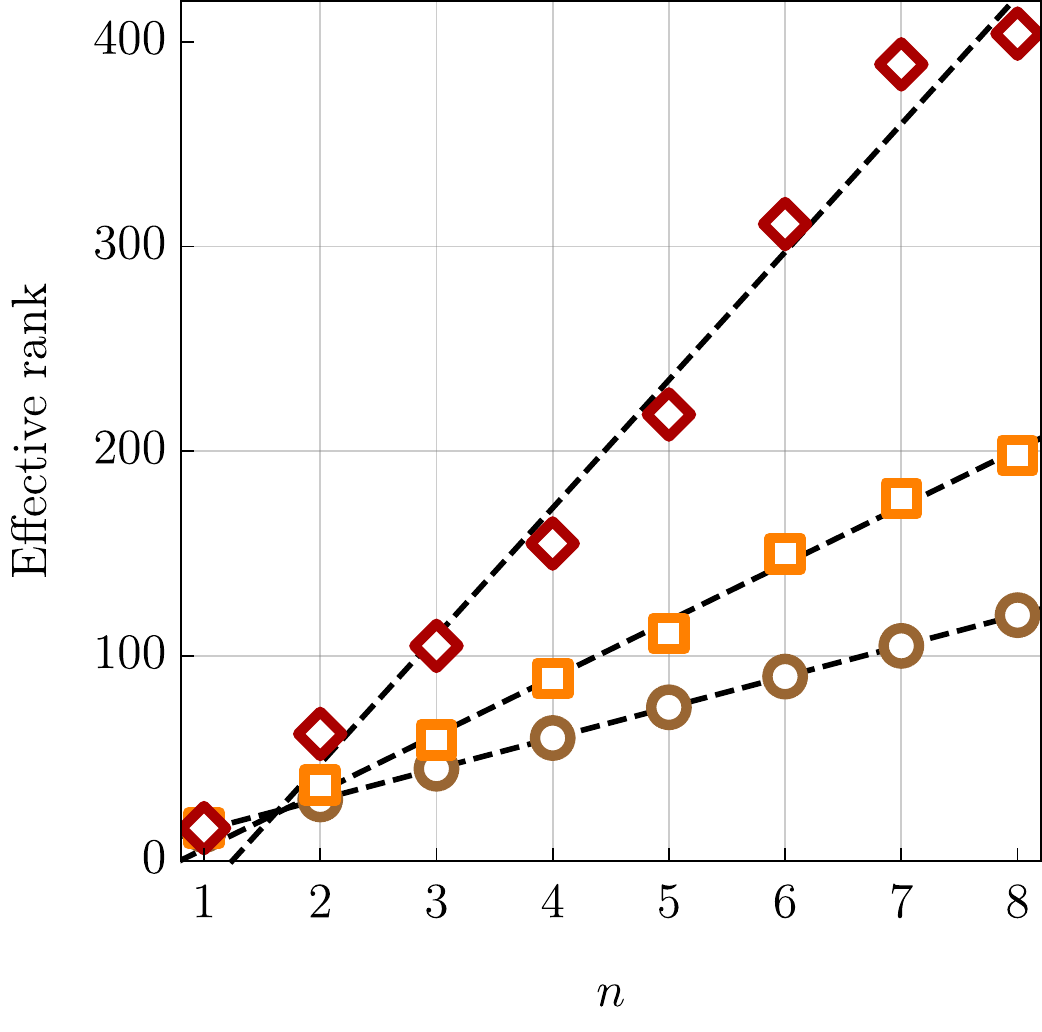} \vspace{-0.8cm}
\caption{\label{fig:z-scaling} Effective rank of the $Z_{ij}^{ab}$ intermediate for the linear 
alkanes 
C$_n$H$_{2n+2}$ (top panel) and water clusters $\big($H$_2$O$\big)_n$ (bottom panel) extracted 
from the CCSD/cc-pVDZ calculations. The brown circles, orange squares and red diamonds indicate the 
effective rank obtained with the thresholds $\varepsilon=10^{-3}$, $10^{-4}$, and $10^{-5}$, 
respectively. The black dashed lines were obtained by least-squares fitting to the corresponding 
data points.}
\end{figure}

\newpage
\newpage

\section{\label{sec:oz-alkenes} Effective rank of $O^{ij}_{kl}$ and $Z_{ij}^{ab}$ intermediates -- linear alkenes}
\vspace{0.5cm}

\begin{figure}[ht]
\includegraphics[scale=0.60]{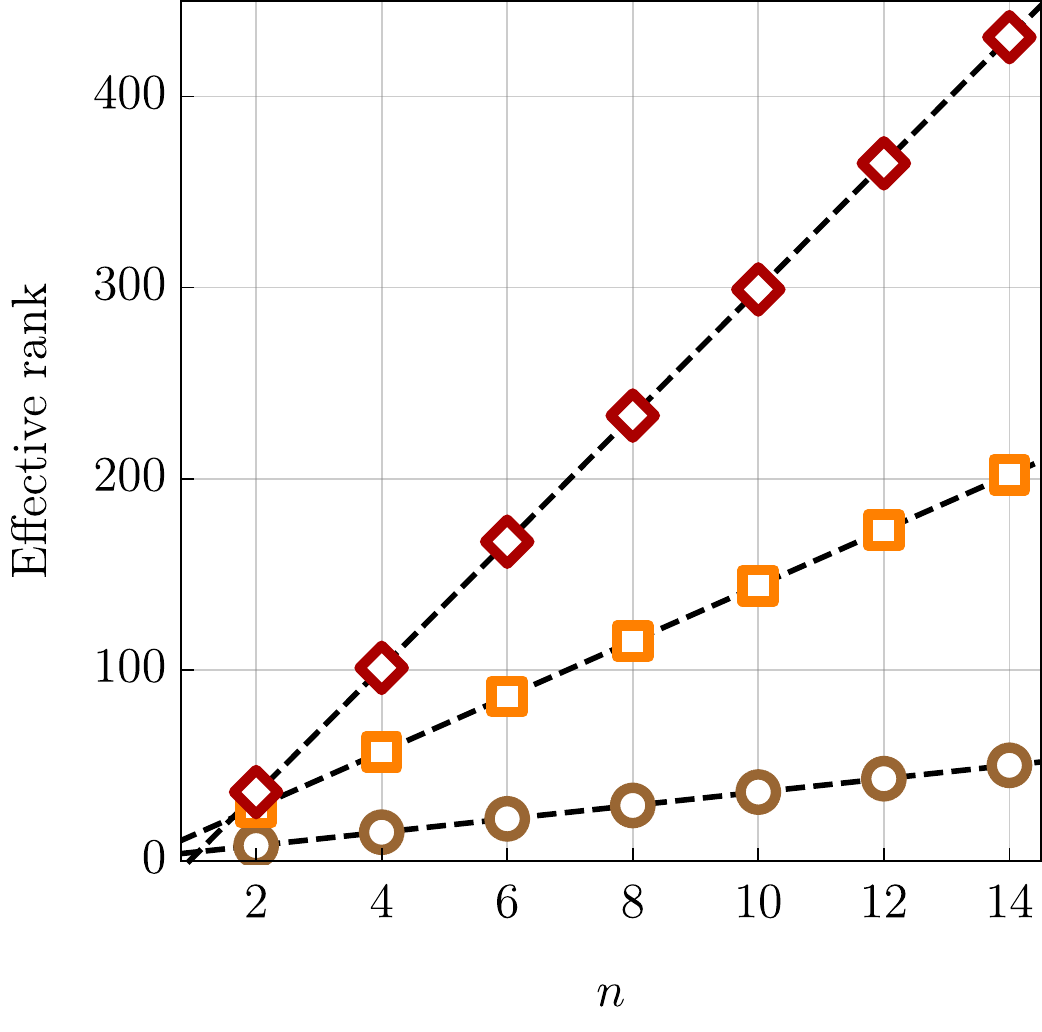} \hspace{0.5cm}
\includegraphics[scale=0.60]{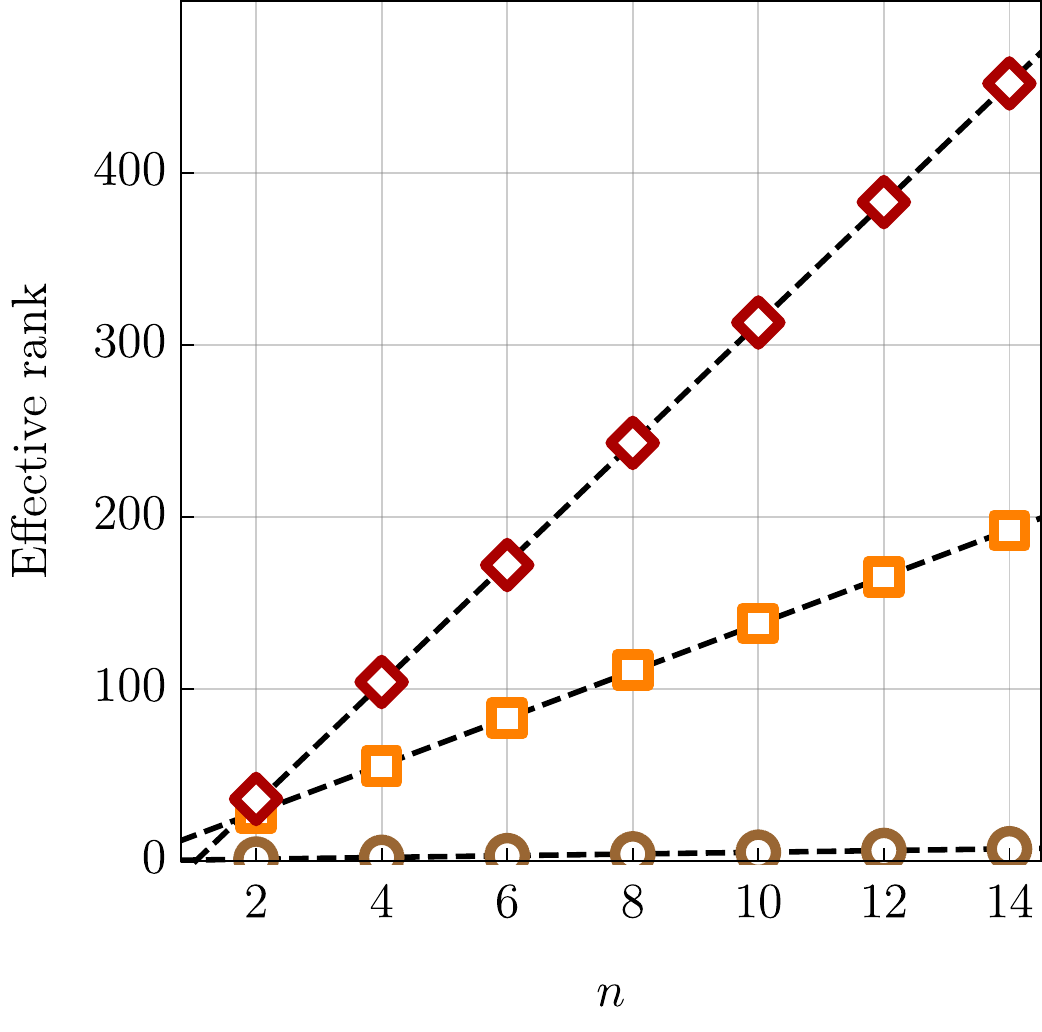} \vspace{-0.8cm}
\caption{\label{fig:oz-scaling-alkenes} Effective rank of the $O_{ij}^{ab}$ intermediate (left panel) and $Z_{ij}^{ab}$ 
intermediate (right panel) for the linear alkenes C$_n$H$_{2n}$ extracted from the CCSD/cc-pVDZ calculations. The 
brown circles, orange squares and red diamonds indicate the effective rank obtained with the thresholds 
$\varepsilon=10^{-2}$, $10^{-3}$, and $10^{-4}$, respectively. The black dashed lines were obtained by least-squares 
fitting to the corresponding data points.}
\end{figure}

\newpage
\newpage

\section{\label{sec:oz-dynamic} Dynamic and fixed approach to the determination of the effective rank of the 
$O^{ij}_{kl}$ intermediate}

\begin{figure}[ht]
\includegraphics[scale=0.60]{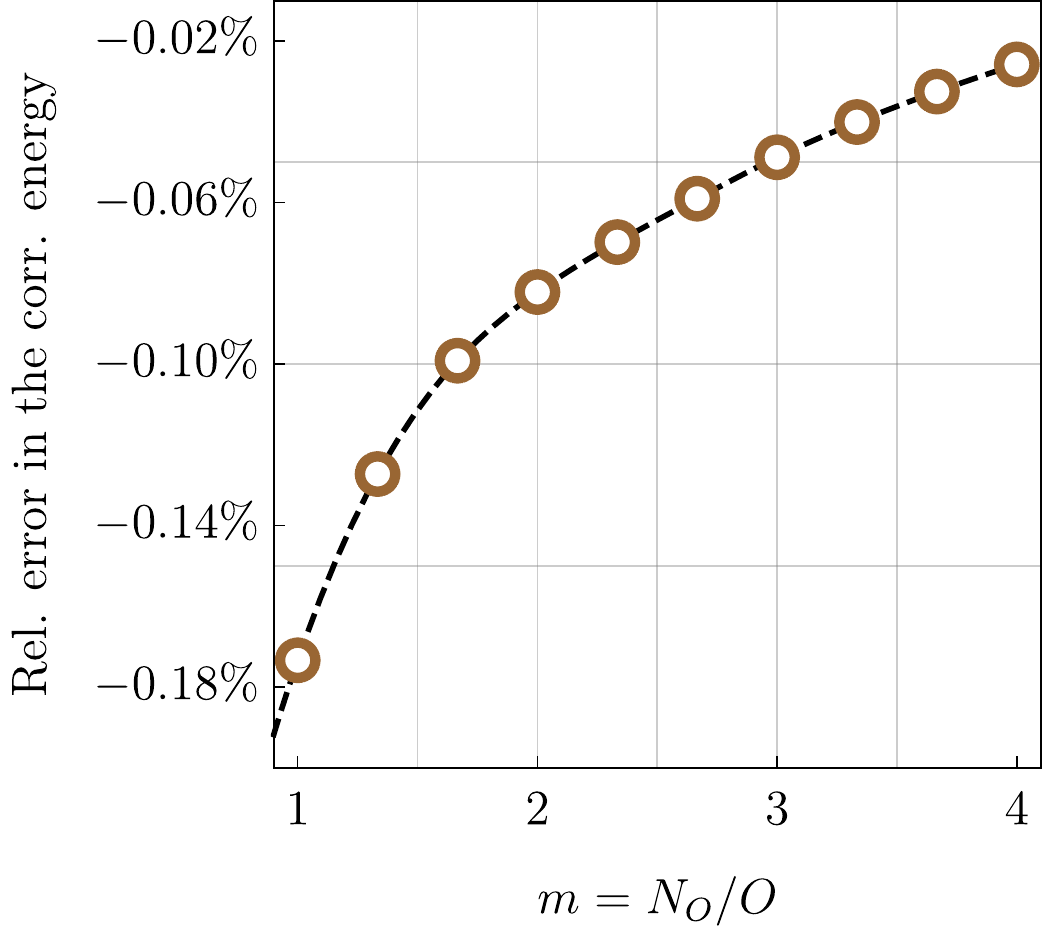}\vspace{-0.5cm}
\caption{\label{fig:oz-thr-1} Relative error in the RR-CCSD correlation energy (in percent) resulting from truncation 
of the expansion of the $O^{ij}_{kl}$ intermediate as a function of the expansion parameter $m=N_O/O$ (aniline 
molecule, cc-pVDZ basis set). The black dashed lines were obtained by interpolation of the corresponding data points.}
\end{figure}

\begin{figure}[ht]
\includegraphics[scale=0.60]{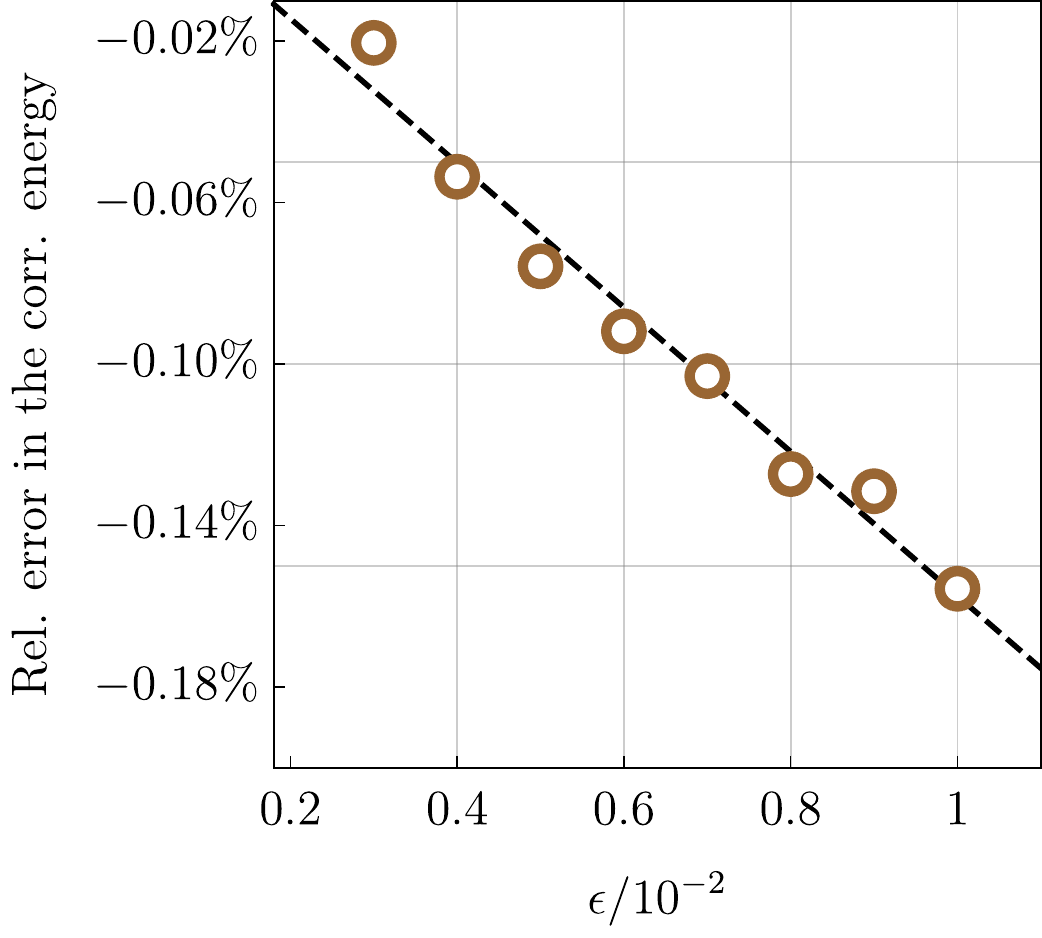} \hspace{0.5cm}
\includegraphics[scale=0.62]{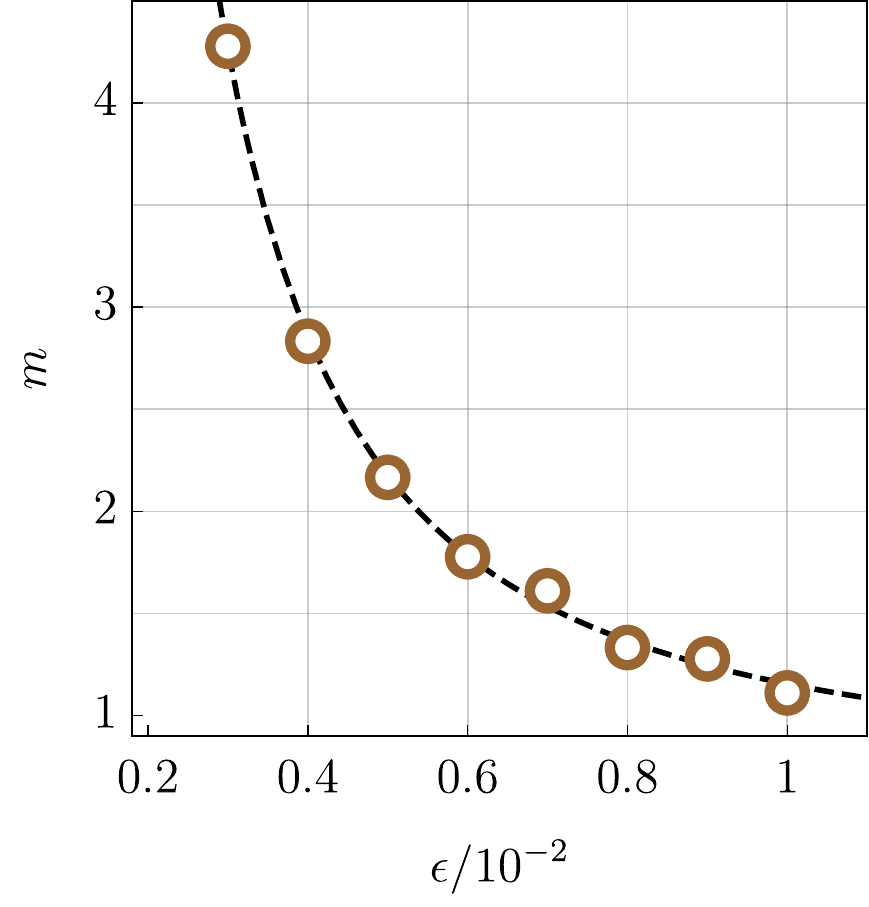} \vspace{-0.5cm}
\caption{\label{fig:oz-thr-2} Left panel: relative error in the RR-CCSD correlation energy (in percent) resulting 
from truncation of the expansion of the $O^{ij}_{kl}$ intermediate as a function of the threshold $\epsilon$ used 
for dropping eigenpairs. Right panel: the parameter $m=N_O/O$ as a function of the threshold $\epsilon$. All results 
are given for aniline molecule in the cc-pVDZ basis set.}
\end{figure}

\newpage
\newpage

\section{\label{sec:rr-avdz} Approximate treatment of the $O^{ij}_{kl}$ and $Z_{ij}^{ab}$ intermediates in augmented 
basis sets}

\begin{table}[h!]
\caption{\label{tab:oz-trunc1-aug}
Statistical measures of relative errors (in percent) in the RR-CCSD/aug-cc-pVDZ correlation energy 
(for $\neig=\nmo$) resulting from truncation of the expansions of the $O^{ij}_{kl}$ and $Z_{ij}^{ab}$ intermediates, see Eqs. (27) and (34) from the main text, at length $N_{\mathrm{O}}=N_{\mathrm{Z}}=mO$, where $O$ is the number of occupied orbitals in the system. The value of the parameter $m$ is given in the first column. The statistics comes from 
RR-CCSD/aug-cc-pVDZ calculations for 70 molecules contained in the Adler-Werner benchmark set.
}
\begin{ruledtabular}
\begin{tabular}{lcccc}
 $m$ & mean  & mean abs. & standard  & max. abs. \\
     & error & error     & deviation & error     \\
\hline
$1$ & $-0.173$ & $0.173$ & $0.045$ & $0.344$ \\
$2$ & $-0.072$ & $0.080$ & $0.042$ & $0.153$ \\
$3$ & $-0.032$ & $0.041$ & $0.030$ & $0.133$ \\
$4$ & $-0.014$ & $0.018$ & $0.014$ & $0.062$ \\
\end{tabular}
\end{ruledtabular}
\end{table}

\begin{figure}[h!]
\includegraphics[scale=0.70]{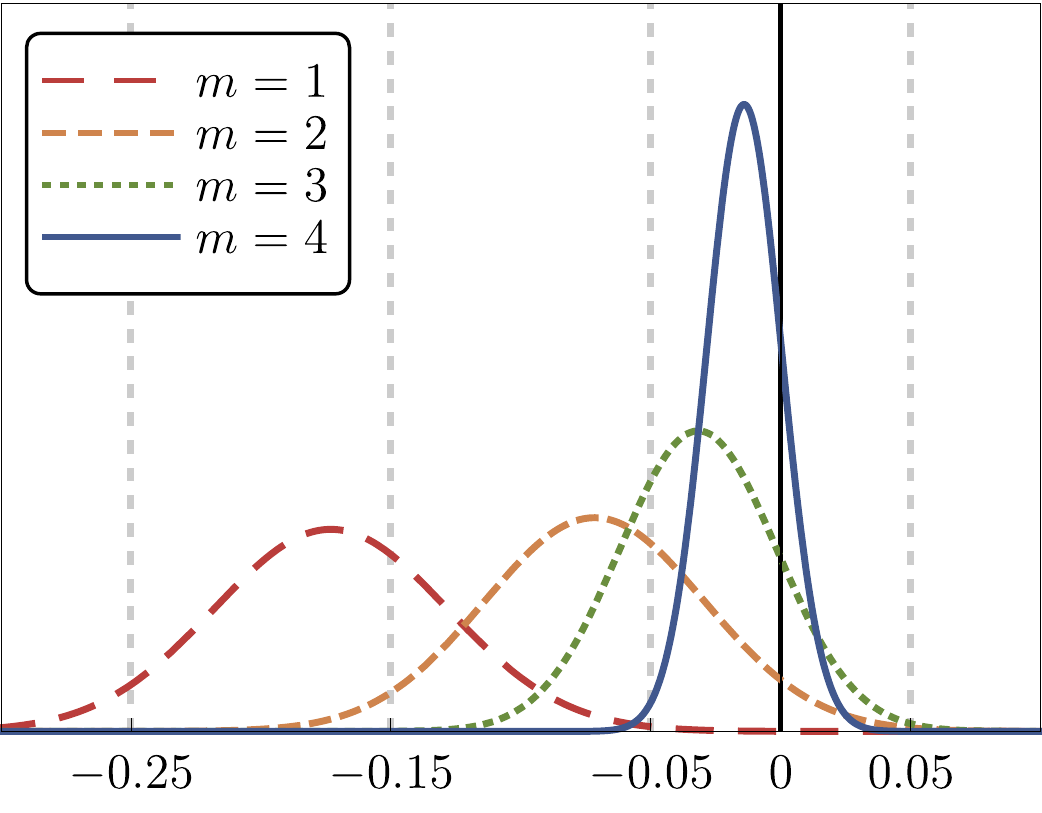}
\caption{\label{fig:oz-trunc1-aug} 
Distribution of relative error (in percent) resulting from 
the truncation of the $O^{ij}_{kl}$ and $Z_{ij}^{ab}$ intermediates, see Eqs. (27) and (34) from the main text, at length $N_{\mathrm{O}}=N_{\mathrm{Z}}=mO$, where $O$ is the number of occupied orbitals in the system and the value of the parameter $m$ is given in the legend. The results were obtained with $\neig=N_{\mathrm{MO}}$. The statistics comes from RR-CCSD/cc-pVDZ calculations for 70 molecules contained in the Adler-Werner benchmark set.
}
\end{figure}

\newpage
\section{\label{sec:rr-vdz} RR-CCSD/cc-pVDZ error distributions}

\begin{figure}[ht!]
\includegraphics[scale=1.00]{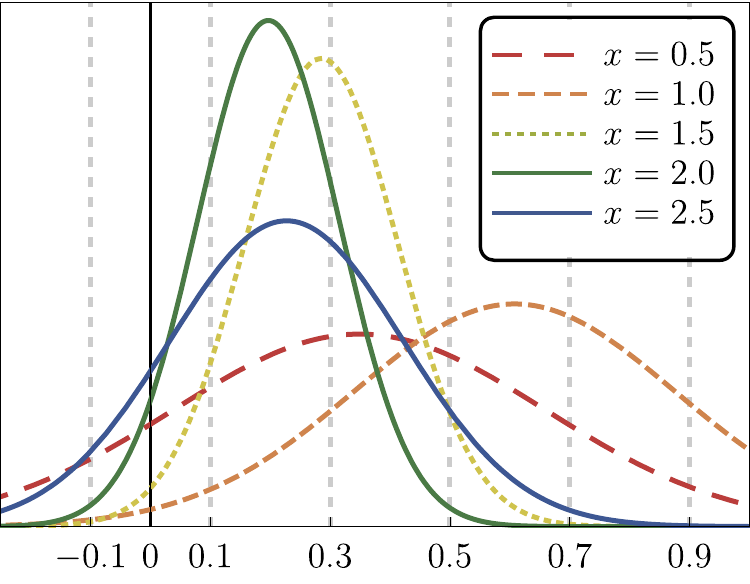}
\\\vspace{0.5cm}
\includegraphics[scale=1.00]{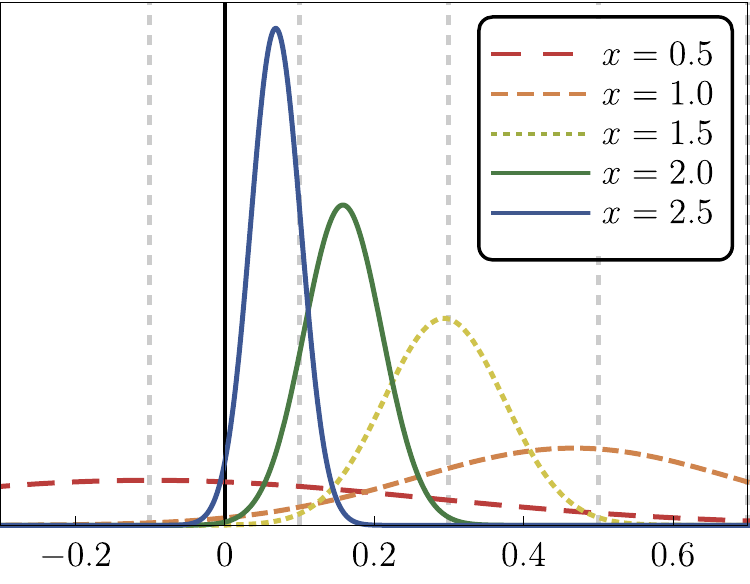}
\caption{\label{fig:final-accuracy-vdz} Distribution of relative error (in percent) in the 
RR-CCSD/cc-pVDZ correlation energy with respect to the exact CCSD/cc-pVDZ results. The dimension 
of the excitation subspace ($\neig$) is expressed as $\neig=x\cdot N_{\mathrm{MO}}$, where 
$N_{\mathrm{MO}}$ is the total number of orbitals in the system. The excitations subspace
was obtained by diagonalization of MP2 amplitudes (top panel) or MP3 amplitudes (bottom panel). 
The statistics comes from calculations for 70 molecules contained in the Adler-Werner benchmark 
set.
}
\end{figure}
\newpage

\section{\label{sec:e5st} Explicit formula for the $E_{\mathrm{ST}}^{[5]}$ correction}

The notation for all quantities is the same as in the main text. Intermediates:
\begin{align}
\begin{split}
 &I_{ik}^C = t_i^c\,V_{kc}^C, \;\;\; J_A=t_i^a\,V_{ia}^A, \;\;\;
 K_{QA} = B_{kc}^Q\,V_{kc}^A, \;\;\; Z_{ji}^{QA} = B_{ja}^Q\,V_{ia}^A.
\end{split}
\end{align}
The last intermediates requires $\propto O^2V\naux\ntri$ operations to compute, remaining ones 
$N^4$ or less.

Explicit expression for the $E_{\mathrm{ST}}^{[5]}$ correction (computational cost of the 
rate-determining contraction step is given below each individual term):
\begin{align}
\begin{split}
 E_{\mathrm{ST}}^{[5]} &= 
 \underbrace{\Big[ \big( B_{ja}^Q\,Z_{kj}^{QB}\big)\,I_{ik}^C \Big] 
 V_{ia}^A\,t_{ABC}}_{O^2V\naux\ntri} +
 \underbrace{\Big[ \big( B_{ja}^Q\,K_{QC} \big) V_{ia}^A \Big] I_{ij}^B\,t_{ABC}}_{O^2V\ntri^2} \\
 &+ \underbrace{\big( Z_{jk}^{QC}\,Z_{kj}^{QB} \big) \big( J_A\,t_{ABC} \big)}_{O^2\naux\ntri^2}
 + \underbrace{\big( K_{QB}\,K_{QC} \big) \big( J_A\,t_{ABC} \big)}_{\naux\ntri^2}.
\end{split}
\end{align}

\section{\label{sec:e4t} Explicit formula for the $E_{\mathrm{T}}^{[4]}$ correction}

Intermediates:
\begin{align}
 I_{XA} = U_{ia}^X\,V_{ia}^A, \;\;\;
 J_{QA} = B_{ia}^Q\,V_{ia}^A, \;\;\;
 K_{XA}^Q = \bar{D}_{ia}^{QX}\,V_{ia}^A, \;\;\;
 t_{ia,jb,C} = \Big( t_{ABC}\,V_{ia}^A \Big) V_{jb}^B.
\end{align}
The last two intermediates require $\propto OV\naux\ntri\neig$ and $\propto O^2V^2\ntri^2$ 
operations to compute, respectively, remaining ones $N^4$ or less.

Explicit expression for the $E_{\mathrm{T}}^{[4]}$ correction (computational cost of the 
rate-determining step is given below each individual term):
\begin{align}
\begin{split}
 E_{\mathrm{T}}^{[4]} &= 
 \underbrace{\Big( t_{ia,jb,C}\,U_{ja}^X \Big) \Big( B_{kc}^C\,\bar{D}_{ic}^{QX} 
 \Big) B_{kb}^Q}_{O^2V\neig\naux\ntri} + 
 \underbrace{\Big( t_{ia,jb,C}\,U_{ja}^X \Big) K_{XC}^Q\,B_{ib}^Q}_{O^2V^2\ntri\neig} \\
 &+ \underbrace{t_{ABC} \Big( \bar{D}_{ic}^{QX}\,I_{XC} \Big) \Big( B_{ka}^Q\,V_{ia}^A \Big) 
  V_{kc}^B}_{O^2V\naux\ntri^2}
  + \underbrace{\Big( t_{ia,jb,C}\,U_{ja}^X \Big) \Big( \bar{D}_{ib}^{QX}\,J_{QC} 
  \Big)}_{O^2V^2\ntri\neig} \\
 &+ \underbrace{t_{ABC}\,K_{XA}^Q\,I_{XB}\,J_{QC}}_{\naux\neig\ntri^2} 
  + \underbrace{\Big( t_{ia,jb,C}\,U_{ja}^X \Big) \Big( V_{kc}^C\,B_{ic}^Q \Big) 
\bar{D}_{kb}^{QX}}_{O^2V\neig\naux\ntri}.
\end{split}
\end{align}

\newpage

\section{\label{sec:trip2-vdz} Error of the rank-reduced perturbative triples corrections: cc-pVDZ 
basis set}

\begin{figure}[ht!]
\includegraphics[scale=1.05]{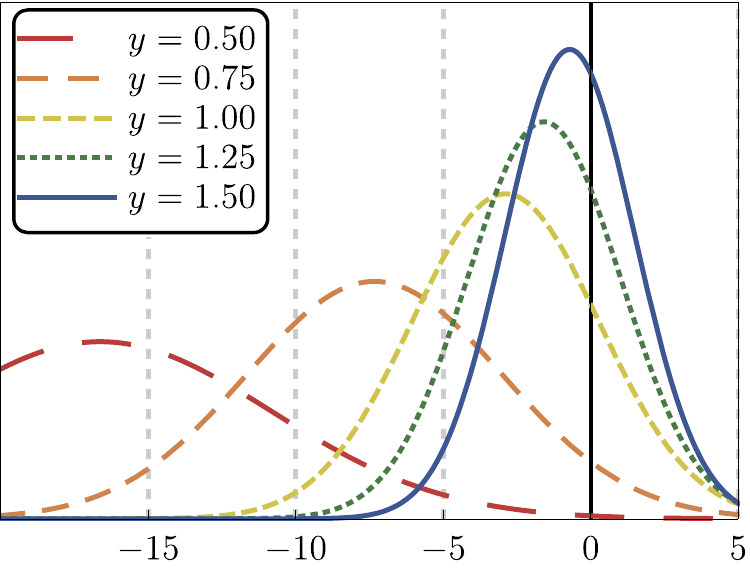}
\\\vspace{0.5cm}
\includegraphics[scale=1.00]{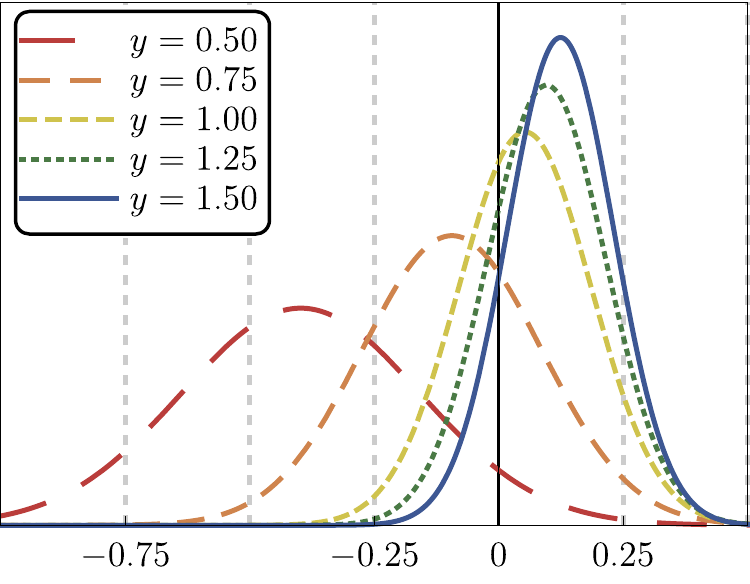}
\caption{\label{fig:trip2-vdz} Distribution of relative error (in percent) in the $E_{\mathrm{(T)}}$ correction (top 
panel) and total RR-CCSD(T)/cc-pVDZ correlation energy (bottom panel) with respect to the exact CCSD(T)/cc-pVDZ method. 
The dimension of the triples excitation subspace ($\ntri$) is expressed as $\ntri=y\cdot \nmo$, where 
$\nmo$ is the total number of orbitals in the system. The statistics comes from calculations for 70 
molecules contained in the Adler-Werner benchmark set.
}
\end{figure}

\newpage

\section{\label{sec:iso34} Raw isomerization energies for the ISO34 bechmark set}

\newcolumntype{P}[1]{>{\centering\arraybackslash}p{#1}}
\setlength{\LTcapwidth}{0.95\textwidth}

\begin{center}
\begin{longtable}{P{3.0cm}P{4.0cm}P{4.0cm}P{4.0cm}}
\caption{\label{tab:iso34}
ISO34 benchmark set isomerization energies calculated with the RR-CCSD(T)/cc-pVTZ and the exact CCSD(T)/cc-pVTZ 
methods. The recommended values of the parameters ($\neig=2\nmo$, $N_{\mathrm{O}}=N_{\mathrm{Z}}=4O$, and $\ntri=\nmo$) 
were employed in the RR-CCSD(T) computations. All results are given in kJ/mol.} \\

\hline\hline
reaction & RR-CCSD(T) & exact CCSD(T) & difference \\
\hline 
\endfirsthead

\multicolumn{3}{c}%
{{\tablename\ \thetable{} -- continued from previous page}} \\
\hline\hline reaction & RR-CCSD(T) & exact CCSD(T) & difference \\ \hline 
\endhead

\hline \multicolumn{3}{|r|}{{Continued on next page}} \\ \hline
\endfoot

\hline \hline
\endlastfoot

1  & \phantom{00}5.24 & \phantom{00}5.10 & \phantom{$-$}0.14 \\
2  & \phantom{0}98.77 & \phantom{0}98.27 & \phantom{$-$}0.50 \\
3  & \phantom{0}31.40 & \phantom{0}31.64 & $-$0.24 \\
4  & \phantom{00}4.72 & \phantom{00}4.66 & \phantom{$-$}0.06 \\
5  & \phantom{00}4.70 & \phantom{00}4.73 & $-$0.03 \\
\hline
6  & \phantom{00}9.79 & \phantom{00}9.94 & $-$0.15 \\
7  & \phantom{0}47.75 & \phantom{0}47.95 & $-$0.20 \\
8  & \phantom{0}94.54 & \phantom{0}94.97 & $-$0.43 \\
9  & \phantom{0}26.37 & \phantom{0}26.90 & $-$0.53 \\
10 & \phantom{0}15.26 & \phantom{0}15.69 & $-$0.43 \\
\hline
11 & \phantom{00}8.35 & \phantom{00}8.24 & \phantom{$-$}0.11 \\
12 & 187.98           & 188.55           & $-$0.57 \\
13 & 152.12           & 152.89           & $-$0.77 \\
14 & 101.32           & 101.32           & \phantom{$-$}0.00 \\
15 & \phantom{0}33.78 & \phantom{0}33.60 & \phantom{$-$}0.18 \\
\hline
16 & \phantom{0}45.21 & \phantom{0}45.61 & $-$0.40 \\
17 & 116.45           & 116.03           & \phantom{$-$}0.42 \\
18 & \phantom{0}48.78 & \phantom{0}49.17 & $-$0.39 \\
19 & \phantom{0}19.98 & \phantom{0}19.50 & \phantom{$-$}0.48 \\
20 & \phantom{0}75.57 & \phantom{0}75.26 & \phantom{$-$}0.31 \\
\hline
21 & \phantom{00}4.35 & \phantom{00}4.39 & $-$0.04 \\
22 & \phantom{00}7.60 & \phantom{00}7.54 & \phantom{$-$}0.06 \\
23 & \phantom{0}21.86 & \phantom{0}21.55 & \phantom{$-$}0.31 \\
24 & \phantom{0}49.03 & \phantom{0}48.79 & \phantom{$-$}0.24 \\
25 & 111.93           & 111.79           & \phantom{$-$}0.14 \\
\hline
26 & \phantom{0}69.33 & \phantom{0}69.07 & \phantom{$-$}0.46 \\
27 & 267.43           & 266.99           & \phantom{$-$}0.44 \\
28 & 130.53           & 130.24           & \phantom{$-$}0.29 \\
29 & \phantom{0}53.28 & \phantom{0}53.43 & $-$0.15 \\
30 & \phantom{0}38.92 & \phantom{0}38.35 & \phantom{$-$}0.57 \\
\hline
31 & \phantom{0}62.62 & \phantom{0}62.97 & $-$0.35 \\
32 & \phantom{0}25.76 & \phantom{0}26.50 & $-$0.74 \\
33 & \phantom{0}33.61 & \phantom{0}33.34 & \phantom{$-$}0.27 \\
34 & \phantom{0}27.79 & \phantom{0}28.33 & $-$0.54 \\
\end{longtable}
\end{center}

\begin{center}
\begin{longtable}{P{2.0cm}P{2.0cm}P{2.0cm}P{2.0cm}P{2.0cm}P{2.0cm}P{2.0cm}}
\caption{\label{tab:iso34-2}
Individual contributions to the ISO34 benchmark set isomerization energies: Hartree-Fock (HF), MP2 (including the HF 
contribution), $\Delta(\mbox{CCSD})=E_{\mathrm{CCSD}}-E_{\mathrm{MP2}}$, and 
$\Delta(\mbox{T})=E_{\mathrm{CCSD(T)}}-E_{\mathrm{CCSD}}$ (see the previous Table for details of the calculations).
All results are given in kJ/mol.} \\
\hline\hline
reaction & HF & MP2 & \multicolumn{2}{c}{$\Delta(\mbox{CCSD})$} & \multicolumn{2}{c}{$\Delta(\mbox{T})$} \\
\hline
         &    &     & exact & RR & exact & RR \\
\hline 
\endfirsthead

\multicolumn{3}{c}%
{{\tablename\ \thetable{} -- continued from previous page}} \\
\hline\hline
reaction & HF & MP2 & \multicolumn{2}{c}{$\Delta(\mbox{CCSD})$} & \multicolumn{2}{c}{$\Delta(\mbox{T})$} \\
\hline
         &    &     & exact & RR & exact & RR \\
\hline 
\endhead

\hline \multicolumn{3}{|r|}{{Continued on next page}} \\ \hline
\endfoot

\hline \hline
\endlastfoot

1  & \phantom{00}6.83 & \phantom{0}19.24 & $-$12.98 & $-$12.95 & $-$1.16 & $-$1.06 \\
2  & 112.75           & \phantom{0}99.70 & $-$0.11 & $-$0.07 & $-$1.33 & $-$0.87 \\
3  & \phantom{0}40.28 & \phantom{0}20.84 & \phantom{$-$}9.65 & \phantom{$-$}9.48 & \phantom{$-$}1.15 & \phantom{$-$}1.09 
\\
4  & \phantom{00}6.69 & \phantom{00}4.64 & \phantom{$-$}0.26 & \phantom{$-$}0.26 & $-$0.24 & $-$0.19 \\
5  & \phantom{00}1.93 & \phantom{00}5.63 & $-$1.44 & $-$1.42 & \phantom{$-$}0.55 & \phantom{$-$}0.48 \\
\hline
6  & \phantom{00}9.92 & \phantom{0}10.79 & $-$0.94 & $-$1.04 & 0.09 & \phantom{$-$}0.04 \\
7  & \phantom{0}64.16 & \phantom{0}38.88 & \phantom{$-$}7.96 & \phantom{$-$}7.72 & 1.11 & \phantom{$-$}1.16 \\
8  & \phantom{0}90.44 & \phantom{0}93.43 & \phantom{$-$}1.55 & \phantom{$-$}1.58 & 0.00 & $-$0.50 \\
9  & \phantom{0}25.30 & \phantom{0}29.35 & $-$4.11 & $-$4.32 & 1.66 & \phantom{$-$}1.35 \\
10 & \phantom{00}0.98 & \phantom{0}19.74 & $-$6.53 & $-$6.61 & 2.47 & \phantom{$-$}2.13 \\
\hline
11 & $-$45.88         & \phantom{0}17.60 & $-$18.29 & $-$17.59 & ??? & ??? \\
12 & 233.49           & 198.88           & $-$10.50 & $-$10.87 & 0.17 & $-$0.03 \\
13 & 158.40           & 177.18           & $-$28.02 & $-$28.21 & 3.73 & \phantom{$-$}3.16 \\
14 & \phantom{0}82.94 & 112.50           & $-$12.68 & $-$12.60 & 1.39 & \phantom{$-$}1.42 \\
15 & \phantom{0}30.26 & \phantom{0}34.61 & $-$1.17 & $-$1.12 & 0.16 & \phantom{$-$}0.28 \\
\hline
16 & \phantom{0}55.99 & \phantom{0}35.93 & \phantom{$-$}8.60 & \phantom{$-$}8.46 & \phantom{$-$}1.08 & \phantom{$-$}0.82 
\\
17 & 112.23           & 121.29           & $-$4.78 & $-$4.72 & $-$0.48 & $-$0.14 \\
18 & \phantom{0}54.71 & \phantom{0}48.74 & \phantom{$-$}1.19 & \phantom{$-$}1.18 & $-$0.75 & $-$1.14 \\
19 & \phantom{0}28.30 & \phantom{0}15.42 & \phantom{$-$}4.89 & \phantom{$-$}4.90 & $-$0.82 & $-$0.34 \\
20 & \phantom{0}86.51 & \phantom{0}76.81 & \phantom{$-$}1.09 & \phantom{$-$}1.20 & $-$2.64 & $-$2.43 \\
\hline
21 & \phantom{00}5.03 & \phantom{00}4.63 & $-$0.24 & $-$0.26 & \phantom{$-$}0.00 & $-$0.03 \\
22 & \phantom{00}5.34 & \phantom{0}11.92 & $-$3.14 & $-$3.05 & $-$1.25 & $-$1.28 \\
23 & \phantom{0}19.37 & \phantom{0}23.30 & $-$1.34 & $-$1.25 & $-$0.41 & $-$0.20 \\
24 & \phantom{0}41.90 & \phantom{0}51.79 & $-$3.00 & $-$2.92 & \phantom{$-$}0.00 & \phantom{$-$}0.16 \\
25 & 128.97           & 107.59           & \phantom{$-$}4.70 & \phantom{$-$}4.68 & $-$0.50 & $-$0.34 \\
\hline
26 & \phantom{0}67.98 & \phantom{0}72.46 & $-$2.39 & $-$2.18 & $-$1.00 & $-$0.95 \\
27 & 291.89           & 282.25           & $-$7.64 & $-$8.23 & $-$7.62 & $-$7.59 \\
28 & 144.04           & 131.81           & $-$2.78 & $-$2.83 & \phantom{$-$}1.21 & \phantom{$-$}1.55 \\
29 & \phantom{0}47.93 & \phantom{0}59.75 & $-$3.54 & $-$3.30 & $-$2.78 & $-$3.17 \\
30 & \phantom{0}46.35 & \phantom{0}37.92 & \phantom{$-$}2.36 & \phantom{$-$}2.52 & $-$1.93 & $-$1.53 \\
\hline
31 & \phantom{0}51.19 & \phantom{0}70.76 & $-$6.61 & $-$6.41 & $-$1.18 & $-$1.43 \\
32 & \phantom{0}15.20 & \phantom{0}30.48 & $-$3.11 & $-$3.65 & $-$0.87 & $-$1.07 \\
33 & \phantom{0}44.13 & \phantom{0}33.58 & \phantom{$-$}3.65 & \phantom{$-$}3.87 & $-$3.89 & $-$3.84 \\
34 & \phantom{0}19.32 & \phantom{0}30.52 & $-$4.19 & $-$4.24 & \phantom{$-$}2.00 & \phantom{$-$}1.52 \\
\end{longtable}
\end{center}

\newpage

\section{\label{sec:cart-phenol} Isomers of fluorophenol: molecular structures in Cartesian 
coordinates}

Attached to this document as \texttt{fluorophenol.tar}. The \texttt{xyz} files 
included in the tarball are named \texttt{subst-fph-ang.xyz}, where \texttt{subst} stands for 
ortho/meta/para substitution pattern and \texttt{ang} denotes the value of the torsional angle 
$\tau$ in degrees ($\tau=0,15^\circ,\dots,180^\circ$). All coordinates are given in 
\AA{}ngstr\"{o}ms.

\section{\label{sec:torsional-phenol} Raw torsional energies for three isomers of fluorophenol}

\renewcommand{\arraystretch}{1.00}
\begin{table}[h!]
\caption{\label{tab:torsional-phenol}
Torsional energies for the \emph{ortho/meta/para} isomers of fluorophenol computed 
using the RR-CCSD(T)/cc-pVTZ method (``RR'') and the exact CCSD(T)/cc-pVTZ method (``exact''). For each isomer 
relative energies with respect to its $\tau=0$ conformation are given. The torsional angle $\tau$ 
is given in degrees and the energies~in~kJ/mol.
}
\begin{ruledtabular}
\begin{tabular}{lcccccc}
  & 
 \multicolumn{2}{c}{\emph{ortho}} & 
 \multicolumn{2}{c}{\emph{meta}}  &
 \multicolumn{2}{c}{\emph{para}} \\\cline{2-3}\cline{4-5}\cline{6-7}
 $\tau$ & RR & exact & RR & exact & RR & exact \\
\hline
 15  & \phantom{0}1.27 & \phantom{0}1.28 & 0.89 & \phantom{0}0.88 & 
 \phantom{0}0.71 & \phantom{0}0.72 \\
 30  & \phantom{0}4.73 & \phantom{0}4.73 & 3.36 & \phantom{0}3.35 & 
 \phantom{0}2.69 & \phantom{0}2.72 \\
 45  & \phantom{0}9.42 & \phantom{0}9.46 & 6.85 & \phantom{0}6.84 & 
 \phantom{0}5.47 & \phantom{0}5.52 \\
 60  & 14.29 & 14.39 & 10.51 & 10.54 & 8.33 & \phantom{0}8.45 \\
 75  & 18.42 & 18.59 & 13.39 & 13.48 & 10.57 & 10.76 \\
 90  & 21.02 & 21.23 & 14.62 & 14.74 & 11.55 & 11.76 \\
 105 & 21.56 & 21.77 & 13.71 & 13.83 & 10.57 & 10.76 \\
 120 & 20.17 & 20.34 & 10.97 & 11.03 & 8.33 & 8.45 \\
 135 & 17.53 & 17.64 & \phantom{0}7.16 & \phantom{0}7.18 & \phantom{0}5.47 & \phantom{0}5.52 \\
 150 & 14.55 & 14.62 & \phantom{0}3.31 & \phantom{0}3.32 & \phantom{0}2.69 & \phantom{0}2.72 \\
 165 & 12.24 & 12.27 & \phantom{0}0.49 & \phantom{0}0.50 & \phantom{0}0.71 & \phantom{0}0.72 \\
 180 & 11.36 & 11.40 & $-$0.53 & $-$0.53 & \phantom{0}0.00 & \phantom{0}0.00 \\
\end{tabular}
\end{ruledtabular}
\end{table}